%% file: sdme-v302.tex
\documentclass[epj]{svjour}
\usepackage{graphics}
\usepackage{amssymb}
\usepackage{epsfig}
\usepackage{axodraw}
\usepackage{amsmath}
\usepackage{color}          
\usepackage{rotating}

\newcommand{\bdm}{\begin{displaymath}}
\newcommand{\edm}{\end{displaymath}}

\begin{document}

\hugehead

\title{Spin Density Matrix Elements in Exclusive $\rho^0$
Electro\-production on $^1$H and $^2$H Targets at 27.5 GeV Beam~Energy}

\author{ 
The HERMES Collaboration \medskip \\
A.~Airapetian,$^{16}$
N.~Akopov,$^{27}$
Z.~Akopov,$^{27}$
A.~Andrus,$^{15}$
E.C.~Aschenauer,$^{7}$
W.~Augustyniak,$^{26}$
R.~Avakian,$^{27}$
A.~Avetissian,$^{27}$
E.~Avetissian,$^{11}$
S.~Belostotski,$^{19}$
% ...
}

\institute{ 
$^1$Physics Division, Argonne National Laboratory, Argonne, Illinois 60439-4843, USA\\
$^2$Istituto Nazionale di Fisica Nucleare, Sezione di Bari, 70124 Bari, Italy\\
$^3$School of Physics, Peking University, Beijing 100871, China\\
$^4$Department of Modern Physics, University of Science and Technology of China, Hefei, Anhui 230026, China\\
$^5$Nuclear Physics Laboratory, University of Colorado, Boulder, Colorado 80309-0390, USA\\
$^6$DESY, 22603 Hamburg, Germany\\
$^7$DESY, 15738 Zeuthen, Germany\\
$^8$Joint Institute for Nuclear Research, 141980 Dubna, Russia\\
$^9$Physikalisches Institut, Universit\"at Erlangen-N\"urnberg, 91058 Erlangen, Germany\\
$^{10}$Istituto Nazionale di Fisica Nucleare, Sezione di Ferrara and Dipartimento di Fisica, Universit\`a di Ferrara, 44100 Ferrara, Italy\\
$^{11}$Istituto Nazionale di Fisica Nucleare, Laboratori Nazionali di Frascati, 00044 Frascati, Italy\\
$^{12}$Department of Subatomic and Radiation Physics, University of Gent, 9000 Gent, Belgium\\
$^{13}$Physikalisches Institut, Universit\"at Gie{\ss}en, 35392 Gie{\ss}en, Germany\\
$^{14}$Department of Physics and Astronomy, University of Glasgow, Glasgow G12 8QQ, United Kingdom\\
$^{15}$Department of Physics, University of Illinois, Urbana, Illinois 61801-3080, USA\\
$^{16}$Randall Laboratory of Physics, University of Michigan, Ann Arbor, Michigan 48109-1040, USA \\
$^{17}$Lebedev Physical Institute, 117924 Moscow, Russia\\
$^{18}$National Institute for Subatomic Physics (Nikhef), 1009 DB Amsterdam, The Netherlands\\
$^{19}$Petersburg Nuclear Physics Institute, Gatchina, Leningrad region,
188300 Russia\\
$^{20}$Institute for High Energy Physics, Protvino, Moscow region, 142281 Russia\\
$^{21}$Institut f\"ur Theoretische Physik, Universit\"at Regensburg, 93040 Regensburg, Germany\\
$^{22}$Istituto Nazionale di Fisica Nucleare, Sezione Roma 1, Gruppo Sanit\`a and Physics Laboratory, Istituto Superiore di Sanit\`a, 00161 Roma, Italy\\
$^{23}$TRIUMF, Vancouver, British Columbia V6T 2A3, Canada\\
$^{24}$Department of Physics, Tokyo Institute of Technology, Tokyo 152, Japan\\
$^{25}$Department of Physics, Vrije Universiteit, 1081 HV Amsterdam, The Netherlands\\
$^{26}$Andrzej Soltan Institute for Nuclear Studies, 00-689 Warsaw, Poland\\
$^{27}$Yerevan Physics Institute, 375036 Yerevan, Armenia\\
} 

\date{Received: January 15, 2009 / Revised version: \today }


\abstract{
Spin Density Matrix Elements (SDMEs) describing the angular distribution
of exclusive $\rho^0$ electroproduction and decay are determined
in the HERMES experiment with 27.6 GeV beam
energy and unpolarized hydrogen and deuterium targets. 
Eight (fifteen) SDMEs that are related (unrelated) to the
longitudinal polarization of the beam 
are extracted
in the kinematic region 
$1~\rm{GeV}^2 < Q^2 < 7$~GeV$^2$, $3.0~{\rm GeV} < W < 6.3$~GeV, and 
 $-t < 0.4$ GeV$^2$.
Within the given experimental uncertainties, 
a hierarchy of relative sizes of helicity amplitudes is observed.
Kinematic dependences of all SDMEs on $Q^2$ and $t$ 
are presented, as well as 
the longitudinal-to-transverse $\rho^0$ electroproduction cross section ratio 
as a function of $Q^2$.
A small 
but statistically significant deviation from the hypothesis of
$s$-channel helicity conservation is observed. An indication is seen of a 
contribution of unnatural-parity-exchange amplitudes; these amplitudes are 
naturally generated with a quark-exchange mechanism. 
}

\PACS{
{13.60.-r,13.60.Le,13.88.+e}{}
}

\maketitle



\section{Introduction}

\begin{figure}
  \begin{center}
\epsfig{file=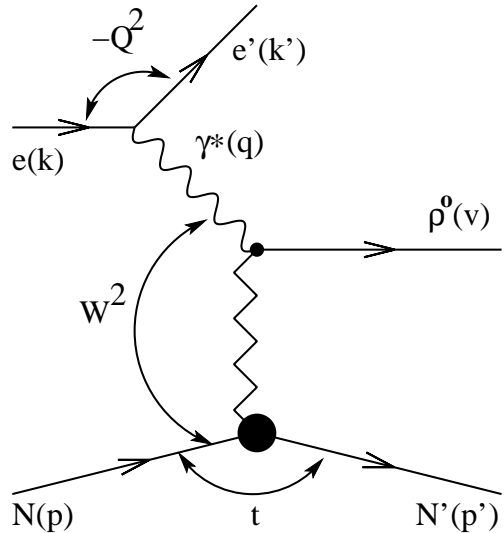,width=0.8\linewidth} \\
\caption{A generic $t$-channel exchange process for 
$\gamma^\star N \rightarrow \rho^0 N'$. Each particle's
four-momentum is denoted in parentheses.
} 
\label{proc}
\end{center}
\end{figure}

In exclusive production of  vector mesons such as
$\rho$, $\omega$ or $\phi$ from deep-inelastic
lepton scattering (see Fig.~\ref{proc}),  
measurements of angular and momentum distributions
of the scattered lepton and vector meson decay products allow
one to study the production mechanism and, in a
model-dependent way, the structure of the nucleon.

\begin{figure}[hbtc!]
  \begin{center}
\epsfig{file=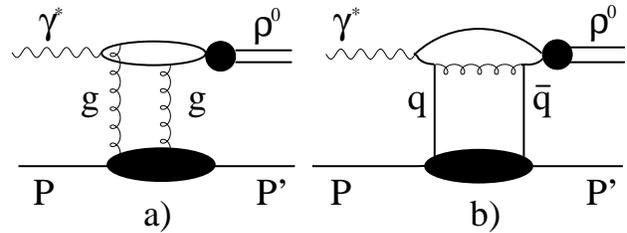,width=8.2cm}
  \caption{Examples of
a) a two-gluon exchange diagram and b) a quark-exchange diagram, shown
for the lowest order in the strong coupling constant $\alpha_s$.
 }
\label{diagram}
\end{center}
\end{figure}

For more than 40 years, many basic features of vector meson production
by a virtual photon have been successfully explained in terms of the
Vector Meson Dominance (VMD) model~\cite{vmd1,vmd2}.
In this model, the virtual photon fluctuates into a vector meson
whose interaction with the nucleon could be described, for example, using  
Regge phenomenology.
More recently, in the context of perturbative Quantum Chromo-Dynamics (pQCD), 
exclusive
meson production at sufficiently large values of the photon virtuality $Q^2$
and the invariant mass of the photon-nucleon system $W$
is assumed to be dominated by so-called handbag-diagrams
(see Fig.~\ref{diagram}) that involve various non-perturbative nucleon
structure functions,
known as Generalized Parton Distributions 
(GPDs)~\cite{gpd1,gpd2,gpd3,strikman}.
\begin{sloppypar}
In pQCD, the common model of the 
production of vector mesons at high $Q^2$ and $W$ can be considered
as three consecutive steps~\cite{brodsk}:
i) dissociation of the virtual photon into a quark-antiquark ($q \bar q$) pair, 
ii) scattering of the pair on a nucleon (nucleus), iii) formation of the 
observed 
vector meson from the $q \bar q$-pair. (A full quantum mechanical 
treatment includes all possible time orderings, which 
may be more important at lower energies.) 
The interaction of the  $q \bar q$-pair
with the nucleon can proceed via two distinct mechanisms.
The first one, two-gluon exchange, 
is described by the Feynman
diagram shown in Fig.~\ref{diagram}a. 
This process transfers the same quantum numbers
as pomeron exchange in the Regge picture, and is anticipated to exhibit a
similar phenomenology.
The second mechanism is described by  the exchange of a
$q \bar q$-pair,  also  possibly  with  additional gluons
connecting them, and is called quark-exchange (Fig.~\ref{diagram}b).
The corresponding process in
Regge phenomenology~\cite{iwing} is  the exchange of ``secondary'' 
reggeons, 
such as  $\rho$, $\omega$, $f_2$ and  $a_2$ in the 
case of natural-parity exchange (NPE), in which 
the spin $J$ and parity $P$ associated with the reggeon
trajectory are $J^P=0^+,1^-,2^+,...$,
or $\pi$, $a_1$, $b_1$ mesons with $J^P=0^-,1^+,...$ 
in the case of ``unnatu\-ral-parity'' exchange (UPE). 
In the GPD formalism, NPE (UPE) processes are described by  
$H$ and $E$  ($\widetilde H$ and $\widetilde E$)  GPDs.
In the intermediate energy range of the HERMES experiment 
($3~\rm{GeV} < W < 6$~GeV) and
the moderate values of photon virtuality ($1~\rm{GeV}^2 < Q^2 < 7$~GeV$^2$)
both Regge phenomenology and pQCD may be applied to describe 
exclusive vector meson production.
The interpretations they offer of the experimental data are
often complementary, although not necessarily consistent.
\end{sloppypar}

\begin{sloppypar}
The main focus of this work is on the measurement of Spin Density Matrix 
Elements
(SDMEs) of the $\rho^0$  meson, which describe the distribution
of final spin states of this produced vector meson. 
These elements
depend on amplitudes for the angle- and momentum-dependent
transition processes between 
initial spin states of the virtual photon
and final spin states of the produced vector meson.
The values of SDMEs serve to establish the hierarchy
of helicity amplitudes that are commonly used to describe exclusive $\rho^0$
production. In this way the relative importance of the various 
$\gamma^* \to \rho^0$ transitions is revealed. 
Two main ordering principles are observed in vector meson
leptoproduction, $s$-channel helicity conservation (SCHC) and the dominance of
NPE over UPE mechanisms. 
SCHC implies that 
only $\gamma^* \rightarrow \rho^0$ transitions
with the same helicities of virtual photon and $\rho^0$ occur in the reaction
when considered in the ``hadronic'' center-of-mass frame (defined below).
These concepts apply both in the 
reggeon-exchange 
picture and in pQCD.
In particular, we note that a 
signal of UPE  is evidence of quark-antiquark exchange (Fig.~\ref{diagram}b), 
as the pomeron
has natural parity.
\end{sloppypar}

At high energies pome\-ron exchange dominates, and secondary-reggeon 
exchanges with
natural parity are suppressed by a 
factor $\sim M/W$~\cite{iwing} in their amplitudes;
$M$ is an energy scale in Regge phenomenology chosen to be equal to
the nucleon mass.
Also suppressed, by a factor $\sim (M/W)^2$~\cite{iwing}, are
the most important unnat\-ural-parity exchanges mediated 
by $\pi$, $a_1$, and $b_1$ reggeons.
Therefore  substantial  UPE contributions can be expected  only
at lower values of $W$.

In the pQCD framework, the  leading-twist contribution describes the transition of
longitudinal photons to longitudinal vector mesons, which is $s$-chan\-nel
helicity conser\-ving and corresponds to natural-pari\-ty exchange.
As it is not agreed how strongly the various other contributions are
suppressed at a given energy, measurements of SDMEs
in the HERMES kinematics help to distinguish these contributions
and are of particular interest. 
Non-conservation of $s$-channel helicity in exclusive $\rho^0$ production 
was already observed at collider energies~\cite{zeussdme,zeussdme2,H1}. 
At lower energies it was observed at HERMES~\cite{tytgat}, and
for exclusive $\omega$ production at CLAS~\cite{clasomega}.

At sufficiently large values of $W$, experiments are typically sensitive to 
partons that carry 
small nucleon momentum fraction $x$, 
where the parton density in the nucleon is dominated by gluons.   
High-energy data of H1 and ZEUS~\cite{zeussdme,zeussdme2,H1,ipivanov} are well 
described by 
two-gluon exchange. At lower values of $W$, larger values of $x$ 
are probed, where
the parton density in the nucleon receives 
significant contributions from quarks.  
Indeed, a contribution from the quark-exchange mechanism has been suggested 
to be necessary to
describe exclusive $\rho^0$ production at intermediate virtual-photon 
energies,  
as in the case of the HERMES 
data~\cite{rho-xsec,sdme-publ,dsa-lipka,twopion} and corresponding 
calculations~\cite{vgg,lehm,kochelev,diehl}.

In leptoproduction, the spin transfer from the virtual photon to the vector meson 
is commonly described by helicity amplitudes, from which  
SDMEs can be constructed. 
The detection of the scattered lepton and the vector meson decay products 
allows one to reconstruct the full reaction kinematics and the three-dimensional 
angular distribution of the production and decay of the $\rho^0$ meson.
For an unpolarized or helicity-balanced lepton beam, 
the expression for
this distribution contains a set of ``unpolarized'' SDMEs as coefficients.
An additional set of ``polarized'' 
SDMEs, which appear in products with the beam polarization in the
expression for the angular distribution with polarized beam,
can be determined if 
information on the longitudinal polarization of the lepton beam is 
available~\cite{wolf,fraas}.
In a very recent new classification scheme of  SDMEs~\cite{mdiehl}, also the 
cases of longitudinal and transverse target polarizations are described.
However,
the analysis in this paper follows  the representation introduced in 
Ref.~\cite{wolf}.

Early theoretical calculations~\cite{vmd2} 
of SDMEs in vector meson production were based on the VMD model.  
More recent calculations combining this model with pQCD
models~\cite{brodsk,divanov,royen,kuraev-theor,ryskin,ivanov-theor} and
with Regge phenomenology~\cite{manaenkov,laget}
are mainly focused on the high-energy kinematics of
the HERA collider data.
A contemporary account of the various theoretical approaches is given in 
Ref.~\cite{ipivanov}.
Recent model calculations based on GPDs present SDMEs for both high
and intermediate energies, considering first only 
two-gluon exchange~\cite{golos},
and recently incorporating quark exchange~\cite{golos2,golos3}.

In this analysis, the beam polarization is used for the first time 
in an SDME extraction, 
thereby making possible the determination of the additional 8 polarized SDMEs. 
The  high-statistics data samples accumulated at HERMES in the years 1996--2005
on both  hydrogen and deuterium targets are used 
to determine $\rho^0$ decay angle distributions with an
accuracy superior to that of the previously published HERMES $^3$He data 
from 1995~\cite{sdme-publ} and 
of the preliminary HERMES results from hydrogen data collected in 
1996--1997~\cite{tytgat,abb-spin01}. 
The improved statistical accuracy permits the study of the nature of the exchange 
mechanism, and  in particular the testing of the  hypothesis of $s$-channel helicity 
conservation.
The availability of both hydrogen and deuterium targets offers 
the possibility to search for significant contributions of secondary reggeon 
exchange with 
isospin $I=1$ and natural parity.

\begin{sloppypar}
The structure of this paper is as follows.
The kinematics,  SDME formalism, and
HERMES experiment are described in the next three sections.
The analysis procedure including event selection 
and background subtraction is discussed in section 5. 
The extraction of the SDMEs from the data using 
a Monte Carlo based maximum likelihood method is described in section 6.
The experimental results on SDMEs integrated over the entire
observed kinematic region are 
presented in section 7, and their kinematic dependences 
are shown in  section 8.
An indication of the contribution of unnatural-parity-exchange 
amplitudes is discussed in section 9. 
Contributions of helicity-flip and UPE amplitudes to the   
cross section are estimated in section 10.
The ratio of longitudinal to transverse $\rho^0$ 
electroproduction cross-sections  is presented in section 11. 
The results are summarized in section 12.
\end{sloppypar}

\section{Kinematics}

Figure~\ref{proc} identifies 
the kinematic variables of $\rho^0$ leptoproduction,
\begin{equation}
\gamma^* +N \rightarrow \rho ^0 + N',
\label{exclus}
\end{equation}
where $N(N')$ denotes the initial (scattered) nucleon.
The four-momenta of the incoming and outgoing lepton are denoted by $k$ and
$k^{\prime}$, the difference of which defines the four-momentum $q$
of the virtual photon $\gamma^*$. 
In the  laboratory ($lab$) frame, $\vartheta$ is the scattering angle between 
the incoming and outgoing lepton, whose incoming and outgoing  energies are denoted by
$E$  and $E^{\prime}$.
The photon virtuality is given by:
\begin{equation}
Q^2 = -q^2=-(k-k^{\prime})^2\stackrel{lab}{\approx}4\,E\,
E^{\prime}\sin^2 \frac{\vartheta}{2},
\end{equation}
which is positive in leptoproduction. In this equation the 
electron rest mass is neglected.
The four-momenta of the target nucleon and of the recoiling baryon are 
denoted by $p$ and $p^{\prime}$, respectively, and both have rest mass $M$ 
of the nucleon, irrespective of target. 

The Bjorken scaling variable $x_B$ is defined as\footnote{ 
This kinematic observable is to be distinguished from the variable
$x$ of the quark parton model, which represents in the GPD formalism
the average longitudinal momentum fraction of the probed parton in the initial 
and final states.}
\begin{equation}
x_B = \frac{Q^2}{2\,p\cdot q}  =\frac{Q^2}{2\,M\,\nu},
\end{equation}
with
\begin{equation}
\nu = \frac{p\cdot q}{M}\stackrel{lab}{=}E-E^{\prime},
\end{equation}
so that $\nu$ represents the energy transfer from
the incoming lepton to the virtual photon in the laboratory frame. 
The squared invariant mass
of the photon-nucleon system is given by:
\begin{equation} \label{wdef}
W^2=(q+p)^2 = M^2+2\,M\,\nu-Q^2.
\end{equation}
The squared four-momentum  transfer 
from virtual photon to vector meson equals that between the momenta of
the initial and final nucleons or nuclei,
\begin{equation}\label{eqt}
t=(q-v)^2 =  (p - p')^2,
\end{equation}
where $v$ is the four-momentum of the vector meson.
The variables $t$, $t_0$, and 
\begin{equation}\label{eqndeftpr}
t'=t-t_0
\end{equation}
are always negative, where  $-t_0$ represents  
the smallest kinematically allowed value of $-t$ at 
fixed $\nu$ and $Q^2$.
In the photon-nucleon center-of-mass frame considered here, 
the condition  $t = t_0$
corresponds to the case where the momentum of the produced $\rho^0$ is 
collinear with that of the $\gamma^*$.
Typically for exclusive processes at intermediate and high energies, 
$|t_0|$ is much smaller than $|t|$ and therefore $t'\approx t$.

At very low $t$, the approximation 
$-t' \approx v_T^2$ holds, where $v_T$ is the  transverse momentum of the  vector meson
with respect to the direction of the virtual photon, {\it i.e.,} 
the subtraction of  $t_0$ removes the contribution of the longitudinal 
component of the momentum transfer. 

\pagebreak
The variable $\epsilon$ 
represents the  ratio of fluxes of 
longitudinal and transverse virtual photons:
\begin{eqnarray}\label{expreps}
\epsilon &=& \frac{1-y - y^2\frac{Q^2}{4\nu^2}}{1-y+ \frac{1}{4}y^2
(\frac{Q^2}{\nu^2} + 2)} \\
&\stackrel{lab}{\approx}&
\frac{1}{1+2(1+\frac{\nu^2}{Q^2})\tan^2\frac{\vartheta}{2}}\nonumber
\end{eqnarray}
with $y = p\cdot q / p\cdot k   \stackrel{lab}{=} \nu / E$.

\begin{figure}
  \begin{center}
\epsfig{file=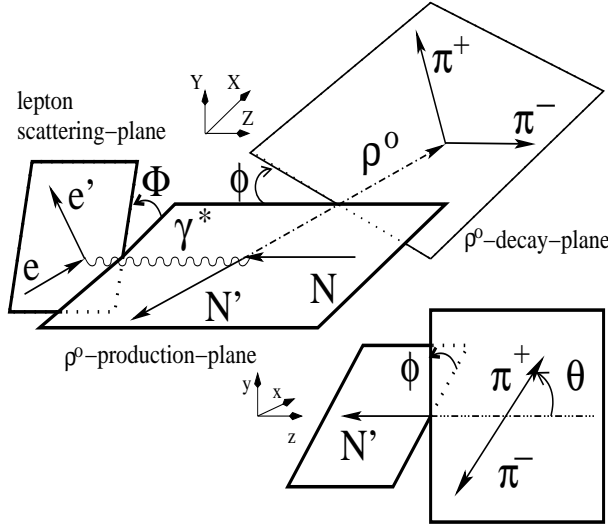,width=8.0cm,height=7.cm}
\vspace{0.5cm}
\caption{ {\small
Definition of angles in the process 
$\gamma^\star N \rightarrow \rho^o N' \rightarrow \pi^+ \pi^- N'$~\cite{DESY2}. 
Here $\Phi$ is the angle between the $\rho^0$ production plane 
and the lepton scattering plane  in the ``hadronic'' center-of-mass system 
of virtual photon and target nucleon.
$\Theta$ and $\phi$ are polar and azimuthal angles of the decay $\pi^+$ in the 
vector meson rest frame.
}}
\label{hellen}
\end{center}
\end{figure}

The ``exclusivity'' of $\rho^0$ production is characterized by the variable 
\begin{equation}\label{deltae}
\Delta E = \frac{M_{X}^{2} -M^{2}}{2 M } \stackrel{lab} = E_V - (E_{\pi^+}+E_{\pi^-}),
\end{equation}
where $M_X$ is the invariant mass of  the recoiling  system,
$E_V= \nu + t/(2 M )$ is the energy of 
the exclusively  produced  $\rho ^0$  meson,  and
 $(E_{\pi^+}+E_{\pi^-})$ is the sum of the energies of
the two pions.
For exclusive vector meson production  $(\ref{exclus})$, $M_X =M$ holds and
hence $\Delta E=0$, given perfect detector and beam energy resolution. 

\begin{sloppypar}
Angles used for the description of the
process $\gamma^\star N \rightarrow \rho^o N'   \rightarrow \pi^+ \pi^- N'$
are defined according to Ref.~\cite{DESY2} and presented in Fig.~\ref{hellen}. 
The helicity amplitudes are defined
in the ``hadronic'' center-of-mass system of virtual photon and target nucleon,  
where the $Z$-axis is directed along the virtual photon 
three-momentum $\vec{q}$. The $Y$-axis 
of the right-handed  system is parallel to $\vec{q} \times \vec{v}$.
It is the normal to the $\rho^0$ production plane spanned by the
three-momenta $\vec{q}$ and $\vec{v}$, of the virtual photon  and $\rho^0$-meson,
respectively.    The angle  $\Phi$   
between the $\rho^0$-production plane 
and the lepton-scattering plane in the ``hadronic'' center-of-mass system 
is specified by: 
\begin{eqnarray} 
\cos \Phi = \frac{ (\vec{q} \times \vec{v}) \cdot (\vec{k} \times \vec{k}') }
{ | \vec{q} \times \vec{v} | \cdot |\vec{k} \times \vec{k}'| } ,  \\ \nonumber
\sin \Phi = 
 \frac{ [ (\vec{q} \times \vec{v} )\times (\vec{k} \times \vec{k}' )] \cdot \vec{q} }
 { |\vec{q} \times \vec{v}| \cdot |\vec{k} \times \vec{k}'| \cdot |\vec{q}| }.
\label{phicap-def}
\end{eqnarray}
\end{sloppypar}

The angle  $\phi$
between the $\rho^0$-production plane  and  $\rho^0$-decay plane  is defined by:
\begin{eqnarray} 
\cos \phi = 
\frac{ (\vec{q} \times \vec{v} )\cdot (\vec{v} \times \vec{p}_{\pi^+} ) }
{ | \vec{q} \times \vec{v}| \cdot |\vec{v} \times \vec{p}_{\pi^+} | },  \\ \nonumber 
 \sin \phi = 
 \frac{[ (\vec{q} \times \vec{v} )\times \vec{v} ] \cdot ( \vec{p}_{\pi^+}  \times \vec{v} ) } 
{ | (\vec{q} \times \vec{v} )\times \vec{v} | \cdot |\vec{p}_{\pi^+} \times \vec{v} | },
\label{phismall-def}
\end{eqnarray} 
where $\vec{p}_{\pi^+}$ is the three-momentum of the positive decay pion in the ``hadronic''
center-of-mass system.

The polar angle  $\Theta$ of the decay $\pi^+$ in the 
vector meson rest frame, with the $z$-axis aligned opposite to the outgoing nucleon
momentum $\vec{P}'$  and the $y$-axis parallel to $Y$ and directed along
$\vec{P}' \times \vec{q}$, is defined by:
\begin{eqnarray} 
\cos \Theta  = \frac{ -\vec{P}' \cdot \vec{P}_{\pi^+} }{| \vec{P}' | 
\cdot |\vec{P}_{\pi^+}|},\\ \nonumber 
\label{theta-def}
\end{eqnarray} 
where $\vec{ P}_{\pi^+}$ is the three-momentum of the positive decay pion.

Note that the relation between this notation and 
the notations of the so-called 
``Trento convention''~\cite{trento} and Ref.~\cite{mdiehl} is: 
$\Phi = -\phi_{\rm [25]} = -\phi_{h{\rm [38]}}$,  
$\phi = \varphi_{\rm [25]}$,
$\Theta  = \vartheta_{\rm [25]}$.


\section{Formalism}

\subsection{Helicity Amplitudes} 

Exclusive vector meson leptoproduction $(\ref{exclus})$ is 
commonly described by
helicity amplitudes 
$F_{\lambda_{V}\lambda '_N;\lambda_{\gamma}\lambda _N}$,
defined in the ``hadronic'' center-of-mass system 
of virtual photon and target nucleon \cite{wolf}
(see Fig.~\ref{hellen}). 
Helicity indices $\lambda_{\gamma}$ and $\lambda_{V}$ describe the spin states of 
virtual photon and $\rho$ meson, 
respectively, while $\lambda _N$  ($\lambda '_N$) is the helicity of the initial 
(scattered) nucleon. The helicity 
amplitude can be expressed 
as the scalar product of the matrix element of the electromagnetic 
current vector $J^{\mu}$ and the virtual-photon
polarization vector $e_{\mu}^{(\lambda_{\gamma})}$:
\begin{equation} 
F_{\lambda_{V} \lambda '_{N}; \lambda_{\gamma}  \lambda_{N} } = (-1)^{\lambda_{\gamma}}
\langle v \lambda_{V}; p' \lambda '_{N} | J ^{\mu} | p \lambda_{N} \rangle e_{\mu}^{(\lambda_{\gamma})},
\label{jacobwick}
\end{equation}
where $e_{\mu}^{(\pm 1)}$ describes the transverse and $e_{\mu}^{(0)}$ the longitudinal 
polarization of the virtual photon. The ket vector
$| p \lambda_{N} \rangle$ corresponds to the incident nucleon and the bra vector 
$\langle v \lambda_{V}; p' \lambda '_{N} |$ 
describes the final state of the $\rho ^0$ meson and scattered nucleon.
The amplitudes depend on $Q^2$, $W$ and $t$.
For convenience, these dependences may be omitted in the following.

The amplitudes obey the relation~\cite{wolf}
\begin{eqnarray}
\lefteqn{F_{-\lambda_{V} -\lambda '_{N}; -\lambda_{\gamma}  -\lambda_{N} } =}\hspace*{1.0cm} \nonumber \\
& & (-1)^{(\lambda_{V}-\lambda '_{N})- (\lambda_{\gamma}-\lambda _{N}) }
F_{\lambda_{V} \lambda '_{N}; \lambda_{\gamma}  \lambda_{N} }, 
 \label{parity}
\end{eqnarray}
which is a consequence of parity conservation in the strong and 
electromagnetic interactions.

\subsection{Natural and Unnatural-Parity-Exchange Amplitudes}

A helicity amplitude $F$ 
can be decomposed into an amplitude $T$ for natural-parity exchange 
and an amplitude $U$ for unnatural-parity exchange: 
\begin{equation}
F_{\lambda_{V} \lambda '_{N}; \lambda_{\gamma}  \lambda_{N} } =
T_{\lambda_{V} \lambda '_{N}; \lambda_{\gamma}  \lambda_{N} }+
U_{\lambda_{V} \lambda '_{N}; \lambda_{\gamma}  \lambda_{N} },
\label{decomp}
\end{equation}
with
\begin{eqnarray} 
T_{\lambda_{V} \lambda '_{N}; \lambda_{\gamma}  \lambda_{N} }= \nonumber\hspace*{3.5cm} \\   
\frac{1}{2} \bigl( F_{\lambda_{V} \lambda '_{N}; \lambda_{\gamma}  \lambda_{N} } 
+(-1)^{-\lambda_{V}+ \lambda_{\gamma} } 
F_{-\lambda_{V} \lambda '_{N}; -\lambda_{\gamma}  \lambda_{N} }\bigr ), 
 \label{defnpe} 
\end{eqnarray} 
\begin{eqnarray}
U_{\lambda_{V}\lambda '_{N};\lambda_{\gamma}\lambda_{N} }=\nonumber \hspace*{3.5cm}  \\    
\frac{1}{2} \bigl
(F_{\lambda_{V} \lambda '_{N}; \lambda_{\gamma}  \lambda_{N} }-(-1)^{-\lambda_{V}+ \lambda_{\gamma} }
F_{-\lambda_{V} \lambda '_{N}; -\lambda_{\gamma}  \lambda_{N} }\bigr ).
  \label{defupe}
\end{eqnarray}
From  definitions $(\ref{defnpe})$, $(\ref{defupe})$ and 
relation $(\ref{parity})$ 
the amplitudes $T$ and $U$  obey the symmetry relations~\cite{wolf}: 
\begin{eqnarray}
T_{\lambda_{V} \lambda'_{N};\lambda_{\gamma} \lambda_{N}} &=&  
(-1)^{-\lambda_{V}+\lambda_{\gamma}}
T_{-\lambda_{V} \lambda'_{N};-\lambda_{\gamma} \lambda_{N}}\nonumber \\ 
& = &(-1)^{\lambda'_{N}-\lambda_{N}}
T_{\lambda_{V} -\lambda'_{N};\lambda_{\gamma} -\lambda_{N}},
 \label{symmnat}
\end{eqnarray} 
\begin{eqnarray}
U_{\lambda_{V} \lambda '_{N};\lambda_{\gamma} \lambda _{N}} &=& 
-(-1)^{-\lambda_{V}+\lambda_{\gamma}}
U_{-\lambda_{V} \lambda'_{N};-\lambda_{\gamma} \lambda _{N}} \nonumber  \\ 
&=& -(-1)^{\lambda '_{N}-\lambda_{N}}
U_{\lambda_{V} -\lambda '_{N};\lambda_{\gamma} -\lambda _{N}}\;. 
  \label{symmunn}
\end{eqnarray}

\begin{sloppypar}
For convenience, we introduce the abbreviation  
$\widetilde {\sum} \equiv 
\frac{1}{2}{\sum}_{\lambda'_N \lambda_N}$ for the 
summation  over the  final 
nucleon  helicity indices and averaging over the initial spin states  
of the nucleon. In the following 
the nucleon 
helicity indices of the amplitudes are implicit, but will be included
when required for clarity.
If  $T_{\lambda _V \lambda _{\gamma}}$
appears without the symbol $\widetilde{\sum}$, 
all nucleon helicity indices are equal to $1/2$. 
\end{sloppypar}

\begin{sloppypar}
For NPE amplitudes, transitions diagonal in nucleon spin
($\lambda ' _N = \lambda _N $) are dominant. Furthermore,
since for scattering off an unpolarized target 
there is no interference between nucleon spin-flip and non-spin-flip amplitudes,
the fractional contribution of nucleon spin-flip NPE amplitudes
to SDMEs is of the order of 
$-t'/(4M^2)$, which is small at low $t'$.
In this case, 
neglecting the small nucleon spin-flip amplitudes 
$T_{\lambda _V \pm 1/2;\lambda _{\gamma} \mp 1/2}$
and using  (\ref{symmnat})  reduces 
the summation and averaging $\widetilde  {\sum}$ to one term:
\begin{eqnarray} \label{tsum}
\nonumber
\lefteqn{\widetilde  {\sum}T_{\lambda_V \lambda_{\gamma}}
T^*_{\lambda'_V \lambda'_{\gamma} }
\equiv  \frac{1}{2}{\sum_{\lambda_N \lambda'_N} }T_{\lambda_V \lambda'_N;
\lambda_{\gamma} \lambda_N}
T^*_{\lambda'_V \lambda'_N;\lambda'_{\gamma}\lambda_N}} \\
\nonumber
&=
&T_{\lambda_V 1/2;\lambda_{\gamma}1/2}T^*_{\lambda'_V 1/2;\lambda'_{\gamma}1/2} \\
\nonumber
&&\hspace*{1.2cm} +T_{\lambda_V -1/2;\lambda_{\gamma}1/2}T^*_{\lambda'_V -1/2;\lambda'_{\gamma}1/2} \\
&\approx& \rule{0cm}{14pt} T_{\lambda_V 1/2;\lambda_{\gamma}1/2}T^*_{\lambda'_V 1/2;\lambda'_{\gamma}1/2}
\equiv  T_{\lambda_V \lambda_{\gamma}}T^*_{\lambda'_V \lambda'_{\gamma}},
\end{eqnarray}
where $T^*$ represents the complex conjugate quantity. 
\end{sloppypar}

For  UPE amplitudes in general,  the dominance of diagonal transitions 
($\lambda_N = \lambda_N'$) cannot be proven, so that
no relation similar to (\ref{tsum}) can be derived and
therefore  $\widetilde {\sum}$  is always used.

For unpolarized targets, there is no 
interference between NPE and UPE amplitudes 
\cite{wolf} as
\begin{equation} \label{tusum}
\widetilde{\sum} T_{\lambda _V \lambda _{\gamma} }
U^*_{\lambda '_V \lambda '_{\gamma} }=0\, ,
\end{equation}
following from
relations (\ref{symmnat}) and (\ref{symmunn}) without additional assumptions.

\subsection{Spin Density Matrices of Photon and Vector Meson} 

The photon spin density matrix normalized to unit flux of transverse photons
comprises the unpolarized ($U$) and
 polarized ($L$)  matrices\footnote{The adjectives ``(un)polarized'' 
are used here with the same meaning as when applied to SDMEs.},
with $P_{beam}$ being the longitudinal polarization of the beam:
\begin{equation} \label{matr}
\varrho^{U+L}_{\lambda_{\gamma} \lambda '_{\gamma }} =
\varrho_{\lambda_{\gamma}
\lambda '_{\gamma }}^{U} +
P_{beam} \; \varrho_{\lambda_{\gamma} \lambda '_{\gamma}}^{L},
\end{equation}
\begin{equation}
\hspace*{-4.9cm}\varrho^{U}_{\lambda_{\gamma} \lambda '_{\gamma}}(\epsilon, \Phi) =
\label{matr-unpol}
\end{equation}
\begin{displaymath}  
\frac{1}{2} \! \left(\!\! \begin{array}{ccc}
1 & \sqrt{\epsilon(1+\epsilon)} e^{-i\Phi} & -\epsilon e^{-2i\Phi} \\
\sqrt{\epsilon(1+\epsilon)} e^{i\Phi} & 2 \epsilon & - \sqrt{\epsilon(1+\epsilon)}
e^{-i\Phi} \\
- \epsilon  e^{2 i\Phi} & - \sqrt{\epsilon(1+\epsilon)} e^{i\Phi} & 1 \\
\end{array} \! \right) \!,
\end{displaymath}

\begin{displaymath}
\hspace*{-4.9cm}\varrho^{L}_{\lambda_{\gamma} \lambda '_{\gamma}}(\epsilon, \Phi) =
 \label{matr-lpol}
\end{displaymath}
\begin{equation} 
 \frac{\sqrt{1-\epsilon}}{2} \left( \begin{array}{ccc}
\sqrt{1+\epsilon}  & \sqrt{\epsilon} e^{-i\Phi} & 0 \\
\sqrt{\epsilon} e^{i\Phi} &  0  & \sqrt{\epsilon} e^{-i\Phi} \\
0 & \sqrt{\epsilon} e^{i\Phi} & -\sqrt{1+\epsilon}  \\
\end{array} \right ).
\end{equation}

The spin density matrix $\rho_{\lambda_{V} \lambda '_{V}}$
of the produced vector meson  is 
related to that of the virtual photon, 
$\varrho^{U+L}_{\lambda_{\gamma} \lambda '_{\gamma }}$, 
through the von~Neumann formula: 
\begin{eqnarray} \label{neumann}
\rho_{\lambda_{V} \lambda '_{V}} =  
 \nonumber  \hspace*{6.0cm}  \\ 
\frac{1}{2 \mathcal{N}} \sum_{\lambda_{\gamma} 
\lambda '_{\gamma}\lambda_N \lambda '_N}
\!\!\!\!   F_{\lambda_{V}\lambda '_N;\lambda_{\gamma}\lambda _N}\;
 \varrho^{U+L}_{\lambda_{\gamma} \lambda '_{\gamma }}\;
  F_{\lambda '_{V} \lambda '_N;\lambda '_{\gamma}\lambda _N}^{*}, \hspace*{0.5cm}
 \end{eqnarray}
where $F_{\lambda_{V}\lambda '_N;\lambda_{\gamma}\lambda _N}$ denotes the helicity 
amplitude of the $\gamma^* N \to \rho^0 N$ transition
defined in (\ref{jacobwick}). 
The normalization factor is given by 
\begin{eqnarray}
\mathcal{N} =   \mathcal{N}_T + \epsilon \mathcal{N}_L,
\label{ntotal}
\end{eqnarray}
with
 \begin{eqnarray} 
\lefteqn{ \mathcal{N}_{T}=  \widetilde{\sum} 
 ( |T_{11}|^2+|T_{01}|^2+|T_{-11}|^2} \nonumber \\
& & \mbox{\hspace*{1.0cm}} + |U_{11} |^2+|U_{01}|^2+|U_{-11}|^2 ),
\label{sigmatrans}
\end{eqnarray}
\begin{eqnarray}
\mathcal{N}_L = \widetilde{\sum} ( |T_{00}|^2+2|T_{10}|^2+2|U_{10}|^2 ).
\label{sigmalong}
\end{eqnarray}
Equation (\ref{sigmalong}) is obtained by using symmetry
relations $(\ref{symmnat})$ and $(\ref{symmunn})$.

If the spin density matrix of the photon is decomposed into the standard set 
of nine 
hermitian matrices $\Sigma ^{\alpha}$  
($\alpha=0,\;1,\;...,\;8$), a set of nine matrices 
$\rho ^{\alpha}_{\lambda_{V} \lambda_{V}'}$ 
is obtained for the  vector meson \cite{wolf}: 
\begin{eqnarray}  \label{rhomatr}
\rho_{\lambda_{V} \lambda'_{V}}^{\alpha} 
&=& \frac{1}{2 \mathcal{N}_{\alpha}}  \!\!  \sum_{\lambda_{\gamma} \lambda'_{\gamma} \lambda'_N\lambda _N} \!\!\!  F_{\lambda_{V}\lambda '_N;\lambda_{\gamma}\lambda _N}
 \Sigma_{\lambda_{\gamma} \lambda '_{\gamma }}^{\alpha} 
  F_{\lambda '_{V} \lambda' _N;\lambda '_{\gamma}\lambda _N}^{*} \nonumber \\
&\equiv&  \frac{1}{\mathcal{N}_{\alpha}}\widetilde{\sum}_{\lambda_{\gamma} \lambda '_{\gamma}}
  F_{\lambda_{V}\lambda_{\gamma}}
 \Sigma_{\lambda_{\gamma} \lambda '_{\gamma }}^{\alpha}
  F_{\lambda '_{V}\lambda '_{\gamma}}^{*}.
\end{eqnarray}
The four matrices  $\rho^{\alpha}$ for $\alpha = 0, 1, 2, 3$ in 
(\ref{rhomatr})
describe vector meson production by transverse virtual photons:
unpolarized, linearly polarized in two orthogonal directions, and circularly 
polarized, respectively. 
For these cases $\mathcal{N}_{\alpha} = \mathcal{N}_T$.
Vector meson production by longitudinal virtual photons 
corresponds to $\alpha = 4$ 
in (\ref{rhomatr})  and $\mathcal{N}_{\alpha} = \mathcal{N}_L$.
The interference between the amplitudes of vector meson production 
by transverse and  longitudinal virtual photons
is described by (\ref{rhomatr}) 
for $\alpha =5, 6, 7,$ and~$8$ 
with  $\mathcal{N}_{\alpha} = \sqrt{\mathcal{N}_T \mathcal{N}_L}$.

\subsection{Cross Sections}

The  differential cross section of the 
reaction $\gamma^* N \to \rho^0 N  \to \pi^+ \pi^- N$ is given by
\begin{eqnarray} \label{xsect-matr}
\lefteqn{\frac{d \sigma_{full}(W,Q^2)}{dt \; d\Phi \;  d\phi \; d\cos\Theta}
= \frac{f(W,Q^2)}{4\pi}} \nonumber    \\
& \times&\hspace*{-1.2cm}\sum_{\lambda_{\gamma} \lambda'_{\gamma} 
\lambda_V \lambda'_V \lambda_N \lambda'_N}
F_{\lambda_{V}\lambda'_N;\lambda_{\gamma}\lambda_N}\;
  \varrho^{U+L}_{\lambda_{\gamma} \lambda'_{\gamma }}(\epsilon,\Phi) \;
  F_{\lambda'_{V} \lambda'_N;\lambda'_{\gamma}\lambda_N}^* \nonumber   \\
& \mbox{\hspace*{1.5cm}} &\times \; Y_{1 \lambda_V}(\phi,\cos \Theta) \;\;  
Y_{1 \lambda '_V}^*(\phi,\cos \Theta),
\end{eqnarray}
in terms of $\varrho^{U+L}_{\lambda_{\gamma} \lambda_{\gamma}'}$,
the virtual-photon spin density matrix,
the helicity amplitudes $F_{\lambda_{V}\lambda'_N;\lambda_{\gamma}\lambda_N}$
describing the transition of the virtual photon
with helicity $\lambda_{\gamma}$ to the vector meson with helicity $\lambda_{V}$,
and the spherical harmonics  $Y_{1 m}(\phi,\cos \Theta), m = \pm 1,0$ 
(defined as 
in~\cite{wolf,ipivanov,mdiehl})
that describe the angular distribution of the pions from the decay 
$\rho ^0 \rightarrow \pi^+ +\pi^-$. 
It is assumed here that the branching ratio of the $\rho^0$-meson decay
into $\pi^+ \pi^-$ is 100\%.
The kinematic factor
\begin{eqnarray}
f(W,Q^2)=\frac{1}{16 \pi (\nu^2 + Q^2 )}
\label{kinfactor}
\end{eqnarray}
\noindent in (\ref{xsect-matr}) accounts for the fact that the flux 
of transverse photons in electroproduction is not unity (see Ref.~\cite{wolf}
for the relation of the differential virtual-photon 
cross section to the differential
electroproduction cross section).

The singly differential cross section 
$\frac{d\sigma_{full}}{dt}$ for  $\rho^0$ meson production is
obtained by  integrating (\ref{xsect-matr})  over $\Phi, \; \phi, \;\cos\Theta$.
The integration over $\Phi$ eliminates the interference between contributions 
of transverse and longitudinal 
photons and makes the photon density matrix diagonal. For this 
case, the full differential cross section becomes the linear 
combination of the cross sections $\frac{d \sigma_{T}}{dt}$ 
and $\frac{d \sigma_{L}}{dt}$ 
of vector meson production with transverse and longitudinal photons, 
respectively:
\begin{eqnarray} \label{dsigmadt}
\frac{d \sigma_{full}}{dt} = \epsilon \frac{d \sigma_{L}}{dt}+\frac{d \sigma_{T}}{dt},
\end{eqnarray}
where
\begin{eqnarray} 
\frac{d \sigma_{i}}{dt}(W,Q^2,t) = f(W,Q^2) \mathcal{N}_{i}(W,Q^2,t),
\label{sigmalsigmat}
\end{eqnarray} 
for $i=L,T$, where 
$\mathcal{N}_T$ and $\mathcal{N}_L$ are defined in (\ref{sigmatrans})
and (\ref{sigmalong}), respectively.

The ``differential''  longi\-tu\-di\-nal-to-transverse cross section ratio
is defined as:
\begin{eqnarray}
R (W,Q^2,t) \equiv \frac{d\sigma_L}{dt}/\frac{d\sigma_T}{dt}
= \frac{ \mathcal{N}_L}{ \mathcal{N}_T}\;.
\label{eqr}
\end{eqnarray}
The complete representation for $R$ in terms of helicity amplitudes is obtained by
inserting (\ref{sigmalong}) and (\ref{sigmatrans}) into (\ref{eqr}).
Approximate expressions for $R$ 
related to SCHC or NPE  will be discussed in section 11.

\subsection{Accessible Spin Density  Matrix Elements }

\begin{sloppypar}
For an unpolarized target and a longitudinally polarized beam,
the 3-dimensional angular distribution of $\rho^0$ production and decay
is described by 26 matrix elements 
$\rho_{\lambda_{V} \lambda_{V}'}^{\alpha}$~\cite{wolf}.
If the experiment can be performed only at one beam energy,
the matrix elements $\rho^0_{\lambda_{V}\lambda '_{V}}$ and  
$\rho^4_{\lambda_{V}\lambda '_{V}}$
cannot be disentangled, so that only 23 elements are  accessible.
It is customary to extract from the experimental data the following elements:
\begin{equation}  
r^{04}_{\lambda_{V}\lambda '_{V}} = (\rho^{0}_{\lambda_{V}\lambda '_{V}}
+ \epsilon R \rho^{4}_{\lambda_{V}\lambda '_{V}})/( 1 + \epsilon R ), 
\nonumber
\end{equation}
\begin{equation}
r^{\alpha}_{\lambda_{V}\lambda'_{V}} =
\begin{cases}
\frac{  \rho^{\alpha}_{\lambda_{V}\lambda'_{V}}}{(  1 + \epsilon R )},
\; \alpha = 1,2,3, \\
\frac{ \sqrt{R} \rho^{\alpha}_{\lambda_{V}\lambda '_{V}}}
{(1 + \epsilon R )}, \; \alpha = 5,6,7,8.
\end{cases}
  \hspace*{0.25cm}
\label{rmatr}
\end{equation}
From now on, we will designate  $r^{04}_{\lambda_{V}\lambda '_{V}}$ and
$r^{\alpha}_{\lambda_{V}\lambda '_{V}}$  ($\alpha =$ 1-3, 5-8)
as the {\it Spin Density  Matrix Elements} (SDMEs).
\end{sloppypar}

In Appendix~A,  ~$(\ref{a1})$-$(\ref{a17})$,
the SDMEs  are expressed in terms of
NPE and UPE amplitudes, as obtained by 
combining (\ref{rhomatr}) and (\ref{rmatr}).

\subsection{Extraction of SDMEs from Measured Angular Distributions}

\begin{sloppypar}
Measurement of the 3-dimensional $\rho^0$ production and decay angular distribution
\begin{eqnarray} \label{def-w}
\lefteqn{\mathcal{W}^{U+L}(W, Q^2,t, \Phi, \phi, \cos \Theta)} \hspace*{1.75cm}\nonumber \\
& & \equiv  \frac{d \sigma_{full}}{dt \; d\Phi \;  d\phi \; d\cos\Theta}/\frac{d \sigma_{full}}{dt}
\end{eqnarray}
reveals the helicity structure of the $\gamma^* N \to \rho^0 N$ transition.
Its integral 
over  the variables  $\Phi$,  $\phi$, and  $\cos\Theta$ 
is equal to unity.
The  $W,Q^2$ and $t$ dependences of $\mathcal{W}^{U+L}$ are contained in the 
corresponding dependences of 
the SDMEs $r^{\alpha}_{\lambda_V \lambda '_V}$.
The full angular dependence of
$\mathcal{W}^{U+L}(\Phi,  \phi, \cos{\Theta})$, 
as a linear function of the SDMEs 
$r^{\alpha}_{\lambda_V \lambda '_V}$, is given
in (\ref{eqang1}-\ref{eqang3}) as derived in Ref.~\cite{wolf}.
(Note that these formulae are on the next page.)
 \begin{figure*}[hbtc!]
\begin{eqnarray}
\mathcal{W}^{U+L}(\Phi,\phi,\cos{\Theta}) =  \mathcal{W}^{U}(\Phi,\phi,\cos{\Theta}) + 
\mathcal{W}^{L}(\Phi,\phi,\cos{\Theta}), 
\label{eqang1} \hspace*{7.2cm}
\end{eqnarray}
\begin{eqnarray}
\mathcal{W}^{U}(\Phi,\phi,\cos{\Theta})  &=& \frac{3} {8 \pi^{2}} \Bigg[
         \frac{1}{2} (1 - r^{04}_{00}) + \frac{1}{2} (3 r^{04}_{00}-1) \cos^2{\Theta}
- \sqrt{2} \mathrm{Re} \{ r^{04}_{10} \} \sin 2\Theta 
\cos \phi - r^{04}_{1-1}  \sin ^{2} \Theta \cos 2 \phi \hspace*{1.0cm}
\nonumber \\ 
&-& \epsilon \cos 2 \Phi \Big( r^{1}_{11} \sin ^{2} \Theta  + r^{1}_{00} \cos^{2}{\Theta}
  - \sqrt{ 2}  \mathrm{Re} \{r^{1}_{10}\} \sin 2  \Theta  \cos  \phi
    - r^{1}_{1-1} \sin ^{2} \Theta \cos 2 \phi   \Big)   \nonumber  \\
&-& \epsilon \sin 2 \Phi \Big( \sqrt{2} \mathrm{Im} \{r^{2}_{10}\} \sin 2 \Theta \sin \phi +
       \mathrm{Im} \{ r^{2}_{1-1} \} \sin ^{2} \Theta \sin 2 \phi  \Big)  \nonumber \\
&+& \sqrt{ 2 \epsilon (1+ \epsilon)}  \cos \Phi
\Big(  r^{5}_{11} \sin ^2 {\Theta} +
 r^{5}_{00} \cos ^{2} \Theta - \sqrt{2} \mathrm{Re} \{r^{5}_{10}\} \sin 2 \Theta \cos \phi -
 r^{5}_{1-1} \sin ^{2} \Theta cos 2 \phi  \Big)  \nonumber \\
&+& \sqrt{ 2 \epsilon (1+ \epsilon)}  \sin \Phi
\Big( \sqrt{ 2} \mathrm{Im} \{ r^{6}_{10} \} \sin 2 \Theta \sin \phi
+ \mathrm{Im} \{r^{6}_{1-1} \} \sin ^{2} \Theta \sin 2 \phi \Big) \Bigg],
\label{eqang2} \\
\mathcal{W}^{L}(\Phi,\phi,\cos \Theta)  &=& \frac{3}{8 \pi^{2}} P_{beam} \Bigg[
  \sqrt{ 1 - \epsilon ^{2} }  \Big(  \sqrt{ 2}  \mathrm{Im} \{ r^{3}_{10} \}
\sin 2 \Theta \sin \phi +
   \mathrm{Im} \{ r^{3}_{1-1}\} \sin ^{2} \Theta \sin 2 \phi  \Big)  \nonumber  \\
&+& \sqrt{ 2 \epsilon (1 - \epsilon)} \cos \Phi
\Big( \sqrt{2} \mathrm{Im} \{r^{7}_{10}\} \sin 2 \Theta \sin \phi
+  \mathrm{Im} \{ r^{7}_{1-1} \}  \sin ^{2} \Theta \sin 2 \phi   \Big)  \nonumber \\
&+& \sqrt{ 2 \epsilon (1 - \epsilon)} \sin \Phi
\Big( r^{8}_{11} \sin ^{2} \Theta + r^{8}_{00} \cos ^{2}
\Theta -  \sqrt{2} \mathrm{Re}\{ r^{8}_{10}\} \sin 2 \Theta \cos \phi 
- r^{8}_{1-1} \sin ^{2} \Theta \cos 2\phi \Big)  \Bigg]. 
\label{eqang3}
\end{eqnarray}
 \end{figure*}
\end{sloppypar}

\subsection{$s$-Channel Helicity Conservation.} 

The measurement of SDMEs allows the determination of the extent 
to which $s$-channel helicity is 
conserved for a given process and kinematic conditions.
SCHC implies that
the contributions from all non-diagonal transitions  
$F_{\lambda_{V} \lambda '_{N}; \lambda_{\gamma}  \lambda_{N} }$ with 
$\lambda_{\gamma} \neq \lambda_{V}$
are zero.
In terms of NPE and UPE amplitudes,  only
$T_{00}$, $T_{11}$, and $U_{11}$ remain.
As a consequence, all spin density matrix elements vanish  
except the unpolarized SDMEs $r_{00}^{04}$, $r^{1}_{1-1}$,
Im$\{r^{2}_{1-1}\}$,  Re$\{r^{5}_{10}\}$, Im$\{r^{6}_{10}\}$,
and the polarized ones Im$\{r^{7}_{10}\}$ and Re$\{r^{8}_{10}\}$, as can be seen 
from (\ref{a1}-\ref{a17}) 
of Appendix~A. If SCHC holds, 
SDMEs are not independent, as 
the following relations apply:
\begin{eqnarray} 
r^{1}_{1-1}&=&-\mathrm{Im}\{r^{2}_{1-1}\},
\label{schc01} \\
\mathrm{Re}\{r^{5}_{10}\}&=&-\mathrm{Im}\{r^{6}_{10}\}, 
\label{schc02} \\
\mathrm{Re}\{r^{8}_{10}\}&=&\mathrm{Im}\{r^{7}_{10}\}.  
\label{schc03}
\end{eqnarray}

The measurement of SDMEs also allows the determination of the
extent to which the unnatural-parity-exchange mechanism is relevant for a 
given process
and kinematic conditions. If natural-parity exchange dominates, so that
 the amplitude $U_{11}$
can be neglected, an additional  relation is obtained:
\begin{eqnarray}
1-r_{00}^{04}=2r^{1}_{1-1}=-2\mathrm{Im}\{r^{2}_{1-1}\}.
\label{schcnpe}
\end{eqnarray}

\section{The HERMES Experiment }

The HERMES experiment at DESY used a 27.6 GeV
longitudinally polarized positron or electron beam impinging on pure hydrogen 
or deuterium gas targets internal to the HERA storage ring.
Parts of the data set were collected with longitudinally or transversely
polarized targets, the polarization of which was flipped approximately every minute. 
The average over the target polarization values was confirmed to be
consistent with zero, as required for the extraction of SDMEs in this analysis.
The lepton beam was transversely self-polarized by the emission of synchrotron 
radiation~\cite{sokolov}. Longitudinal polarization at the interaction point
was achieved by spin rotators located upstream and downstream of the HERMES
apparatus. 
For both  positive and negative beam helicities,
the beam polarization was continuously measured by two Compton 
polarimeters~\cite{tpol,lpol}. 
The average beam polarization for the hydrogen 
(deuterium) data set was 0.45 (0.47) after requiring 
$0.15 < P_{beam} < 0.8$, and 
the fractional systematic uncertainty 
of the beam polarization was 3.4\% (2.0\%)~\cite{tpol,lpol}.

The data sample recorded with a longitudinally polarized  hydrogen (deuterium) 
target, 
representing 14\% (38\%), 
of the total statistics, has a residual polarization of 
$0.0221 \pm  0.0001$ ($-0.0036 \pm  0.0009$).
The data sample recorded with a
transversely polarized hydrogen target, 
representing 35\%, has a residual polarization of
$0.0028 \pm 0.0001$. The systematic uncertainty of the target polarization 
measurement is typically 0.04.

\begin{sloppypar}
The HERMES spectrometer is described in detail in ~Ref.~\cite{herspec}.
It was a forward spectrometer in which both scattered lepton
and produced hadrons were detected within an
angular acceptance $\pm$170 mrad horizontally, and $\pm$(40 - 140) mrad
vertically.
The scattered-lepton trigger was formed from a coincidence between 
three scintillator hodoscopes and a lead-glass calorimeter.
The trigger required an energy of more than 3.5 GeV deposited in the
calorimeter. The tracking system had a momentum resolution of 
$\approx 1.5\%$ and an angular resolution of $ \approx  1$~mrad. 
Lepton identification was accomplished using a lead-glass
calorimeter, a
preshower detector consisting of a scintillator hodoscope preceded by
a lead sheet, and a transition-radiation detector.
Until 1998 the particle identification system 
included  a gas threshold 
$\breve {\mathrm C}$e\-ren\-kov counter, which was replaced in 1999
with a dual-radiator ring-imaging  $\breve {\mathrm C}$erenkov  detector 
(RICH)~\cite{rich}. Combining the responses of these detectors in a
likelihood method leads to  an average lepton identification
efficiency of 98\% with a hadron contamination of less than 1\%. 
\end{sloppypar}

\section{Data Analysis}
\subsection{Exclusive  $\mathbf{\rho^0}$  Events }

\begin{figure}[t!]
 \begin{center}
 \includegraphics[height=7.5cm,width=8.cm]{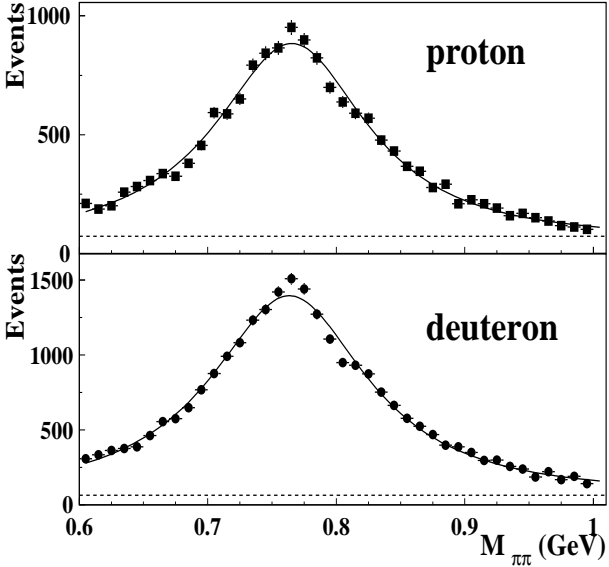}
\caption{\small  
 Two-pion invariant mass distribution in the spectrometer acceptance, 
fitted with 
a skewed Breit-Wigner function (solid line).
The dotted line represents a constant background contribution.
} 
\label{fig-minv}
\end{center}
\end{figure}

\begin{figure}[t!]
\begin{center}
\includegraphics[height=7.5cm,width=8.cm]{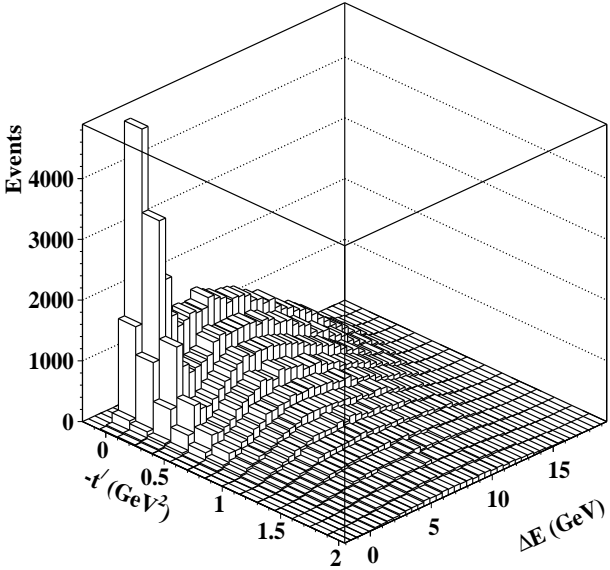}
\caption{\small  
The distribution of $\rho^0$ mesons in 
$\Delta E$, and $t'$ from the 1996-2005 
hydrogen data sample in the range $1~\rm{GeV}^2 < Q^2 < 7$ GeV$^2$.
}
\label{fig-tpr-dele}
\end{center}
\end{figure}

\begin{sloppypar}
Events accepted for the analysis are required to fulfill the following criteria
(see Refs.~\cite{tytgat,craig} for details):
\begin{itemize}
\item three  tracks originate from the target and are recorded  in the spectrometer;
\item two oppositely charged 
hadrons and one lepton with the same charge as the beam
are identified through the likelihood 
analysis of the combined 
responses of the four particle-identification detectors~\cite{herspec};
\item the reconstructed virtual photon satisfies the kinematic constraint  
$1~\rm{GeV}^2 < Q^{2} < 7$ GeV$^2$;
\item the $\rho^0$ meson is selected by requirements on the invariant mass of the 
two hadrons of opposite charge:
$0.6~{\rm GeV}<M_{\pi^{+}\pi^{-}}<1.0$~GeV when both hadrons are assumed to be pions, 
and the veto constraint 
$M_{K^{+} K^{-}} \ge 1.06$ GeV, the latter under the 
hypothesis that both 
hadrons are kaons.  The veto constraint excludes contamination from
 $\phi \to K^{+} K^{-}$ decay.
Two-pion invariant mass distributions 
in the HERMES acceptance for proton and deuteron data are
presented in Fig.~\ref{fig-minv}. A detailed description of
the invariant mass peak of exclusive $\rho^0$ events is published in
Refs.~\cite{rho-xsec,twopion} and also presented in Ref.\cite{tytgat}.
The distribution of these events in both $\Delta E$ and $t'$ is presented in
Fig.~\ref{fig-tpr-dele}. 
\item exclusive $\rho^0$  events are selected by the requirement  
$ -1.0~{\rm GeV} < \Delta E < 0.6 $ GeV (called the ``exclusive region'' in the remainder of the text).
The applicability of such a constraint was explained in detail in 
Ref.~\cite{twopion}, as well as in Refs.~\cite{rho-xsec,sdme-publ,tytgat}.  
In the $\Delta E$ 
spectrum the resolution due to instrumental effects ranges between 0.26 and 0.38 GeV
depending on the spectrometer configuration.
\item  the
``final event sample''  of $\rho^0$ events is obtained from the sample of
exclusive events by the additional requirement 
  $-t' < 0.4 $ GeV$^2$.
This requirement  ensures that the semi-inclusive background 
does not exceed a level of about 10\% in the 
kinematic bins of $Q^2$ and $t'$ presented below.
\end{itemize}
\end{sloppypar}

After the application of all the above requirements, 
the entire kinematic region contains
16362 $\rho^0$ events from hydrogen 
and 25940 events from deuterium, which are used in the subsequent 
physics analysis.

\subsection{Backgrounds for Exclusive $\rho^0$ Events \label{sec-bg} }

In exclusive vector meson production, the target nucleon
remains intact. At HERMES the recoiling target nucleon was not
detected and hence, given the experimental resolution, also a certain number
of non-exclusive events will satisfy the requirements for exclusive 
events. They appear as background remaining underneath the $\Delta E$ peak. 

\subsubsection*{a) Background from Semi-Inclusive Deep-Inelastic Scattering
 \label{sec-sidisbg}}

Background events originate mainly from fragmentation processes in 
Semi-Inclusive Deep-Inelastic Scattering (SIDIS), in which the final state 
contains a pair of oppositely charged hadrons in the spectrometer.
Only a small fraction of this 
background passes the above-described $\Delta E$ and $-t'$ requirements
for exclusive  $\rho^0$ production.

 \begin{figure}[t!]
 \begin{center}
\includegraphics[height=7.5cm,width=8.cm]{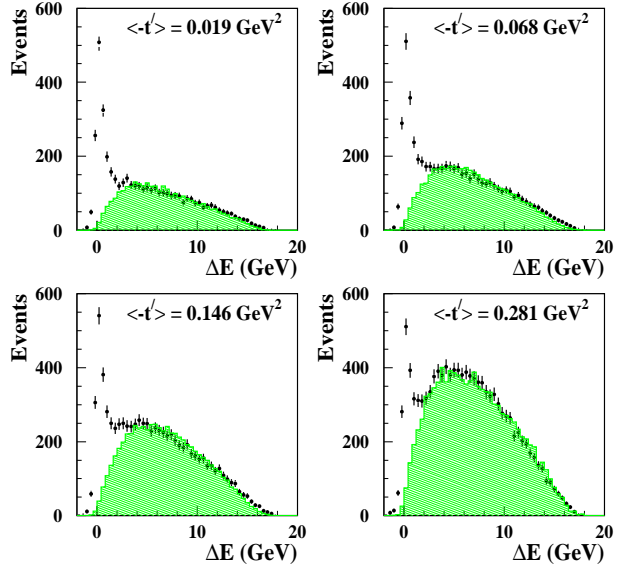}
\caption{\small
Distribution of $\Delta E$ for the 1996-2005 hydrogen data sample
shown for four intervals in $t'$ after application of all event selection requirements,
except the one for  $\Delta E$.  
The shaded areas represent the SIDIS background calculated by PYTHIA,
normalized to the data in the region $2~{\rm GeV} < \Delta E < 20$ GeV. 
}
\label{pyt-q2}
\end{center}
\end{figure}

\begin{sloppypar}
The amount of  SIDIS background and its angular distributions 
in the exclusive region can not be determined with the present apparatus. Therefore, 
the PYTHIA  code~\cite{sjo03} tuned for HERMES
kinematic conditions~\cite{lieb04,vanm05,ulie06} is used.  
Exclusive processes were excluded from the simulated sample. 
The simulated SIDIS events were passed through the same chain of kinematic 
requirements as the experimental ones.
Very good agreement between the experimental data and the simulation is
observed for the shape of the $\Delta E$ 
distributions in the region $\Delta E > 2$ GeV for each kinematic
interval in $Q^2$, $x$ or $t'$,
and for both targets.
This agreement in shape is demonstrated in 
Fig.~\ref{pyt-q2}, which  shows four intervals in $t'$ as an example.
Since no absolute normalization between data and simulation is required to determine the SDMEs
(as shown in the next section),  
for every kinematic interval the fractional background contribution $f_{bg}$ in the exclusive 
region can be obtained by normalizing
the simulation to the data in the region  $2~{\rm GeV} < \Delta E < 20$~GeV. That is,
\begin{eqnarray}
f_{bg} = (N_{excl}^{MC}/N_{excl}^{data}) \; (N_{2-20}^{data}/N_{2-20}^{MC}),
\label{eq-fbg}
\end{eqnarray}
where  $N^{data}_{2-20}$, $N^{MC}_{2-20}$  and $N^{data}_{excl}$, $N^{MC}_{excl}$
are the total number of measured and
simulated events at $2~{\rm GeV} < \Delta E < 20$~GeV and in the exclusive
region, respectively.
This contribution amounts to 8\% for the entire kinematic region and
ranges between 3 and 12\% in the different kinematic intervals.
These values will be used for the subtraction of the SIDIS background angular distributions,
as described in section 6.2.
\end{sloppypar}
 
\subsubsection*{b) Background from 
Non-resonant Exclusive  $\pi^{+} \pi^{-}$ Pairs }

\begin{sloppypar}
The contribution of
non-resonant  $\pi^{+} \pi^{-}$ production and its interference~\cite{soeding}
with resonant $\rho^{0} \to \pi^+$  $\pi^-$ production  are
determined using a modified Breit-Wigner fit to the  invariant mass distribution.
We do not distinguish
between resonant and non-resonant contributions in $\rho^0$
production, following the practice of
previous experimental~\cite{zeussdme,compasssdme} and
theoretical  publications~\cite{ryskin,manaenkov,golos,golos2}.
Therefore the present data were  not corrected for non-resonant background,
which amounts to $\sim 4$ -- $8$\% depending on the
kinematics~\cite{rho-xsec,dsa-lipka,tytgat}.
Note that the contribution of $\rho^0$-$\omega$ interference has been found to
be negligible~\cite{rho-xsec}.
\end{sloppypar}

\subsubsection*{c) Background from Proton-Dissociative Processes}

\begin{sloppypar}
Another possible background consists of events in which the target
proton is excited to some other baryonic resonance, which then decays to 
a proton and typically a soft pion.  In the absence of a recoil detector,
such events cannot be distinguished, but
their contribution to the
exclusive $\rho^0$ production cross section at HERMES was found to
be small ($ 4 \pm 2\%$)~\cite{rho-xsec}.
No correction has been applied
for this background, as the extracted SDMEs were found to change
negligibly at a 
value of $\Delta E$=0.2 GeV, where this background is strongly suppressed.
This approach is supported by results
from ZEUS, where the decay angle distributions of 
the proton-disso\-ci\-a\-tive reaction  have been found
to be consistent with those of exclusive events~\cite{zeussdme,zeussdme2}.
\end{sloppypar}

\section{Extraction of SDMEs}

\subsection{Maximum Likelihood Method}

\begin{sloppypar}
In each bin of $Q^2$ or $t'$, or the entire acceptance,
the SDMEs are obtained
by minimizing the difference between the
3-dimen\-sion\-al $(\cos\Theta,\phi,\Phi)$ 
production and   decay angle distribution of the experimental events 
and that of a sample of fully reconstructed Monte Carlo
events, using a binned maximum likelihood method. 
For the Monte Carlo simulation, events are generated 
isotropically in ($\Phi, \phi, \cos \Theta$)
using the rhoMC generator~\cite{rho-xsec,sdme-publ,craig}
 for exclusive $\rho^0$ production,
simulated in the instrumental context of the spectrometer,
and passed  through the same reconstruction chain as the experimental  data.
Introduction of estimated misalignments 
of the spectrometer into the Monte Carlo simulation used for the SDME extraction
was found to have a negligible effect on the results.
The variables $\cos\Theta$, $\phi$, and $\Phi$
are binned in $8 \times 8 \times 8$ cells.
The content of each of the 512 cells is weighted using (\ref{eqang1}),
whereby the 23 matrix
elements are treated as free parameters.
The number of events $d_i$ in each cell is assumed to obey a Poisson
distribution:
\begin{equation} \label{eq-ml-input} 
P(d_i,c_Nm^{\prime}_i)=\frac{(c_Nm^{\prime}_i)^{d_i}}{d_i!}e^{-c_Nm^{\prime}_i},
\end{equation}
with mean value $c_N m^{\prime}_i$,
where 
$c_N=(\sum_j d_j)/(\sum_j m^{\prime}_j)$ is a
normalization
factor accounting for the difference in the total number of events in
the data ($d_j$)   and simulated  ($m^{\prime}_j$)  sample, and
  $m^{\prime}_i$ is the (re)weighted
number of simulated events in cell $i$. 
The likelihood function is then defined as~\cite{zech}
\begin{equation}
L(\lambda)=\prod_i^{cells}P\left(d_i,c_N(\lambda)m^{\prime}_i(\lambda)\right),
\end{equation}
where $\lambda$ represents the 23 fit parameters that are the 23 SMDEs.
The best fit parameters were determined by maximizing the logarithm of the
likelihood function,
\begin{eqnarray}
\ln L(\lambda) &=& \sum_i\left[d_i\ln\left(c_N(\lambda)m^{\prime}_i(\lambda)\right)
-c_N(\lambda)m^{\prime}_i\right] \nonumber \\
&\ & \hspace*{0.75cm} + {\it\ constant},
\end{eqnarray}
or equivalently by minimizing $-\ln L(\lambda)$.
\end{sloppypar}

The minimization itself and the uncertainty
calculation are performed using the MINUIT package~\cite{minuit}.
In the fitting procedure the samples with negative and positive beam helicity
are fitted simultaneously.
The values of  $\chi^2$ per degree of freedom  ($\chi^2/d.o.f. $)  
for the 16 kinematic intervals ($Q^2$, $t'$ or $x$),   
calculated after completing their likelihood fits,  range between 
0.6 and~1.2  for $8\times 8 \times 8 - 23$ degrees of freedom.
For every SDME, the averages of the  SDMEs extracted from the two 
separate  beam  helicity samples are found to be consistent with each other and 
the result from the common fit. 

\begin{figure*}
  \begin{center}
\includegraphics[height=7.5cm,width=8.cm]{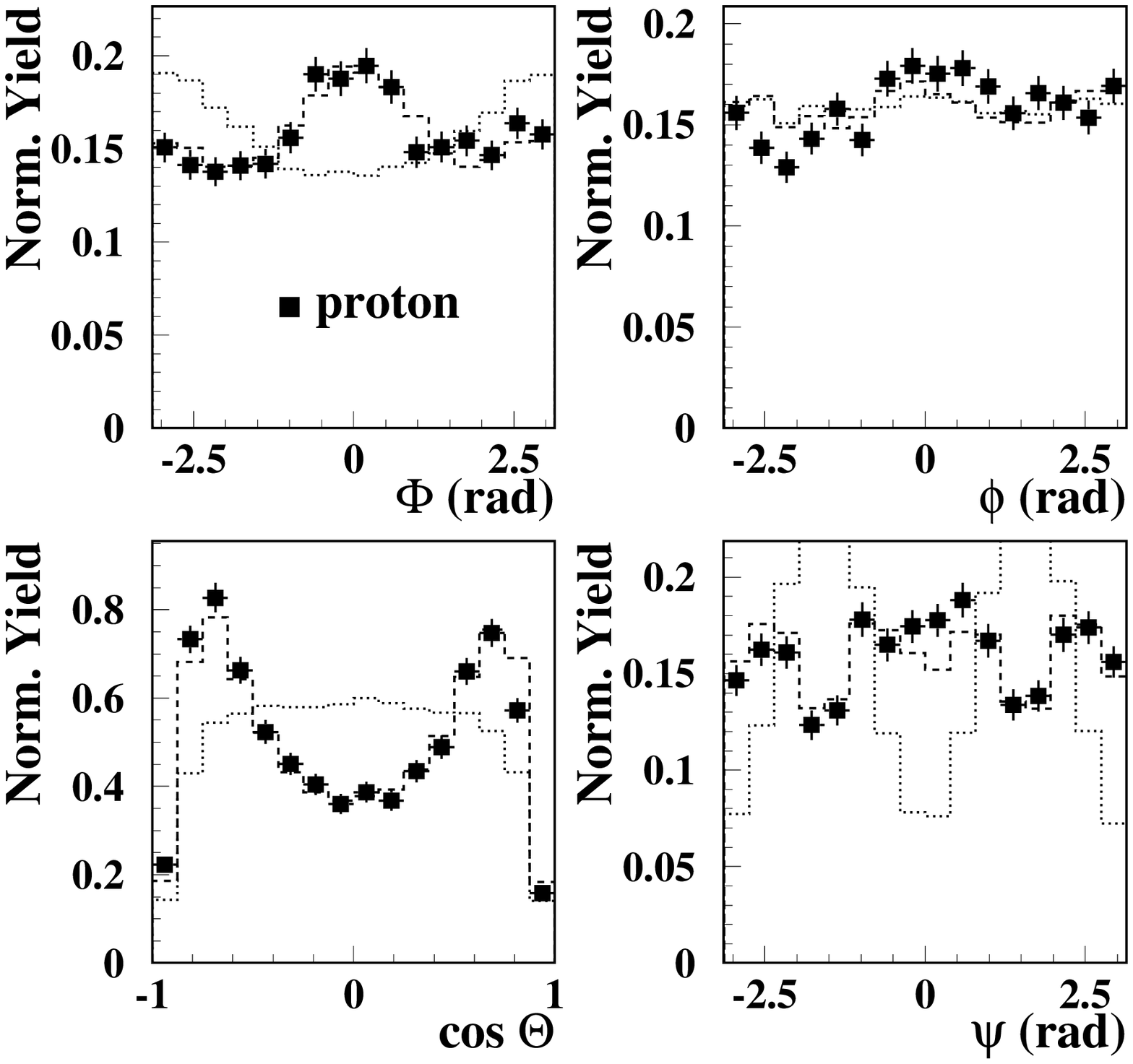}
 \hspace*{0.60cm}
\includegraphics[height=7.5cm,width=8.cm]{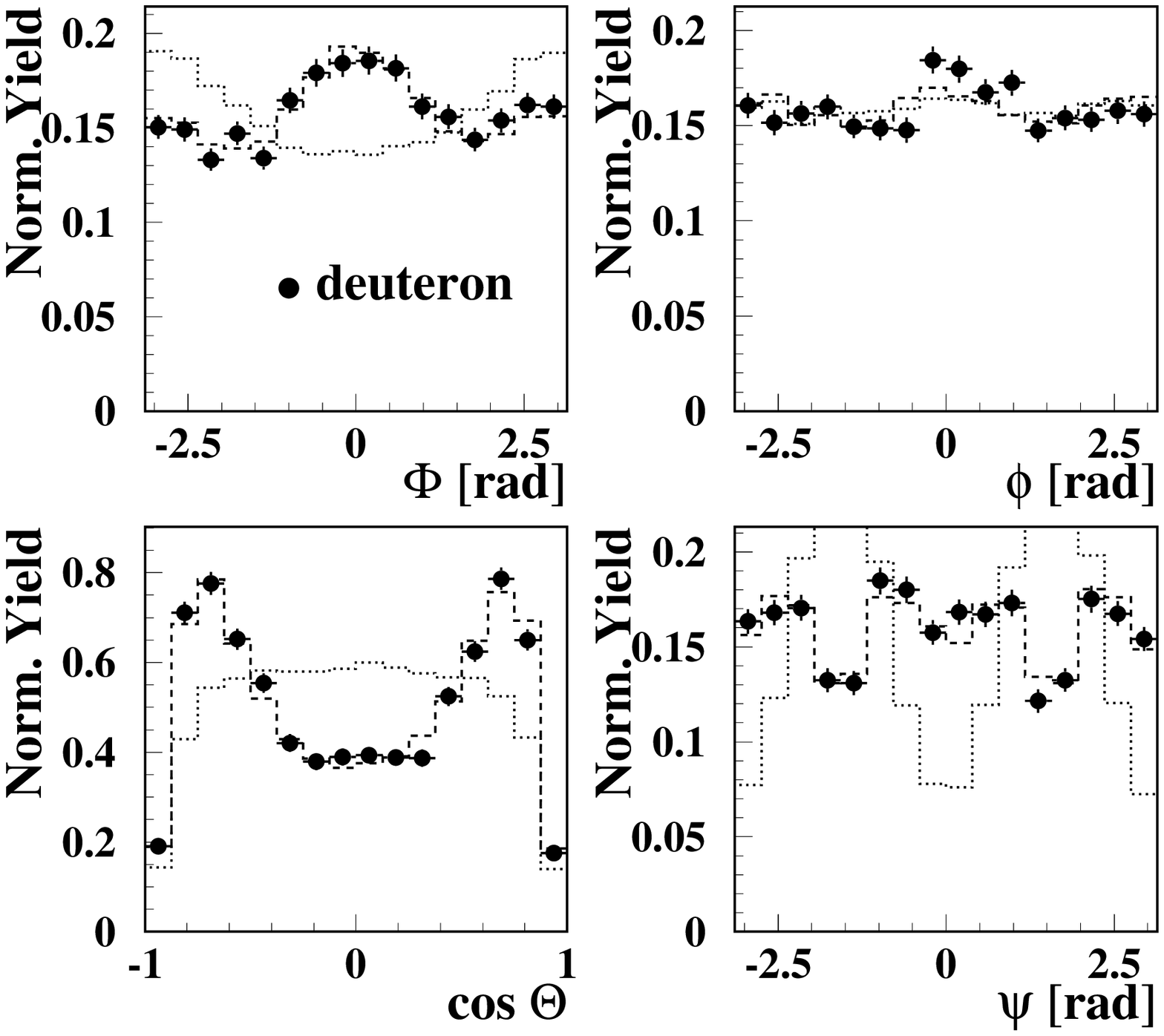}
    \caption{\small 
Angular distributions with common 
arbitrary normalization  for $\rho^0$ meson production and decay.
Data points represent the positive helicity sample of the 
 proton (deuteron)  data in the left  (right) half of the figure. 
The dotted lines represent isotropic input Monte Carlo distributions 
 as modified by  the HERMES acceptance, while
the dashed lines are the results of the 23-parameter fit. Here 16 bins are chosen for a more 
detailed comparison. The data correspond to the full kinematic region of the analysis.
}
\label{angl-shapes}
\end{center}
\end{figure*}

In  Fig.~\ref{angl-shapes} one-dimensional angular distributions
are shown  for the hydrogen  and deuterium data samples, where the 
positive-helicity sample is chosen as representative. In addition 
to distributions in $\cos\Theta$, $\phi$, and $\Phi$, the angular distribution
in $\psi = \phi - \Phi$
is shown, which embodies the entire azimuthal dependence in 
the case of SCHC. For each panel, the data are compared with
the isotropic input distributions as modified
by instrumental effects such as acceptance,
tracking resolution, and reconstruction efficiencies, as well as 
the one-dimensional projections of the
fitted 3-di\-men\-sion\-al angular distribution. These projections
are clearly  in agreement with the data.

\subsection{Background Subtraction}

Before fitting SDMEs to the ($\cos \Theta, \phi, \Phi$) angular
distributions, the SIDIS background in the exclusive region is
subtracted. This subtraction
is performed separately for each interval in  $t'$,  $Q^2$ or $x$.
In a given ($\cos \Theta, \phi, \Phi$) cell,
the number of background events in the exclusive region is
calculated as follows: 
\begin{equation}
N^{bg}_{cell} =  N^{MC}_{cell} \; \frac{N^{data}_{2-20}}{N^{MC}_{2-20}}, 
\label{eq:bg-subt} 
\end{equation}
where the number of SIDIS Monte Carlo events in a given cell is
$N^{MC}_{cell}$, while $N^{data}_{2-20}$ and  $N^{MC}_{2-20}$ are 
defined as in equation (\ref{eq-fbg}).

\subsection{Systematic Uncertainties}

\subsubsection*{a) Background Subtraction} 

The systematic uncertainty of the background subtraction
 procedure is estimated  as  the difference 
between the SDMEs obtained with and without any background correction.

\subsubsection*{b) rhoMC Input Parameters} 

The SDME extraction procedure starts from isotropic distributions 
in $\cos \Theta$, $\phi$, and $\Phi$ generated by rhoMC, as explained above.
The parameterization of the total electroproduction
cross section in rhoMC is chosen in the context  
of a VMD model that incorporates a propagator-type  
$Q^2$ dependence, and also contains a dependence on $R(Q^2)$. As the 
HERMES spectrometer acceptance depends on
$Q^2$, different input parameters result in slightly different 
reconstructed angular
distributions. The corresponding systematic
uncertainty of the resulting
SDMEs is obtained by varying these parameters within
one standard deviation in the total uncertainty
of the parameters given in Refs.~\cite{rho-xsec,sdme-publ}.

The total systematic uncertainty is obtained by adding in quadrature the
uncertainty from the background subtraction procedure
and that due to the uncertainty in the rhoMC input parameters,
which are approximately of equal size.

\subsubsection*{c) Further Systematic Studies}

Several further studies using generated and reconstructed event
samples are performed to estimate possible systematic uncertainties:

i) A consistency check of the extraction method is performed 
by using several known sets of SDMEs 
as input to the rhoMC~\cite{rho-xsec,sdme-publ,craig}
simulation and comparing
the SDMEs extracted from the simulated data with those
used as input to the rhoMC generator.
First, isotropic angular distributions were simulated, corresponding to
all SDMEs vanishing except $r^{04}_{00} = \frac{1}{3}$. 
Alternatively, events were generated assuming SCHC, 
implying that only five unpolarized and two polarized SDMEs  
are non-zero, as explained in section 3.7.  
Finally, the extracted 23 SDMEs are used as input parameters.
In all cases, good agreement is found between input and extracted SDMEs 
at the given level of statistical accuracy.

ii) Several tests are performed to ensure that the choice of the
($\cos \Theta, \phi, \Phi$) cell size does not bias the results of the  
maximum likelihood procedure.
A sample of about 40000 simulated events with angular dependences determined by 
the (normally) extracted 23 SDMEs 
is fitted 
after binning the data in several numbers of angular cells, varying from
$5 \times 5 \times 5$  to $12 \times 12 \times 12$.
The $\chi^2/d.o.f.$ calculated between sets of SDMEs extracted 
with $8 \times 8 \times 8$
and $12 \times 12 \times 12$ binning  is $0.14$. 
Hence the cell size used in the maximum likelihood procedure is
not treated as a source of systematic uncertainty.

iii) Variations of the restrictions on $M_{\pi^+ \pi^-}$, $t'$, and $\Delta E$
result in slightly different amounts of SIDIS background. 
The resulting systematic 
uncertainty is
much smaller than that estimated for the background subtraction procedure, 
and hence is  neglected.

iv) In the SDME extraction procedure, only the shape of the 3-dimensional 
angular 
distribution matters.
As events in which a radiative photon is emitted with an energy
larger than 0.6 GeV are removed from the analysis by the 
constraint $\Delta E < 0.6$ GeV, 
the impact of  radiative effects on the shape of the
3-dimensional angular distribution is strongly reduced.
Two approaches are used to quantify this effect.
First, the DIFFRAD code is used to calculate the radiative effects
for exclusive $\rho^0$ production~\cite{akush,akush2}, as was done also 
in Refs.~\cite{zeussdme,tytgat}.  
As the emission of a real photon by the positron alters the 
direction of the virtual photon,  the
angle  $\Phi$  between lepton scattering plane and $\rho^0$ production plane also changes.
The effect of a small variation ($<2.5$~\%, as suggested in Ref.~\cite{akush2}) 
of the shape  of the $\Phi$-distribution 
is studied by re-weighting  the isotropic input angular distribution. 
The difference between SDMEs obtained 
with and without re-weighting is found to be less than 0.0012
for all SDMEs ($\chi^2/d.o.f. < 0.1$), {\it i.e.,} radiative effects are negligible.

As an independent cross check, radiative effects on the extracted SDMEs 
are studied using a Monte Carlo simulation 
of exclusive $\rho^0$ production with events from the PYTHIA
generator~\cite{sjo03}.
Two statistically independent isotropic angular distributions
are generated, with and without the emission of radiative photons.  
A set of SDMEs is extracted from the (real) data sample
for each of these isotropic input angular distributions.
The difference between the resulting SDMEs is statistically indistinguishable 
($\chi^2/d.o.f. \le 0.2$).

\begin{figure}[ht!]
 \begin{center}
\includegraphics[width=8.0cm]{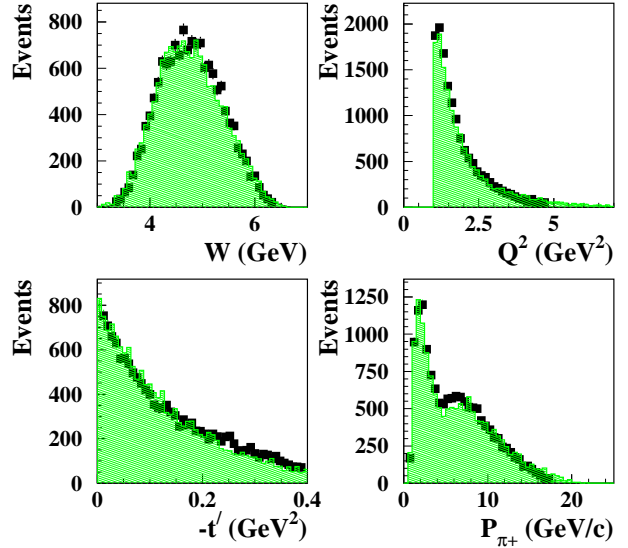}
\caption{\small  
Shape comparison of the distributions in $W$, $Q^2$, $t'$ and $P_{\pi^+}$, the momentum 
of the $\pi^+$ from $\rho^0$ decay in the laboratory system, for the hydrogen data 
sample (squares).
The shaded areas show rhoMC results using the 
extracted 23 SDMEs as input for the simulation, normalized to the data.  
Background has been subtracted from the data. 
\vspace*{-0.50cm}
}
\label{data-mc}
\end{center}
\end{figure}

As a further check we use the extracted 23 SDMEs
as input parameters to rhoMC and compare the shapes of 
the simulated distributions
with the data. In order to restrict the 
comparison to exclusive $\rho^0$ events,
properly normalized SIDIS background distributions from PYTHIA  
are subtracted from the data.
In the maximum likelihood fit method, the extraction of SDMEs only requires 
simulated event distributions normalized to the data.
The shape comparison reveals good agreement 
for all variables, some of which are presented in Fig.~\ref{data-mc}.

\section{Results on SDMEs Integrated over the Entire Kinematic Region }

\subsection{Classification of SDMEs}

The presentation  of the extracted SDMEs will be based on a hierarchy of NPE helicity
amplitudes: 
\begin{equation}
|T_{00}| \sim |T_{11}| \gg |T_{01}|>|T_{10}| \gtrsim|T_{1-1}|.
\label{hierarchy}
\end{equation}
A similar hierarchy was discussed
for the first time in Ref.~\cite{divanov}. In perturbative QCD this is valid   
at asymptotically high $Q^2$. It was experimentally confirmed
at the HERA collider~\cite{zeussdme,zeussdme2,H1}   and
discussed in Refs.~\cite{ipivanov,manaenkov,golos}. 
In the following it will be shown that it applies also at
$Q^2$ values typical of the HERMES experiment.

The extracted 23 SDMEs are categorized into five classes according to the
hierarchy shown above.
Class A comprises SDMEs dominated by the helicity-con\-ser\-ving
amplitudes $T_{00}$ and $T_{11}$
which describe the transitions $\gamma^*_L \to \rho^0_L$ and $\gamma^*_T \to \rho^0_T$, 
respectively.
Class B contains SDMEs that correspond to the interference of the above two 
amplitudes. 
Class C consists of all those SDMEs in which the main term contains a 
contribution linear in the $s$-channel helicity non-conserving  amplitude $T_{01}$,
corresponding to the $\gamma^*_{T} \to \rho^0_{L}$ transition,
except for a term involving 
$r^1_{00}$ for which the $T_{01}$ contribution is quadratic. 
The classes D and E are composed of the SDMEs in which the main terms contain a
contribution linear in the small helicity-flip amplitudes 
$T_{10}$ ($\gamma^*_{L} \to \rho^0_{T}$) 
and $T_{1-1}$  ($\gamma^*_{-T} \to \rho^0_T$), respectively. 
Equations (\ref{a1}-\ref{a17}) in Appendix~A show the representation of all the 
SDMEs in terms of helicity amplitudes ordered 
according to the five classes defined
above.

\subsection{Representation of the Integrated Data}

\begin{figure*}
\begin{center}
\includegraphics[height=16.0cm,width=16.cm]{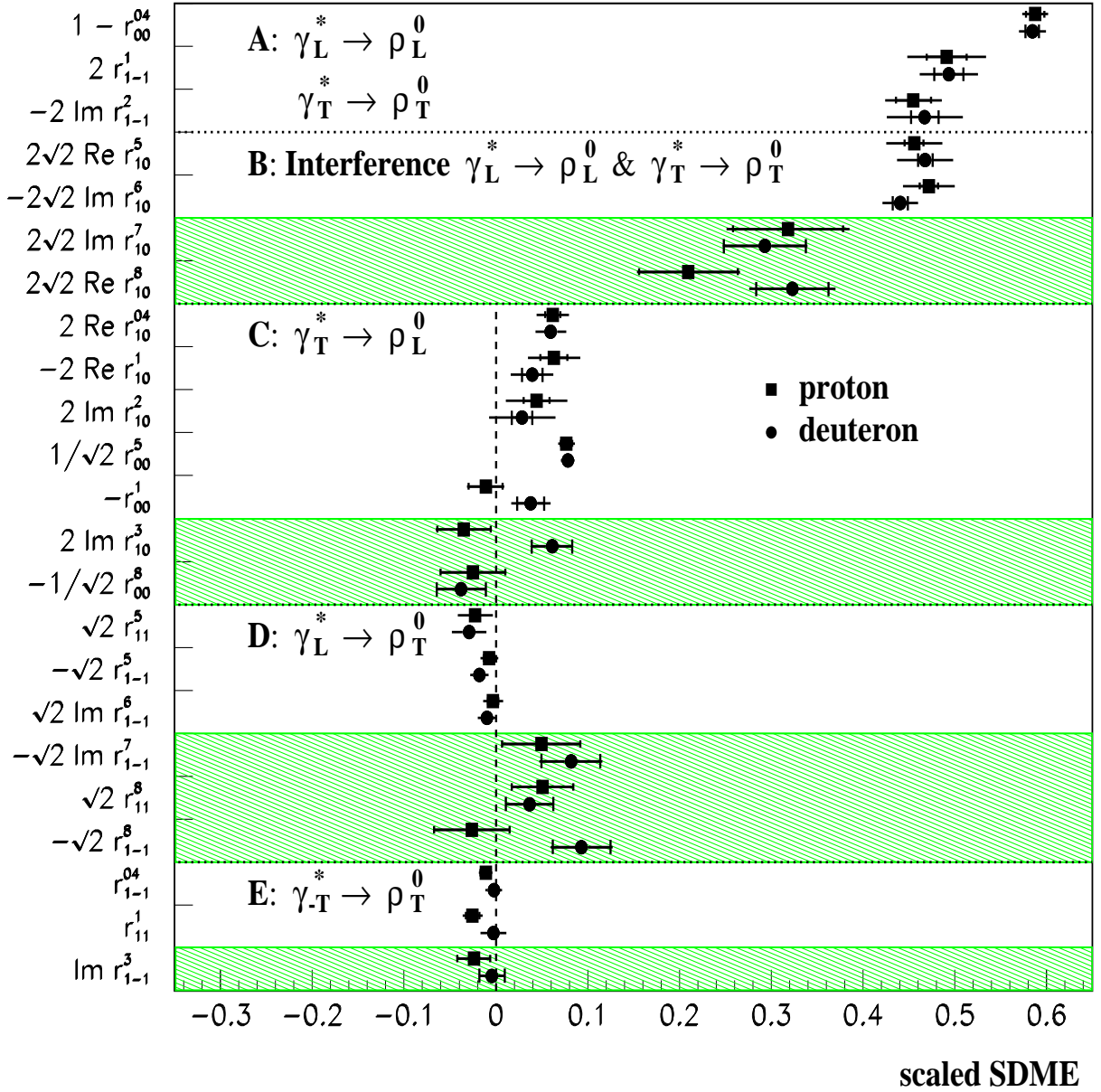}
\caption{\small
The 23 SDMEs extracted from $\rho^0$ data: proton
(squares) and deuteron (circles) in the entire  HERMES kinematics with
$\langle x \rangle  = 0.08, \langle Q^2 \rangle = 1.95$ GeV$^2$, $ \langle -t' \rangle =0.13$ GeV$^2$. 
The SDMEs are multiplied by prefactors in order 
to represent the normalized  leading contribution 
of the corresponding amplitude (see (\ref{a1}-\ref{a17})).
The inner error bars represent the statistical uncertainties, while the outer
ones indicate the statistical and systematic uncertainties added in 
quadrature. SDMEs measured with unpolarized (polarized)
beam are displayed in the unshaded (shaded)  areas. 
The vertical dashed line at zero is indicated for SDMEs expected to be
zero under the hypothesis of SCHC.
}
\label{sdme23-all}
\end{center}
\end{figure*}

Separate maximum likelihood fits to the proton and  deuteron 
data samples are performed in the entire kinematic region: 
$1~{\rm GeV}^2 < Q^{2} < 7$~GeV$^2$, $3~{\rm GeV} < W < 6.3$~GeV 
(corresponding to  $0.03 < x_B < 0.25$), 
and $0~{\rm GeV}^2 < -t' < 0.4$~GeV$^2$. 
The resulting  $\rho ^0$ meson SDMEs  $r^{\alpha}_{\lambda_{V} \lambda '_{V}}$
are listed in Tab.~\ref{tabschccheck} and displayed in Fig.~\ref{sdme23-all},
ordered  according to the classes described above.
The statistical uncertainties are larger for the eight polarized SDMEs (shown in the shaded 
areas of the figure) due to the non-unity of the
beam polarization and the kinematic suppression factor 
$\sqrt{1 - \epsilon}$ (see (\ref{eqang3})).
In order to facilitate  comparison with a recently introduced new representation of
SDMEs~\cite{mdiehl}, the proton SDMEs
in that representation 
are shown in Tab.~\ref{mdiehl} of  Appendix~E.
\begin{table*}[ht!] 
 \renewcommand{\arraystretch}{1.2} 
\begin{center}
\caption{ \label{tabschccheck} 
Values of $\rho^0$ SDMEs
for proton and deuteron data ordered in classes  by horizontal lines
according to the expected hierarchy of helicity amplitudes. Elements 
$r^{\alpha}_{ij}$ with $\alpha = 3,7,8$
are polarized.
The first uncertainty is statistical, the second 
systematic. 
The statistical significance of the magnitude of each SDME
expected to vanish in the case of SCHC is shown in column 
three or five as its absolute value in units of standard deviations of its
total uncertainty, denoted as SDME/$tot$.
Similarly, the four bottom rows of the table show deviations from zero of certain 
combinations  of 
SDMEs that would thereby violate SCHC and NPE dominance (see text). 
}
{\small
\begin{tabular}{|c|c|c|c|c|}
\hline
    & \multicolumn{2}{|c|}{proton} & \multicolumn{2}{|c|}{deuteron} \\
\hline
element & SDME $\pm stat \pm syst$ & SDME/$tot$ & SDME $\pm stat \pm syst$  & SDME/$tot$  \\ 
         
\hline
$r^{04}_{00}$    &  0.412 $\pm$ 0.010 $\pm$ 0.010 &       &  0.416 $\pm$ 0.007 $\pm$ 0.013 &   \\
$r^1_{1-1}$      &  0.246 $\pm$ 0.011 $\pm$ 0.019 &       &  0.247 $\pm$ 0.008 $\pm$ 0.014 & \\
Im $r^2_{1-1}$   & $-0.227$ $\pm$ 0.010 $\pm$ 0.012 &       & $-0.234$ $\pm$ 0.008 $\pm$ 0.019 & \\
\hline
Re $r^5_{10}$    &  0.161 $\pm$ 0.004 $\pm$ 0.010 &       &  0.165 $\pm$ 0.003 $\pm$ 0.010 &  \\
Im $r^6_{10}$    & $-0.167$ $\pm$ 0.003 $\pm$ 0.009 &       & $-0.156$ $\pm$ 0.003 $\pm$ 0.006 &  \\
Im $r^7_{10}$    &  0.112 $\pm$ 0.021 $\pm$ 0.011 &       &  0.104 $\pm$ 0.016 $\pm$ 0.004 &  \\
Re $r^8_{10}$    &  0.074 $\pm$ 0.019 $\pm$ 0.006 &       &  0.114 $\pm$ 0.014 $\pm$ 0.009 &  \\
\hline

Re $r^{04}_{10}$ &  0.031 $\pm$ 0.004 $\pm$ 0.008 & 3.5  &  0.030 $\pm$ 0.003 $\pm$ 0.008 & 3.5 \\
Re $r^1_{10}$    & $-0.032$ $\pm$ 0.007 $\pm$ 0.012 & 2.3  & $-0.020$ $\pm$ 0.006 $\pm$ 0.010 & 1.7 \\
Im $r^2_{10}$    &  0.022 $\pm$ 0.007 $\pm$ 0.015 & 1.3  &  0.014 $\pm$ 0.006 $\pm$ 0.017 & 0.8 \\
$r^5_{00}$       &  0.109 $\pm$ 0.009 $\pm$ 0.009 & 8.7  &  0.111 $\pm$ 0.007 $\pm$ 0.008 & 10.4 \\
$r^1_{00}$       &  0.011 $\pm$ 0.019 $\pm$ 0.008 & 0.6  & $-0.038$ $\pm$ 0.015 $\pm$ 0.016 & 1.7 \\
Im $r^3_{10}$    & $-0.017$ $\pm$ 0.015 $\pm$ 0.004 & 1.1  &  0.031 $\pm$ 0.011 $\pm$ 0.003 & 2.7 \\
$r^8_{00}$       &  0.035 $\pm$ 0.050 $\pm$ 0.010 & 0.7  &  0.053 $\pm$ 0.038 $\pm$ 0.006 & 1.4 \\
\hline
$r^5_{11}$       & $-0.01$ $\pm$ 0.003 $\pm$ 0.013 & 1.2  &  $-0.021$ $\pm$ 0.003 $\pm$ 0.013 & 1.6 \\
$r^5_{1-1}$      &  0.005 $\pm$ 0.004 $\pm$ 0.006 & 0.7  &   0.013 $\pm$ 0.003 $\pm$ 0.006 & 1.9 \\
Im $r^6_{1-1}$   & $-0.002$ $\pm$ 0.004 $\pm$ 0.007 & 0.3  &  $-0.007$ $\pm$ 0.003 $\pm$ 0.006 & 1.0 \\
Im $r^7_{1-1}$   & $-0.035$ $\pm$ 0.030 $\pm$ 0.005 & 1.2  &  $-0.058$ $\pm$ 0.023 $\pm$ 0.009 & 2.3 \\
$r^8_{11}$       &  0.036 $\pm$ 0.024 $\pm$ 0.001 & 1.6  &   0.026 $\pm$ 0.018 $\pm$ 0.003 & 1.4 \\
$r^8_{1-1}$      &  0.019 $\pm$ 0.029 $\pm$ 0.005 & 0.6  &  $-0.066$ $\pm$ 0.022 $\pm$ 0.008 & 2.8 \\
\hline
$r^{04}_{1-1}$   & $-0.011$ $\pm$ 0.005 $\pm$ 0.005 & 1.5  & $-0.002$ $\pm$ 0.004 $\pm$ 0.008 & 0.2  \\
$r^1_{11}$       & $-0.025$ $\pm$ 0.007 $\pm$ 0.008 & 2.3  & $-0.002$ $\pm$ 0.006 $\pm$ 0.013 & 0.1  \\
Im $r^3_{1-1}$   & $-0.024$ $\pm$ 0.018 $\pm$ 0.001 & 1.3  & $-0.004$ $\pm$ 0.014 $\pm$ 0.004 & 0.3  \\
\hline
relation &  \multicolumn{2}{|c|}{ SCHC?}  & \multicolumn{2}{|c|}{  SCHC?  } \\
\hline
$r^{1}_{1-1}$ + Im $r^{2}_{1-1}$  & 0.018   $\pm$  0.012  $\pm$  0.011 & 1.1 & 0.013  $\pm$  0.009  $\pm$  0.014 & 0.8  \\
Re $r^{5}_{10} +$ Im $r^{6}_{10}$ & $-0.006$ $\pm$  0.004  $\pm$  0.001 & 1.5 & 0.010  $\pm$  0.003  $\pm$  0.005 & 1.7 \\
Im $r^{7}_{10} -$ Re $r^{8}_{10}$ &  0.038  $\pm$  0.029  $\pm$  0.006 & 1.3  & $-0.011$ $\pm$  0.022  $\pm$  0.006 & 0.5 \\
\hline
relation &  \multicolumn{2}{|c|}{ SCHC and NPE?}  & \multicolumn{2}{|c|}{  SCHC and NPE?  } \\
\hline
$1-r^{04}_{00}-2r^{1}_{1-1}$ & 0.097  $\pm$  0.017  $\pm$  0.046  & 2.0  & 0.091   $\pm$  0.013  $\pm$   0.038  & 2.3 \\
\hline
\end{tabular} \\[2pt]
}
\end{center}
\end{table*}

\begin{sloppypar}
In Fig.~\ref{sdme23-all} the SDMEs are shown 
multiplied by certain factors in order to allow their
comparison at the level of dominant amplitudes (see (\ref{a1}-\ref{a17})).
For all classes numerical factors are chosen in 
such a way that the coefficient of the
dominant terms is equal to unity.
The plotted representatives for the elements of class A
are chosen so that their main terms are equal to $|T_{11}|^2/\mathcal{N}$;
in particular this requires that the term  $1 - r^{04}_{00}$ be chosen. 
The coefficients for class B are chosen to have the main contribution
to the plotted representatives
for the unpolarized and polarized SDMEs equal to 
$\mathrm{Re}\{T_{11}T^*_{00}\}/\mathcal{N}$ 
and  $\mathrm{Im}\{T_{11}T^*_{00}\}/\mathcal{N}$,
respectively. This corresponds to the general rule that is applicable to  
classes B to E: the dominant contribution of the unpolarized (polarized) 
element presented 
in Fig.~\ref{sdme23-all} is proportional to the real (imaginary) part
of a product of two amplitudes. Class C contains the main terms 
$T_{01}T^*_{00}/\mathcal{N}$ (for $r^5_{00}/\sqrt{2}$ 
and $r^8_{00}/\sqrt{2}$) and 
$T_{01}T^*_{11}/\mathcal{N}$. The dominant contributions for classes D
and E contain terms $T_{10}T^*_{11}/\mathcal{N}$ 
and $T_{1-1}T^*_{11}/\mathcal{N}$, respectively.
\end{sloppypar}

Given the scaled SDMEs in Fig.~\ref{sdme23-all}, it easily can be seen
that the two unpolarized SDMEs of class~B have large values, 
similar to those of class~A.
This suggests the presence of a substantial 
interference between the two dominant amplitudes 
$T_{00}$ and $T_{11}$. 
The two polarized class~B SDMEs are significantly non-zero for 
proton and deuteron as well.
It is also seen that the values of elements in class~C that
contain the dominant term $T_{01}T^*_{11}$ 
are  similar for the unpolarized SDMEs ($\mathrm{Re} \{ r^{04}_{10} \}$, 
$\mathrm{Re}\{r^{1}_{10}\}$, $\mathrm{Im}\{r^{2}_{10}\}$). 
Those unpolarized class~C elements measured with good accuracy, 
$\mathrm{Re} \{ r^{04}_{10} \}$ 
and $r^5_{00}$, are much smaller than the class~B SDMEs, 
whereas the unpolarized class~C elements  are larger 
than the unpolarized class D and class E SDMEs.
This shows that the anticipated hierarchy is supported by the data. 
For class~D SDMEs, slightly positive (negative) values are observed in the
polarized (unpolarized) case. Finally, values of class~E SDMEs for the proton target 
tend to deviate from zero,
while those for the deuteron ones are  consistent with zero.

We note that no significant difference is found between the sets of SDMEs for proton 
and deuteron, as a $\chi^2/d.o.f. = 22.5/23$ is obtained taking into account
the total uncertainties. Hence there appears to be no indication of 
significant contributions of secondary reggeons with 
isospin $I=1$ and natural parity.

\subsection{Test of the SCHC Hypothesis} 

\begin{sloppypar}
As explained in section 3.7, 
only the following seven SDMEs are not restricted to be zero in the case 
of $s$-channel helicity conservation:
$r^{04}_{00}$, $r^{1}_{1-1}$, 
Im$\{ r^{2}_{1-1}\}$,
 Re$\{r^{5}_{10} \}$, Im$\{ r^{6}_{10}\}$
and Im$\{ r^{7}_{10} \}$, Re$\{ r^{8}_{10}\}$. 
All other SDMEs are required by SCHC to be zero. 
The magnitudes of their measured 
offsets from zero, expressed in units of the standard 
deviation of their uncertainty, are shown in one of two 
separate columns of  Tab.~\ref{tabschccheck},
next to the respective SDME. 
Several elements are inconsistent with the hypothesis
of SCHC, in particular several members
of class~C.
\end{sloppypar}

The SDME $r^5_{00}$ is observed to be non-zero at the level of nine (ten)
standard deviations in the
total uncertainty for the proton (deuteron) result, proving $s$-channel 
helicity non-conservation. 
This was already  observed earlier by the 
HERA collider experiments \cite{zeussdme,H1} at a lower significance level,
and with high significance very recently~\cite{zeussdme2}. 
For the first time, HERMES observes  $s$-channel helicity non-conservation
also in other class~C SDMEs, in particular $\mathrm{Re}\{r^{04}_{10}\}$.

The polarized elements $r^8_{00}$ and 
$\mathrm{Im}\{ r^{3}_{10}\}$, related to the
terms Im$\{T_{01}T_{00}^*\}$ and
Im$\{T_{01}T_{11}^*\}$ respectively (\ref{a16},\ref{a21}), 
are extracted 
using a longitudinally polarized lepton beam for the first time.
Both elements are consistent with zero (Figs.~\ref{sdme23-all}, 
\ref{q2-class-c}) within 
the uncertainties.

The relations  imposed by the hypothesis of SCHC 
(\ref{schc01}-\ref{schc03}) are satisfied
within about one standard deviation of the total uncertainty, 
as can be seen from  the corresponding rows of Tab.~\ref{tabschccheck}.
The sensitivity of these relations to  SCHC is low.
In the case of the relation (\ref{schc01})
only the contributions of small double-helicity-flip amplitudes 
(see (\ref{a7},\ref{a9}) ) violate SCHC.
For  the relations (\ref{schc02}-\ref{schc03}), 
equations (\ref{a12}-\ref{a22})
show that the largest SCHC amplitude $T_{00}$ is multiplied by the
smallest $T_{1-1}$ amplitude in the terms that violate SCHC.
The relation  corresponding to the combined  hypotheses of SCHC
and NPE dominance (\ref{schcnpe}) is mar\-gin\-ally violated by
two standard deviations in the total uncertainty.
In evaluating the uncertainties of these relations, correlations between the
corresponding elements (see Tabs.~\ref{tab-corr-h},\ref{tab-corr-d}), 
are taken into account. 

\subsection{Phase Difference between $ \mathbf{T_{11}}$ and $ \mathbf{T_{00}}$}

\begin{figure}[h!]
\includegraphics[height=7.5cm,width=7.5cm]{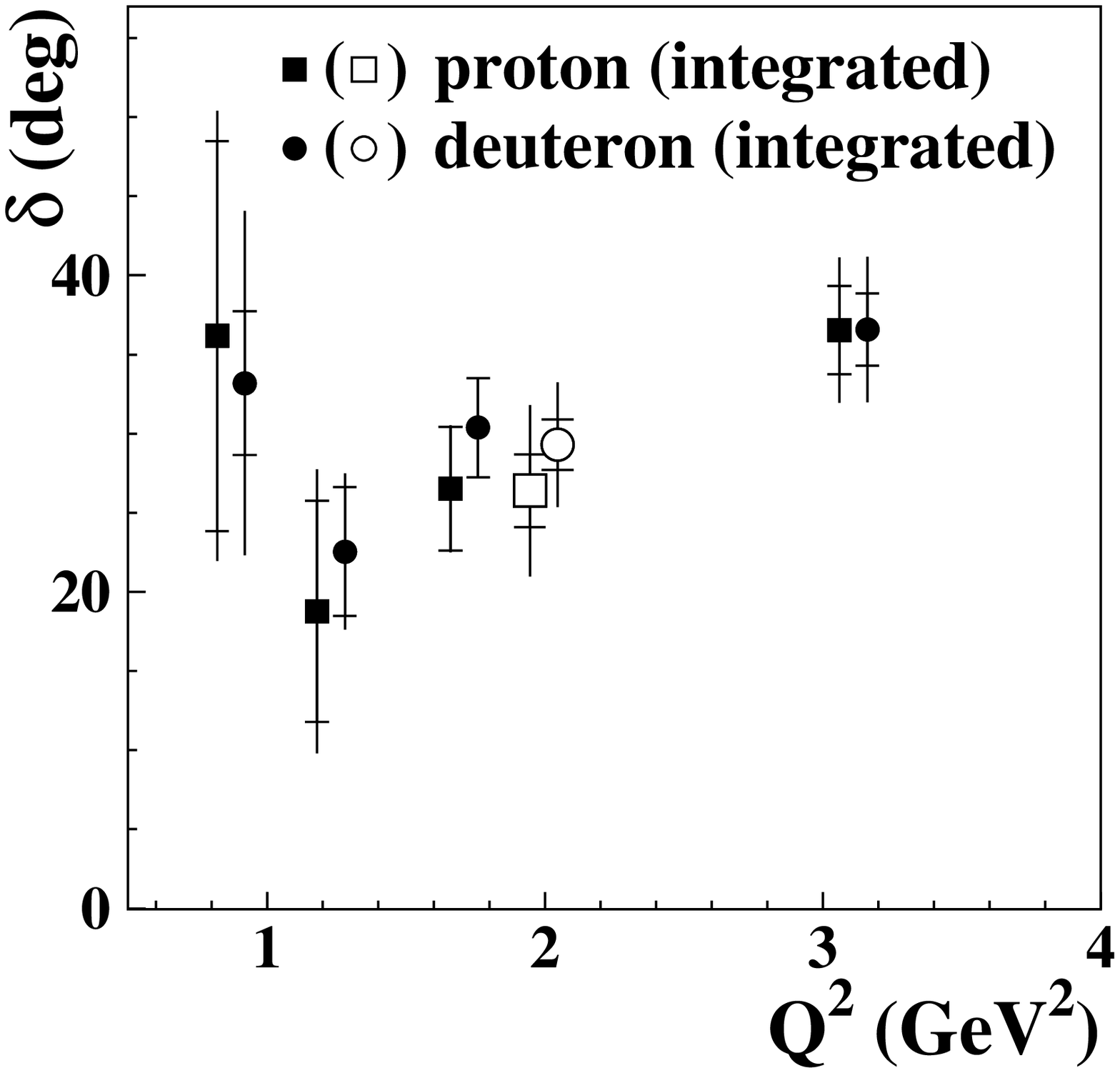}
\caption{\small
The $Q^{2}$ dependence of the phase difference $\delta$ between $T_{11}$ and $T_{00}$ amplitudes   
calculated according to (\ref{eq:cosdelta}) for the 
proton (filled squares) and deuteron (filled circles) data.
The values of $\delta$, for yields
integrated over the range $1~{\rm GeV}^2 < Q^2 < 7$ GeV$^2$, are shown as open
symbols. The inner (outer) bars represent the statistical (total)
uncertainty. 
}
 \vspace*{0.0cm}
\label{deltafig-clb}
\end{figure}

The phase difference $\delta$ between the amplitudes $T_{11}$ and
 $T_{00}$ can be evaluated as follows:
\begin{equation}
\cos \delta= \frac{ 2  \sqrt{\epsilon} (\mathrm{Re}\{r^{5}_{10}\}-\mathrm{Im}\{r^{6}_{10} \})}
{\sqrt{ r^{04}_{00}(1-r^{04}_{00}+r^{1}_{1-1}-\mathrm{Im}\{r^{2}_{1-1}\} )} }   . 
\label{eq:cosdelta}
\end{equation}
This results in
$|\delta|= 26.4 \pm  2.3_{stat} \pm   4.9_{syst}$  degrees for the proton and
$|\delta|= 29.3 \pm   1.6_{stat} \pm  3.6_{syst}$ degrees for the deuteron
(see  Fig.~\ref{deltafig-clb}). 
Using polarized SDMEs, also the sign of $\delta$ can be determined:
\begin{equation}
\sin \delta= \frac{ 2  \sqrt{\epsilon} (\mathrm{Re}\{r^{8}_{10}\}+\mathrm{Im}\{r^{7}_{10}\} ) } 
{\sqrt{  r^{04}_{00}(1-r^{04}_{00}+r^{1}_{1-1}-\mathrm{Im}\{r^{2}_{1-1} \}) }} . 
\label{eq:sindelta}
\end{equation}
Equations (\ref{eq:cosdelta}) and (\ref{eq:sindelta}) are derived in 
Appendix~B.
Second order contributions of
spin-flip amplitudes are neglected in obtaining these formulae.

Using (\ref{eq:sindelta}) it is determined, 
for the first time, 
that the {\it sign} 
of $\delta$ is positive:
$\delta= 30.6 \pm 5.0_{stat} \pm  2.4_{syst}$  degrees for the proton and
$\delta= 36.3 \pm 3.9_{stat} \pm  1.7_{syst}$  for the deuteron.
These values are consistent with each other and their magnitudes are in agreement 
with the results obtained  with (\ref{eq:cosdelta}) for $\cos{\delta}$.

We note that in the GPD-based model of Ref.~\cite{golos3}, the phase
difference between  the amplitudes $T_{11}$ and
$T_{00}$ is found to have a value of about 3 degrees. This appears to be 
inconsistent with the
HERMES results and also, to a lesser extent, with the H1 results~\cite{H1};
the two experimental results agree within their total uncertainties.

\begin{figure*}[hp!]
 \begin{center}
\includegraphics[height=4.5cm,width=4.cm]{./pln/paw-q2dep-1.epsi} \hspace*{0.3cm}
\includegraphics[height=4.5cm,width=4.cm]{./pln/paw-q2dep-2.epsi} \hspace*{0.3cm}
\includegraphics[height=4.5cm,width=4.cm]{./pln/paw-q2dep-3.epsi}

\vspace*{0.3cm}
\includegraphics[height=4.5cm,width=4.cm]{./pln/paw-tdep-1.epsi} \hspace*{0.3cm}
\includegraphics[height=4.5cm,width=4.cm]{./pln/paw-tdep-2.epsi} \hspace*{0.3cm}
\includegraphics[height=4.5cm,width=4.cm]{./pln/paw-tdep-3.epsi}
\caption{ {\small
$Q^{2}$ and $t'$ dependences of class~A
SDMEs describing the dominant transitions 
$\gamma^*_L \to \rho^0_L$ and $\gamma^*_T \to \rho^0_T$.
Filled squares (circles)
correspond to proton (deuteron) data.
Total uncertainties are depicted,  calculated as statistical and systematic uncertainties 
combined in quadrature.
Deuterium data points are presented with a small horizontal
offset here and in Figs.~(\ref{q2-class-b}-\ref{tau-fig}) 
to improve their visibility.
 }\label{q2-class-a}
 }
\end{center}
\vspace*{0.2cm}
 \begin{center}
\includegraphics[height=4.5cm,width=3.9cm]{./pln/paw-q2dep-4.epsi} \hspace*{0.1cm}
\includegraphics[height=4.5cm,width=3.9cm]{./pln/paw-q2dep-5.epsi} \hspace*{0.1cm}
\includegraphics[height=4.5cm,width=3.9cm]{./pln/paw-q2dep-6.epsi} \hspace*{0.1cm}
\includegraphics[height=4.5cm,width=3.9cm]{./pln/paw-q2dep-7.epsi} \\
\vspace*{0.4cm}
\includegraphics[height=4.5cm,width=3.9cm]{./pln/paw-tdep-4.epsi} \hspace*{0.1cm}
\includegraphics[height=4.5cm,width=3.9cm]{./pln/paw-tdep-5.epsi} \hspace*{0.1cm}
\includegraphics[height=4.5cm,width=3.9cm]{./pln/paw-tdep-6.epsi} \hspace*{0.1cm}
\includegraphics[height=4.5cm,width=3.9cm]{./pln/paw-tdep-7.epsi}
\caption{ {\small
 $Q^{2}$ and $t'$ dependences of the class B SDMEs describing
the interference of the  dominant transitions 
 $\gamma^*_L \to \rho^0_L$ and  $\gamma^*_T \to \rho^0_T$.
Filled squares (circles)
correspond to proton (deuteron) data. 
Total uncertainties are depicted,  calculated as statistical and systematic uncertainties 
combined in quadrature. 
 }\label{q2-class-b}
 }
\end{center}
\end{figure*}

\section{The $\mathbf{Q^2}$ and $\mathbf{t'}$ Dependences of the SDMEs}

In the following,  the $Q^2$ dependences are presented in four bins,
where the first bin is defined by $0.5~{\rm GeV}^2 < Q^2 < 1$ GeV$^2$ 
with $\langle Q^2 \rangle = 0.83$ GeV$^2$. 
For the $t'$ dependences, also shown in four bins,
only  data with $Q^2 > 1$ GeV$^2$ are included. The average value
of  $t'$ is almost independent of $Q^2$ and $W$.

\subsection{Class A: Dominant Transitions
$\mathbf{ \gamma^*_L \to \rho^0_L}$ and $\mathbf{ \gamma^*_T \to \rho^0_T}$ }

Class~A comprises SDMEs corresponding to the dominant transitions, 
$\gamma^*_T \rightarrow \rho^0_T$ and $\gamma^*_L \rightarrow \rho^0_L$,
described by the amplitudes  $T_{11}$ and  $T_{00}$.
The $Q^2$ and $t'$ dependences for the class A SDMEs 
$1 - r^{04}_{00}$, $r^{1}_{1-1}$, and $\mathrm{Im}\{r^{2}_{1-1} \}$
are shown in Fig.~\ref{q2-class-a}. 
The three elements exhibit somewhat similar $Q^2$ dependences. 
They are found to be approximately constant over the measured $t'$ range,
as also observed by ZEUS~\cite{zeussdme2} for  $r^{04}_{00}$ 
at average $Q^2$ values of 3 and 10 GeV$^2$.
Such a $t'$ independence indicates similar $t'$-slopes for longitudinal
and transverse components of the vector meson production cross section.

We note that there is good agreement
between the highest $Q^2$ points of the HERMES proton data and the
GPD-based model calculations of Ref.~\cite{golos3}.

\subsection{Class B: Interference of 
$\mathbf{\gamma^*_L \to \rho^0_L}$ and $\mathbf{\gamma^*_T \to \rho^0_T}$
Transitions}

Class B comprises SDMEs describing the interference of the dominant transitions
$\gamma^*_T \rightarrow \rho^0_T$ and $\gamma^*_L \rightarrow \rho^0_L$, {\it i.e.,}
those corresponding to a product of the amplitude
$T_{11}$ and the complex conjugate of $T_{00}$. Polarized 
(unpolarized) SDMEs correspond to the real (imaginary) part of this
product.

\begin{sloppypar}
Figure~\ref{q2-class-b} shows the $Q^2$ and $t'$ dependence of these SDMEs. 
It is apparent that the SCHC 
relations (\ref{schc02}) and (\ref{schc03}) are approximately fulfilled over 
the measured kinematic ranges. 
Considering (\ref{a12}\--\ref{a22}),
this implies that  contributions of single- and
double-heli\-ci\-ty-flip amplitudes are small.
\end{sloppypar}

An indication of a $Q^2$ dependence of the phase difference $\delta$ 
between the amplitudes $T_{11}$ and $T_{00}$ 
(see (\ref{eq:cosdelta})) 
is presented in Fig.~\ref{deltafig-clb}. The result of a fit with a linear 
$Q^2$ dependence has a $\chi^2/d.o.f. =$ 1.41/2 (1.42/2) for the proton (deuteron) data, which is 
smaller than the fit result with no $Q^2$ dependence:  $\chi^2/d.o.f. =$ 4.52/3 (4.38/3).
Note that at the lowest $Q^2$, the
value of $\delta$ has the largest systematic uncertainty due to the rapidly falling acceptance of 
the HERMES spectrometer. No $t'$ dependence of $\delta$ is observed, for either target.

\begin{figure*}[hp!]
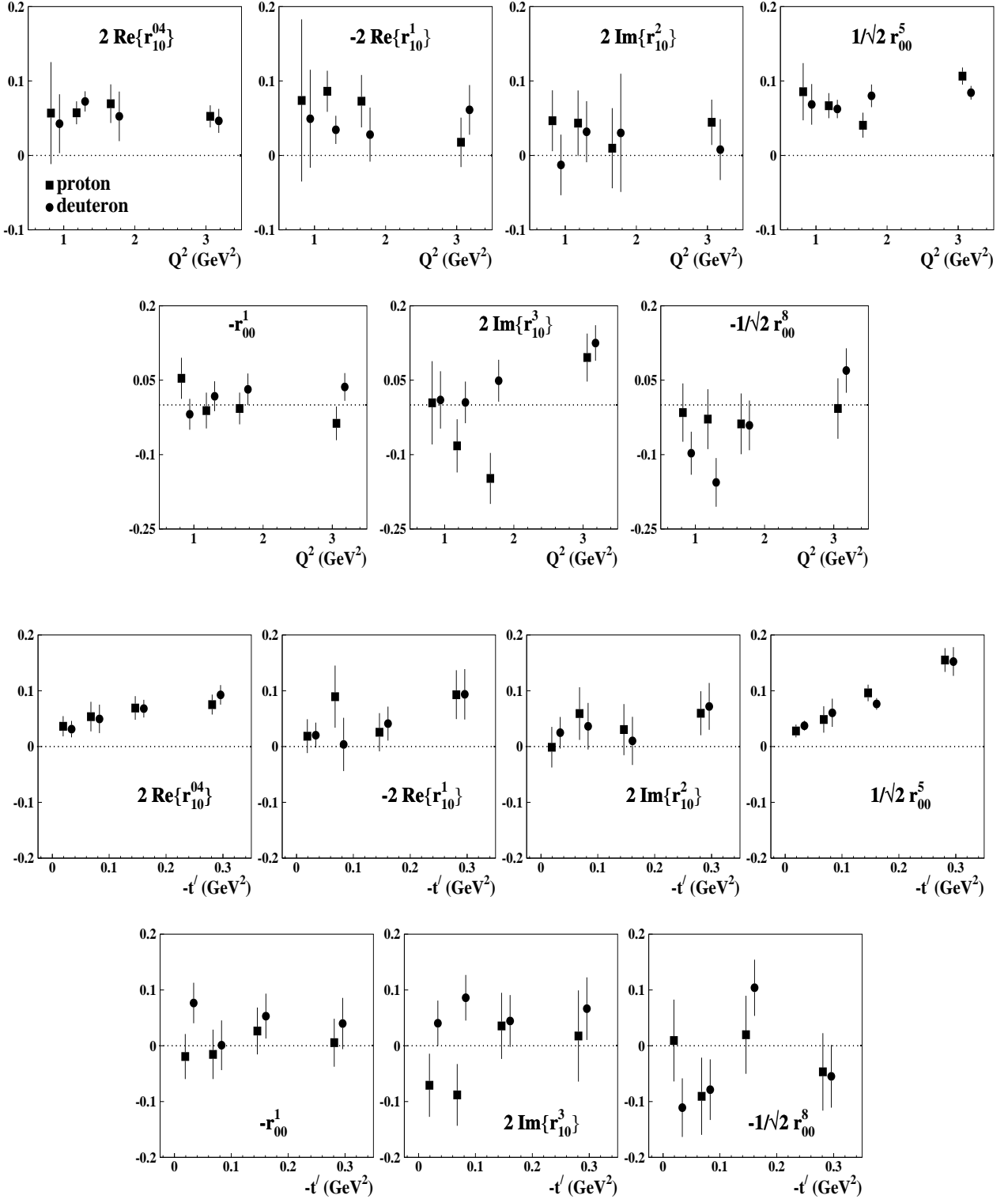

\begin{center}
\includegraphics[height=4.5cm,width=4.cm]{./pln/paw-q2dep-8.epsi} \hspace*{0.cm}
\includegraphics[height=4.5cm,width=4.cm]{./pln/paw-q2dep-9.epsi} \hspace*{0.cm}
\includegraphics[height=4.5cm,width=4.cm]{./pln/paw-q2dep-10.epsi} \hspace*{0.cm}
\includegraphics[height=4.5cm,width=4.cm]{./pln/paw-q2dep-11.epsi}

\vspace*{0.5cm}
\includegraphics[height=4.5cm,width=4.cm]{./pln/paw-q2dep-12.epsi} \hspace*{0.cm}
\includegraphics[height=4.5cm,width=4.cm]{./pln/paw-q2dep-13.epsi} \hspace*{0.cm}
\includegraphics[height=4.5cm,width=4.cm]{./pln/paw-q2dep-14.epsi}

\vspace*{1.0cm}
\includegraphics[height=4.5cm,width=4.cm]{./pln/paw-tdep-8.epsi}
\includegraphics[height=4.5cm,width=4.cm]{./pln/paw-tdep-9.epsi}
\includegraphics[height=4.5cm,width=4.cm]{./pln/paw-tdep-10.epsi}
\includegraphics[height=4.5cm,width=4.cm]{./pln/paw-tdep-11.epsi}

\vspace*{0.5cm}
\includegraphics[height=4.5cm,width=4.cm]{./pln/paw-tdep-12.epsi}
\includegraphics[height=4.5cm,width=4.cm]{./pln/paw-tdep-13.epsi}
\includegraphics[height=4.5cm,width=4.cm]{./pln/paw-tdep-14.epsi}
\end{center}
\caption{\small
$Q^2$ and $t'$ dependence of the class C SDMEs describing the 
interference
of the helicity-flip transition
$\gamma^*_T \to \rho^0_L$ and one of the dominant helicity-conserving transition. 
Filled squares (circles)
correspond to proton (deuteron) data. 
Total uncertainties are depicted,  calculated as statistical and systematic uncertainties 
combined in quadrature. }
\label{q2-class-c}
\end{figure*}

\begin{figure*}[hp!]
 \begin{center}
\includegraphics[height=4.5cm,width=4.cm]{./pln/paw-q2dep-15.epsi} \hspace*{0.1cm} 
\includegraphics[height=4.5cm,width=4.cm]{./pln/paw-q2dep-16.epsi} \hspace*{0.1cm}
\includegraphics[height=4.5cm,width=4.cm]{./pln/paw-q2dep-17.epsi}

\vspace*{0.4cm}
\includegraphics[height=4.5cm,width=4.cm]{./pln/paw-q2dep-18.epsi} \hspace*{0.1cm}
\includegraphics[height=4.5cm,width=4.cm]{./pln/paw-q2dep-19.epsi} \hspace*{0.1cm}
\includegraphics[height=4.5cm,width=4.cm]{./pln/paw-q2dep-20.epsi}

\vspace*{0.80cm}
\includegraphics[height=4.5cm,width=4.cm]{./pln/paw-tdep-15.epsi} \hspace*{0.1cm}
\includegraphics[height=4.5cm,width=4.cm]{./pln/paw-tdep-16.epsi} \hspace*{0.1cm}
\includegraphics[height=4.5cm,width=4.cm]{./pln/paw-tdep-17.epsi}

\vspace*{0.4cm}
\includegraphics[height=4.5cm,width=4.cm]{./pln/paw-tdep-18.epsi} \hspace*{0.1cm}
\includegraphics[height=4.5cm,width=4.cm]{./pln/paw-tdep-19.epsi} \hspace*{0.1cm}
\includegraphics[height=4.5cm,width=4.cm]{./pln/paw-tdep-20.epsi}
\caption{\small
$Q^{2}$ and $t'$ dependences of class D SDMEs describing the interference of the
helicity-flip transition
$\gamma^*_L \to \rho^0_T$ and  
the dominant transition  $\gamma^*_T \to \rho^0_T$.  
Filled squares (circles)
correspond to proton (deuteron) data.
Total uncertainties are depicted,  calculated as statistical and systematic uncertainties 
combined in quadrature.
 }
\label{q2-class-d}
\end{center}
\end{figure*}

\begin{figure*}
 \begin{center}
\includegraphics[height=4.5cm,width=4.cm]{./pln/paw-q2dep-21.epsi}  \hspace*{0.1cm}
\includegraphics[height=4.5cm,width=4.cm]{./pln/paw-q2dep-22.epsi} \hspace*{0.1cm}
\includegraphics[height=4.5cm,width=4.cm]{./pln/paw-q2dep-23.epsi} \hspace*{0.1cm}

\vspace*{0.5cm}
\includegraphics[height=4.5cm,width=4.cm]{./pln/paw-tdep-21.epsi} \hspace*{0.1cm}
\includegraphics[height=4.5cm,width=4.cm]{./pln/paw-tdep-22.epsi} \hspace*{0.1cm}
\includegraphics[height=4.5cm,width=4.cm]{./pln/paw-tdep-23.epsi} \hspace*{0.1cm}
\caption{ {\small
$Q^{2}$ and $t'$ dependences of the class~E 
SDMEs  describing the interference
of the double-helicity-flip transition $\gamma^*_{-T} \to \rho^0_{T}$ and the
dominant transition $\gamma^*_T \to \rho^0_T$. 
Filled squares (circles) correspond to proton (deuteron) data. 
 Total uncertainties are depicted,  calculated as statistical and systematic uncertainties 
combined in quadrature. 
} }
\label{q2-class-e}
\end{center}
\end{figure*}

\subsection{Class C: Helicity-Flip Transition $\mathbf{\gamma^*_T \to \rho^0_L}$ }

Class~C consists of all those SDMEs with the main term containing 
a product of the s-chan\-nel helicity violating  amplitude $T_{01}$
describing the helicity-flip transition  $\gamma^*_T \to \rho^0_L$, 
and the complex conjugate of $T_{00}$,  $T_{11}$ or $T_{01}$,
(see (\ref{a2}-\ref{a21})). 
No clear $Q^2$ dependence is observed for class~C SDMEs (see Fig.~\ref{q2-class-c}).

As suggested by general properties of helicity-flip 
kinematics at low $t'$ values,  
a dependence on $\sqrt{-t'}$  that monotonically increases from zero 
is expected for the amplitude $T_{01}$~\cite{mdiehl}.
This is consistent with the measured SDMEs 
containing this amplitude, as is clearly seen for  $r^5_{00}$
and $\mathrm{Re}\{ r^{04}_{10}\}$ in the third row of Fig.~\ref{q2-class-c}.

\subsection{Class D: Helicity-Flip  Transition  $\mathbf{\gamma^*_L \to \rho^0_T}$ }

Class~D consists of SDMEs for which the main terms in (\ref{a10}-\ref{a23})
contain a product of 
the small helicity-flip amplitude $T_{10}$ with the complex conjugate of $T_{11}$.
Unpolarized (polarized) SDMEs represent the real (imaginary) part
of this product.
Correspondingly, they describe the interference of the helicity-flip transition
$\gamma^*_L \rightarrow \rho^0_T$ with the helicity-conser\-ving transition
$\gamma^*_T \rightarrow \rho^0_T$.  
Figure~\ref{q2-class-d} shows that the class D SDMEs depends only weakly 
on $Q^2$ and $t'$, and are consistent with zero as $-t' \to 0$, as expected.


\subsection{Class E: Double Helicity-Flip  Transition $\mathbf{\gamma^*_{-T} \to \rho^0_T}$}

Class~E consists of the SDMEs with the main term describing the  
interference of the transition
$\gamma^*_{-T} \rightarrow \rho^0_{T}$ with $\gamma^*_T \rightarrow \rho^0_{T}$.
This corresponds to a product of the double spin-flip amplitude  $T_{1-1}$
with the  complex conjugate of the  helicity-con\-serving amplitude  $T_{11}$.
Unpolarized (polarized) SDMEs represent 
the real (imaginary) part of this product.
Their $Q^2$ and $t'$ dependences are presented in Fig.~\ref{q2-class-e},
where it can be seen that the class~D SDMEs depend only weakly 
on $Q^2$ and $t'$, and are consistent with zero as $-t' \to 0$, as expected.

\section{Unnatural-Parity Exchange \label{unpesec}}

For $\rho^0$ production on the proton, or incoherent production
on the deuteron,
the existence of  un\-natural-parity exchange can be tested 
by evaluating the following combinations of SDMEs:
\begin{equation} \label{eq:req-npe}
u_1=1-r_{00}^{04} + 2 r_{1-1}^{04} - 2 r_{11}^{1} - 2 r_{1-1}^{1}, 
\end{equation}
\begin{equation} \label{eq:u2def}
u_2=r^{5}_{11}+r^{5}_{1-1},  
\end{equation}
\begin{equation} \label{eq:u3def}
u_3=r^{8}_{11}+r^{8}_{1-1}.
\end{equation}
If UPE is absent, all three combinations are expected to vanish
without resort to SCHC.
A non-zero result for 
\begin{eqnarray} \label{eq:npe5} 
u_1= \widetilde{\sum} \{ 4 \epsilon |U_{10}|^2+
2|U_{11}+U_{-11}|^2 \}/ \mathcal{N}
\end{eqnarray}
indicates the existence of UPE contributions, while the value for 
\begin{eqnarray} \label{eq:npe6}
\hspace{-0.5cm}  u_2+iu_3=\sqrt{2} \;  \widetilde{\sum}
\{ (U_{11}+U_{-11})^*U_{10} \}/ \mathcal{N}
\end{eqnarray}
can vanish despite the existence of UPE contributions. Such a behavior 
can be explained if a hierarchy exists also for unnatural-parity-exchange 
amplitudes:
\begin{eqnarray} \label{hier}
  \widetilde {\sum} |U_{11}|^2 \gg  \widetilde {\sum}  |U_{10}|^2,  
\widetilde {\sum} |U_{01}|^2, \widetilde {\sum} |U_{-11}|^2.
\end{eqnarray}
This hierarchy is analogous to (\ref{hierarchy}) and can be assumed to be
a general property of UPE amplitudes.

\begin{figure}[ht!]
\begin{center}
\includegraphics[height=4.5cm,width=3.95cm]{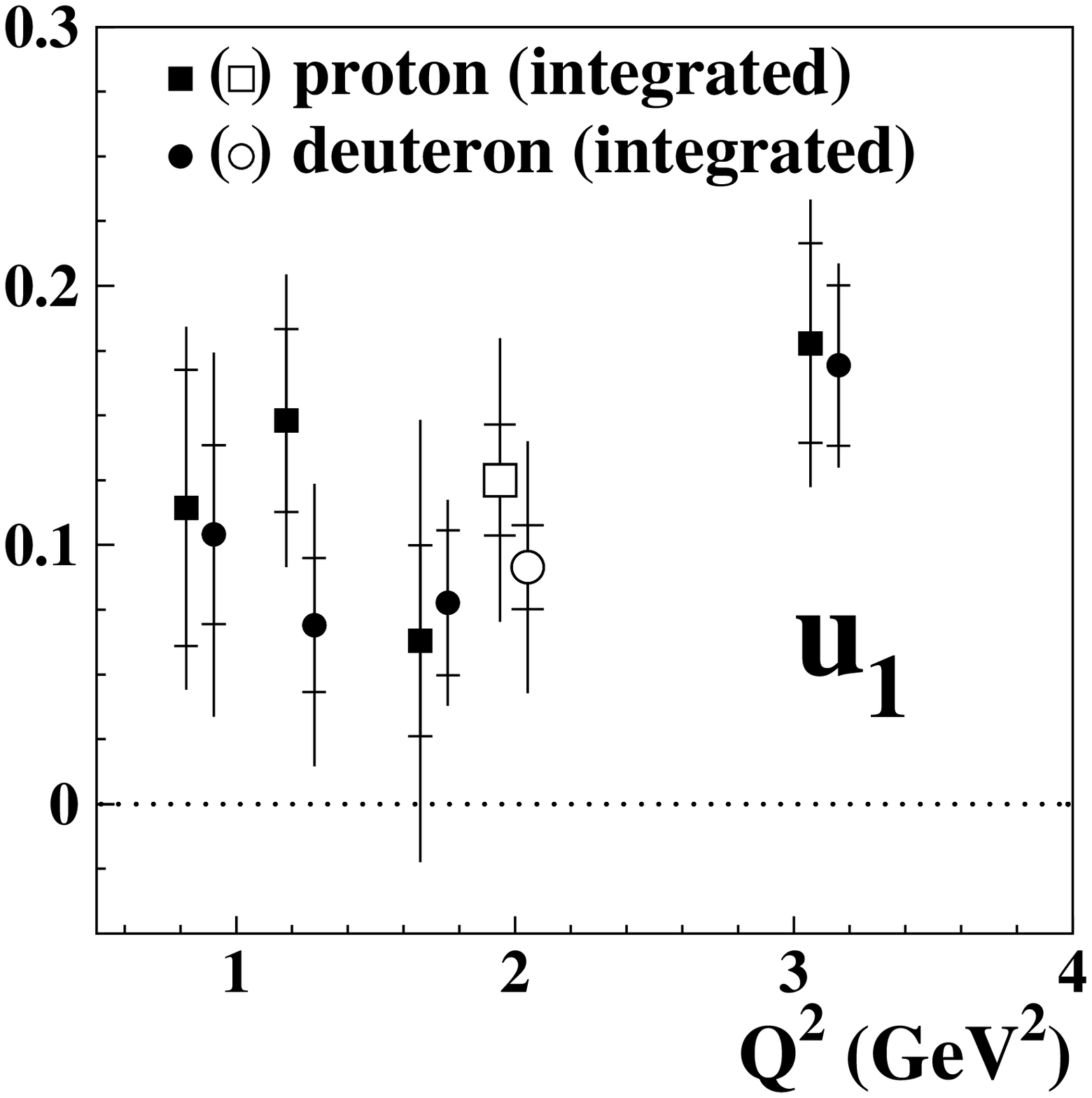} 
\hspace*{0.1cm}
\includegraphics[height=4.5cm,width=3.95cm]{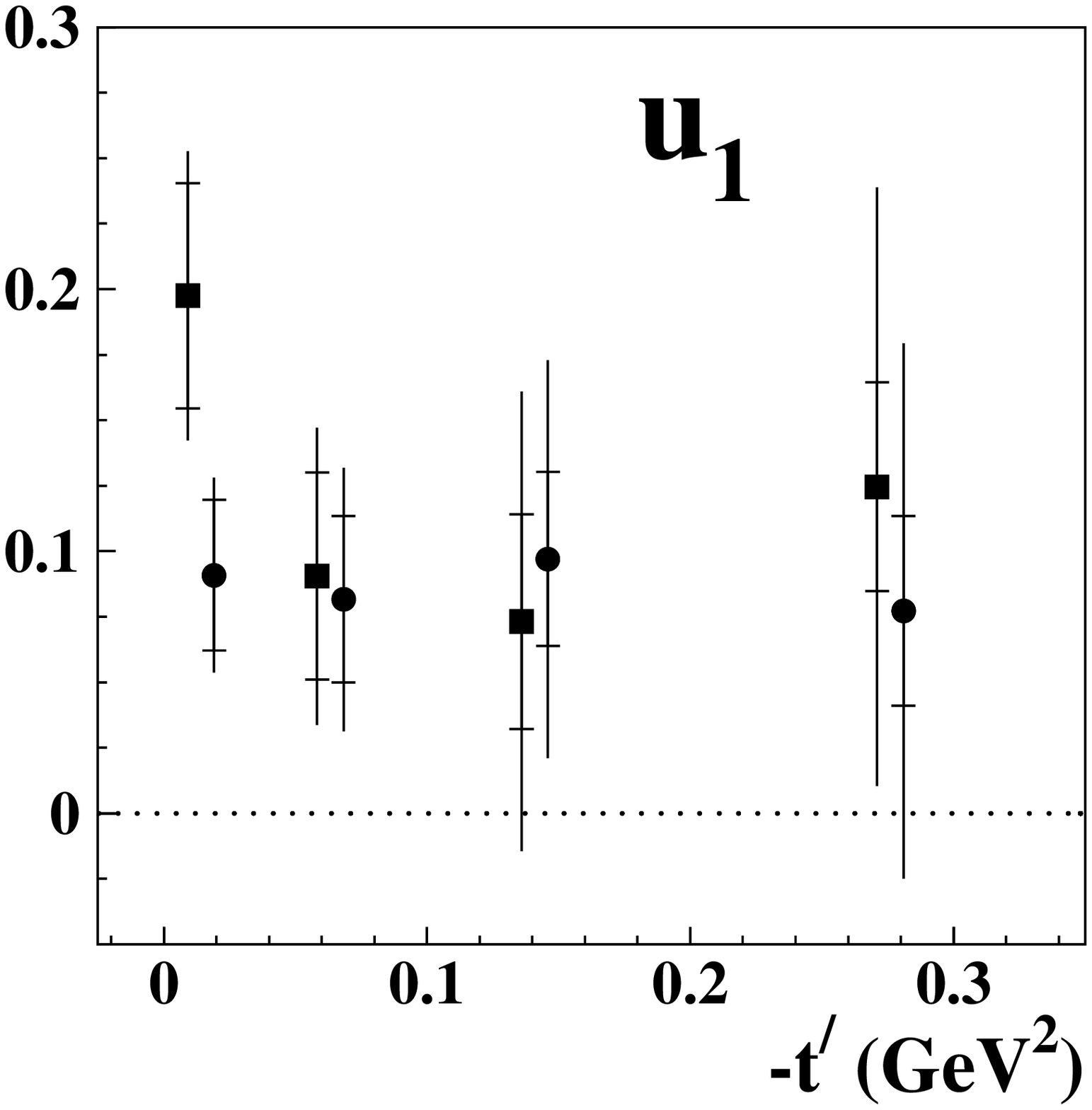}
\end{center}
\caption{ {\small
The $Q^{2}$ and $t'$ dependence of 
$u_1 = 1-r_{00}^{04} + 2 r_{1-1}^{04} - 2 r_{11}^{1} - 2 r_{1-1}^{1}$ for
proton (filled squares) and deuteron (filled circles) data.
The values of $u_1$ for yields integrated over the 
range $1~{\rm GeV}^2 < Q^2 < 7$ GeV$^2$ are shown as open symbols. 
The inner (outer) error bars represent the statistical (total)
uncertainties. 
}}
\label{upedata}
\end{figure}

The proton result 
$u_1=  0.125 \pm 0.021_{stat} \pm  0.050_{syst}$
for the entire HERMES kinematic region
differs from zero at a level of  more than two standard deviations in  
the total uncertainty, 
suggesting the existence of unnatural-parity-exchange contributions.
The deuteron result 
$u_1= 0.091 \pm  0.016_{stat} \pm 0.046_{syst}$ 
also exceeds zero, but with
smaller significance. Note that for both targets, systematic uncertainties strongly
dominate.
The dependences on $Q^2$ and $t'$ of $u_1$ for
the proton and deuteron are presented in Fig.~\ref{upedata} and Tab.~\ref{tab-npe}.
Although the uncertainties are large due to the large number of SDMEs involved 
in relation~(\ref{eq:req-npe}), all measured values of $u_1$ are positive over the
whole kinematic range. For the calculation of uncertainties in (\ref{eq:req-npe}), the
correlations between SDMEs are taken into account (see Tabs.~\ref{tab-corr-h},\ref{tab-corr-d}).

For coherent $\rho^0$ production on the deuteron (isospin zero), only isoscalar 
meson exchange is allowed; 
hence there are no contributions from $\pi$, $a_1$, or $b_1$ exchange.

The incoherent contribution to the cross section on the neutron is expected
to have an unnatural-parity-exchange contribution similar to 
that for the proton. 
The resulting value of $u_1$ for the deuteron is hence expected to be smaller  
than that for the proton due to the admixture of coherent scattering.
A  possible  indication of this behavior is observed in the lowest $t'$ 
bin of the right section of Fig.~\ref{upedata}, where $u_1$ of  
the proton exceeds that of the deuteron.

\begin{figure}
\begin{center}
\includegraphics[height=8.0cm,width=8.0cm]{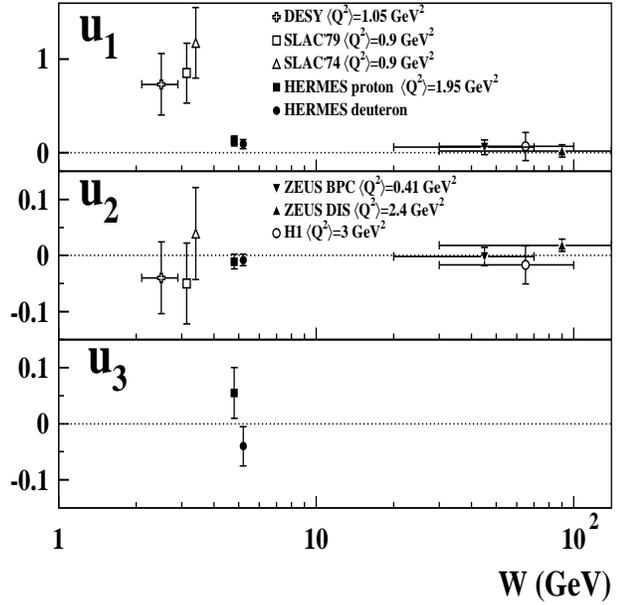} 
\end{center}
\caption{\small
Average values of $u_1$, $u_2$ and $u_3$ calculated 
according to (\ref{eq:req-npe}-\ref{eq:u3def})
from HERMES proton  
(filled squares) and deuteron (filled circles) SDMEs
are shown together with the values 
calculated from published SDMEs from DESY~\cite{DESY2},
SLAC'79~\cite{SLAC1}, SLAC'74~\cite{SLAC3}, ZEUS BPS~\cite{zeussdme}, 
ZEUS DIS~\cite{zeussdme2} and H1~\cite{H1}.
For HERMES (other experiments) 
systematic uncertainties are combined in quadrature with (without) accounting for correlations
between the SDMEs.
The HERMES deuteron and SLAC'74 data points are presented with a small horizontal
offset to improve their visibility. 
}
\label{upe-fig}
\end{figure}

The HERMES results on $u_{1},u_{2}$, and $u_3$ are 
presented in Fig.~\ref{upe-fig} and in the top section of 
Tab.~\ref{npe-tests}. The value of $u_3$ is measured
here for the first time.
The combination of proton and deuteron data
shows the existence of UPE amplitudes on the level of almost
three standard deviations in the total uncertainty: $u_1^{p+d} = 0.106 \pm 0.036$. 
In addition, results on $u_1$ and $u_2$ 
from other experiments are given in Fig.~\ref{upe-fig} and in
the bottom panel of Tab.~\ref{npe-tests}. 
While $u_2$ is measured to be compatible with zero by all experiments,
$u_1$ is found to be consistent with zero only for  high values of $W$, as
expected for $\pi$, $a_1$, and $b_1$ exchanges. For low values of $W$, the averaged result 
from the older measurements,
 $u_1^{low W}= 0.70 \pm 0.16$, is in agreement with 
the conclusion that UPE amplitudes exist at HERMES kinematics.

It is worth recalling that the existence of 
unnatural-parity exchange
in $\rho^0$production
by a virtual photon, with longitudinally polarized beam and target,
results in a double-spin asymmetry~\cite{kochelev}. 
At HERMES~\cite{dsa-lipka} this asymmetry was found to be non-zero for the
proton, with a significance of
1.7 standard deviations of the total uncertainty; the asymmetry for the deuteron
was smaller, as discussed in Refs.~\cite{ipivanov,kochelev}.

We note that there is no agreement between the HERMES measured value of
$u_1$ at $Q^2=3$ GeV$^2$ and values of $u_1$ calculated in
variants of a GPD-based model~\cite{golos3}.

\section{Contribution of the Helicity-Flip and UPE Amplitudes to the Full Cross Section 
\label{tot-crsec}}

\begin{figure}
\begin{center}
\includegraphics[height=8.0cm,width=8.0cm]{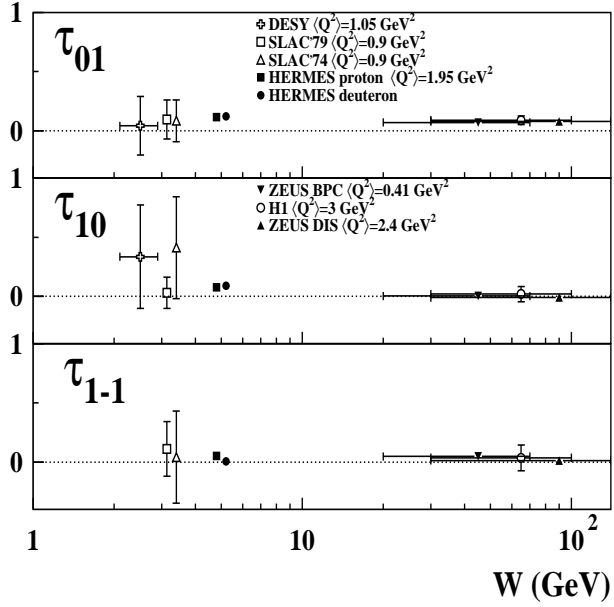} 
\end{center}
\caption{\small
Ratios of certain helicity-flip amplitudes to the 
square root of the sum of all amplitudes squared:  
$\tau_{01}$, $\tau_{10}$, and  $\tau_{1-1}$.
HERMES results on proton (filled squares)  and deuteron (filled circles)
are calculated according to (\ref{tau01f}-\ref{tau1m1f}), while
results from DESY~\cite{DESY2},
SLAC'79~\cite{SLAC1}, SLAC'74~\cite{SLAC3}, ZEUS BPS~\cite{zeussdme}, 
ZEUS DIS~\cite{zeussdme2} and H1~\cite{H1}
are  calculated according  to (\ref{tau01}-\ref{tau1m1}).
For HERMES (other experiments) 
systematic uncertainties are combined in quadrature with (without) accounting for correlations
between the SDMEs. The 
HERMES deuteron and SLAC'74 data points are presented with a small horizontal
offset to improve their visibility. 
}
\label{tau-fig}
\end{figure}

Non-conservation of $s$-channel helicity arises from the existence of non-zero 
helicity-single-flip and/or helicity-double-flip amplitudes.  
It can be quantified by measuring ratios   $\tau_{ij}$, of helicity-flip 
amplitudes $T_{ij}$ to the square root of the sum of all
amplitudes squared,  
\begin{eqnarray}
\tau_{ij} =   \frac{  |T_{ij}|}{\sqrt{\mathcal{N}}},
\label{tau000}
\end{eqnarray}
with $\mathcal{N} = \epsilon \mathcal{N}_L + \mathcal{N}_T$ as defined in Section~3.
The squared ratio $\tau_{ij}^2$ represents the fractional contribution from 
the amplitude $T_{ij}$ to the full cross section. The  $\tau_{ij}$'s
can be expressed in terms of SDMEs, as shown in Appendix~C. 

For the helicity-flip amplitude $T_{01}$,
describing the transition $\gamma^*_T \to \rho^0_L$, 
the quantity $\tau_{01}$
is approximated as:
 \begin{eqnarray}
\tau_{01} \approx \sqrt{\epsilon} \frac{\sqrt{(r_{00}^{5})^2 +(r_{00}^{8})^2}}{\sqrt{2r_{00}^{04}}}.
\label{tau01f}
\end{eqnarray}
For the helicity-flip amplitude $T_{10}$,
describing the   transition  $\gamma^*_L \to \rho^0_T$,
the quantity $\tau_{10}$ is given by
\begin{eqnarray}
\tau_{10} \approx \frac{\sqrt{(r_{11}^{5}+ \mathrm{Im} \{r_{1-1}^{6}\} )^2+( \mathrm{Im} \{ r_{1-1}^{7}\}-r_{11}^{8})^2}}
{\sqrt{2(r_{1-1}^{1}- \mathrm{Im} \{ r_{1-1}^{2} \})}},\;\;\;\; 
\label{tau10f}
\end{eqnarray}
and for the helicity-double-flip amplitude $T_{1-1}$,
describing the transition  $\gamma^*_{-T} \to \rho^0_T$,
the quantity $\tau_{1-1}$ is given by
\begin{eqnarray}
\tau_{1-1}  \approx \frac{\sqrt{(r_{11}^1)^2+( \mathrm{Im} \{r_{1-1}^3 \} )^2}}{ \sqrt{ r_{1-1}^1 - 
\mathrm{Im} \{ r_{1-1}^2 \} }}. \hspace*{1.0cm} 
\label{tau1m1f} 
\end{eqnarray}
The resulting $\tau_{ij}$  values  for
proton and deuteron data are presented in Fig.~\ref{tau-fig} and in the
top section of Tab.~\ref{tau-tests}
for the entire HERMES kinematic region. Non-conservation of 
$s$-channel helicity is
clearly observed for the amplitude $T_{01}$ and, for the first time, for the
amplitude $T_{10}$, although with somewhat lower statistical significance.

Polarized SDMEs cannot be determined from collider data, 
as the collider kinematic conditions imply $\epsilon \approx 1$.
According to (\ref{eqang3}), this suppresses
the contribution of polarized SDMEs to $\mathcal{W}^L$.

In Ref.~\cite{zeussdme} the amplitude ratios are approximated as follows: 
\begin{eqnarray}
\widetilde{\tau}_{01} &\approx& \frac{ r^5_{00} } {\sqrt{ 2 r^{04}_{00} } }, \label{tau01}  \\
\widetilde{\tau}_{10} &\approx& \frac{ \mathrm{Re} \{r^{04}_{10}\} + \mathrm{Re} \{r^1_{10}\} } {\sqrt{ r^{04}_{00}}}, \\ 
\widetilde{\tau}_{1-1} &\approx& \frac{ | r^1_{11} | } {\sqrt{ 2 r^1_{1-1} } }.  \hspace*{1.0cm}
\label{tau1m1}
\end{eqnarray}  
In contrast to (\ref{tau01f}-\ref{tau1m1f}), 
these expressions rely on the assumption of
zero phase difference between the considered amplitude ($T_{01},T_{10}$, or $T_{1-1}$)
and the corresponding dominant amplitude ($T_{00}$ or $T_{11}$).
Results for the quantities   $\widetilde{\tau}_{ij}$   
from ZEUS and other experiments,  calculated from unpolarized  proton SDMEs,
are shown in Fig.~\ref{tau-fig} and in the 
bottom section of  Tab.~\ref{tau-tests}.

\begin{sloppypar}
The combined effect of 
$s$-channel helicity non-con\-servation 
and of a contribution of
UPE to the full cross section
can be estimated 
according to (\ref{dsigmadt},\ref{sigmalsigmat}, \ref{sigmatrans},\ref{sigmalong}) 
as follows. First note that
\begin{eqnarray} 
\frac{d \sigma_{full}}{dt} \approx f \mathcal{N}_0 \bigl( 1 +\tau_T^2 +  \tau_{UPE}^2  \bigr),
 \label{tautotal}
 \end{eqnarray} 
where
\begin{eqnarray} 
 \mathcal{N}_0=\epsilon |T_{00}|^2+|T_{11}|^2
 \label{sigmaschc}
 \end{eqnarray} 
contains only the contributions of $s$-channel helicity conserving 
NPE amplitudes. 
The $s$-channel helicity non-conserving  fractional contribution
of NPE amplitudes
to the cross section is defined as 
\begin{eqnarray}
\tau_T^2 &=&  \bigl( 2\epsilon|T_{10}|^2 + |T_{01}|^2+|T_{1-1}|^2 \bigl) / \mathcal{N}_0 \nonumber \\
&\approx& 2 \epsilon \tau_{10}^{2}+ \tau_{01}^{2}+\tau_{1-1}^{2}. \hspace*{1.0cm}
 \label{tau-t}
 \end{eqnarray}
The HERMES result  for 
$\tau_T^2$ is $0.025 \pm 0.003_{stat} \pm 0.003_{syst}$  and
$0.028 \pm 0.002_{stat} \pm 0.002_{syst}$ for the proton and 
deuteron, respectively.
\end{sloppypar}

Correspondingly, the UPE contribution is defined as: 
\begin{eqnarray}
\tau_{UPE}^2 &=& \nonumber \\
& &\hspace*{-1.5cm}\widetilde{\sum} \bigl( 2\epsilon|U_{10}|^2 + |U_{01}|^2+|U_{1-1}|^2+|U_{11}|^2 \bigl) / \mathcal{N}_0.
\label{tau-u}
 \end{eqnarray}
Because the contributions of amplitudes $U_{01}$ and $U_{1-1}$  
to (\ref{tau-u}) 
are negligibly small,
$\tau_{UPE}^2$ and $u_1$ (\ref{eq:npe5})
can be approximately related to one another as: 
$\tau_{UPE}^2 \approx u_1/2$. Accordingly, the first determination of
the fractional UPE contribution to the full cross section 
$\tau_{UPE}^2$ is $0.063 \pm 0.011_{stat} \pm 0.025_{syst}$ and 
$0.046 \pm  0.008_{stat} \pm  0.023_{syst}$
for the proton and deuteron, respectively.

\section{Longitudinal-to-Transverse Cross Section Ratio \label{lt-ratio}}

In principle the longitudinal-to-transverse cross section ratio $R$ (\ref{eqr}) 
can experimentally be determined directly
from the two cross sections if they can be extracted separately from the data
using, {\it e.g.,} the Rosenbluth decomposition technique~\cite{rosenbluth}.
For given values of $Q^2$ and $W$ (or $Q^2$ and $x_B$),
this requires data sets at different values of $\epsilon$, so that  
measurements at different beam energies 
are necessary~\cite{wolf}.
No data on vector meson production using such a decomposition 
have been published.

\subsection{Approximations for $\mathbf{R}$ } 

A common approximation to the ratio $R$ is experimentally determined
from the measured SDME $r^{04}_{00}$:
\begin{equation}\label{eqr04}
R^{04}  = \frac{1}{\epsilon} \frac{r^{04}_{00}}{1-r^{04}_{00}}.
\end{equation}
The quantity $R^{04}$ represents the ratio of cross sections for longitudinal and
transverse $\rho^0$ polarization, and it
is not identical to  the true $R$ that represents the ratio of the
cross sections with respect to the polarization of the virtual photon.
The relation between $R^{04}$ and $R$ is obtained
by comparing (\ref{eqr04},\ref{a1})  with 
(\ref{eqr},\ref{sigmalong},\ref{sigmatrans}):
\begin{equation}
R  = R^{04} - \frac{\eta(1+\epsilon R^{04})}{ \epsilon (1+\eta)},
\label{truerschc}
\end{equation}
with
\begin{eqnarray} \label{defeta}
\lefteqn{\eta = \frac{(1+ \epsilon R^{04})}{N}} \nonumber \\
& &\times\widetilde{\sum} \{|T_{01}|^2+|U_{01}|^2-
2 \epsilon (|T_{10}|^2+|U_{10}|^2)\}
\end{eqnarray}
(see Appendix D). In the case of SCHC, $\eta = 0$ and  $R^{04} = R$.
The quantity $R^{04}$ can be either smaller or larger than $R$, depending on the
sign of the small parameter $\eta$.
The latter can be calculated from data by neglecting the small 
contributions of the
helicity-flip UPE amplitudes $U_{10}$, $U_{01}$ in (\ref{defeta}):
\begin{equation}
\eta \approx  (1+ \epsilon R^{04}) (\tau_{01}^2 -
2 \epsilon \tau_{10}^2) \;,
\label{appreta}
\end{equation}
where  $\tau_{01}$ and $\tau_{10}$ are given in (\ref{tau01f}-\ref{tau10f}).

\begin{figure}[ht!]
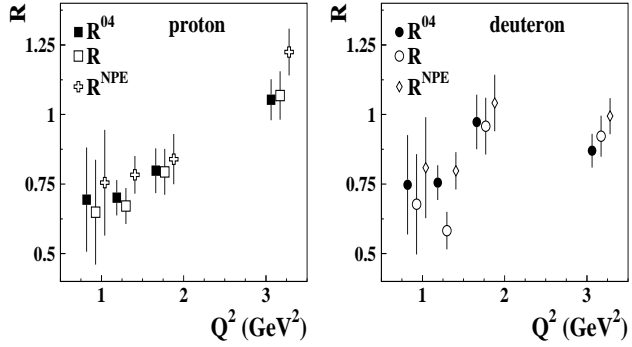

\begin{center}
 
\includegraphics[height=4.5cm,width=3.95cm]{./pln/paw-prot9605-may8.epsi}
\hspace{0.1cm}
\includegraphics[height=4.5cm,width=3.95cm]{./pln/paw-deut9605-may8.epsi}

\end{center}
\caption{\small  
$Q^2$ dependence of the
longitudinal-to-transverse cross section
ratio measured at HERMES. 
Results from proton (deuteron) data are shown in the left (right)
panel.
Filled symbols represent the value of $R^{04}$ 
calculated from $r^{04}_{00}$ (\ref{eqr04}),   
open symbols correspond  to the true value of  $R$ calculated 
according to (\ref{truerschc},\ref{appreta}), and crosses 
(diamonds) represent 
 $R^{NPE}$ (\ref{r-npe}).
Total uncertainties are shown, calculated by 
combining the statistical and systematic 
uncertainties in quadrature. 
The data points for $R$ and $R^{NPE}$ are presented with a small horizontal
offset to improve their visibility.
}
\label{npeltratio}
\end{figure}
 
Regge phenomenology suggests that contributions of 
unnatural-parity exchange are 
more significant at the lower energies 
typical of this experiment, and decrease at
collider energies.
In order to allow a comparison of HERMES
results on $R$ with those at high energy and also with GPD-based calculations,
the ratio  $R^{NPE}$ is determined from
$R^{04}$ by subtracting the contributions of all UPE amplitudes.
The dependence of the difference $\Delta R^{UPE}= R^{04}-R^{NPE}$
on $|U_{ij}|^2$ can be determined in a linear approximation as 
\begin{equation*}
\Delta R^{UPE}=\sum_{ij} \frac {\partial R^{04}}{\partial |U_{ij}|^2} |U_{ij}|^2\; .
\end{equation*}
Assuming the hierarchy (\ref{hier}) of UPE amplitudes, this can be
approximated by retaining only $U_{11}$:
\begin{equation}
\Delta R^{UPE} \approx \frac {\partial R^{04}} {\partial |U_{11}|^2} |U_{11}|^2 \; .
\label{drupe1}
\end{equation}
According to (\ref{truerschc}) and (\ref{defeta}), $R^{04}$ can be approximated
by  $R=\mathcal{N}_L/\mathcal{N}_T$, 
which yields, along with (\ref{sigmatrans},\ref{sigmalong}), 
\begin{eqnarray}
\nonumber
\Delta R^{UPE} &\approx& -\frac{\mathcal{N}_L}{\mathcal{N}_T^2}
  \widetilde{\sum}  \vert U_{11} \vert ^2 \\ \nonumber
&=&  -R \cdot \frac{
\widetilde{\sum}
\vert U_{11} \vert ^2}{\mathcal{N}_T + \epsilon \mathcal{N}_L} \cdot \frac{ \mathcal{N}_T +
\epsilon \mathcal{N}_L}{ \mathcal{N}_T}  \\
  &\approx& -R^{04} \cdot \frac{u_1}{2} \cdot  (1+ \epsilon R^{04})\; . \hspace*{0.5cm}
\label{drupe2}
\end{eqnarray}
Here $u_1 \approx 2  \widetilde{\sum}  \vert U_{11} \vert ^2/(\mathcal{N}_T + \epsilon \mathcal{N}_L)$ 
is used instead of (\ref{eq:npe5}). 
The final approximate formula for $R^{NPE}=R^{04}-\Delta R^{UPE}$ is
\begin{eqnarray}
R^{NPE} \approx R^{04}[1+\frac{u_1}{2} (1+ \epsilon R^{04})]\; .
\label{r-npe}
\end{eqnarray}

\begin{figure*}[ht!]
\begin{center}
\includegraphics[height=7.8cm]{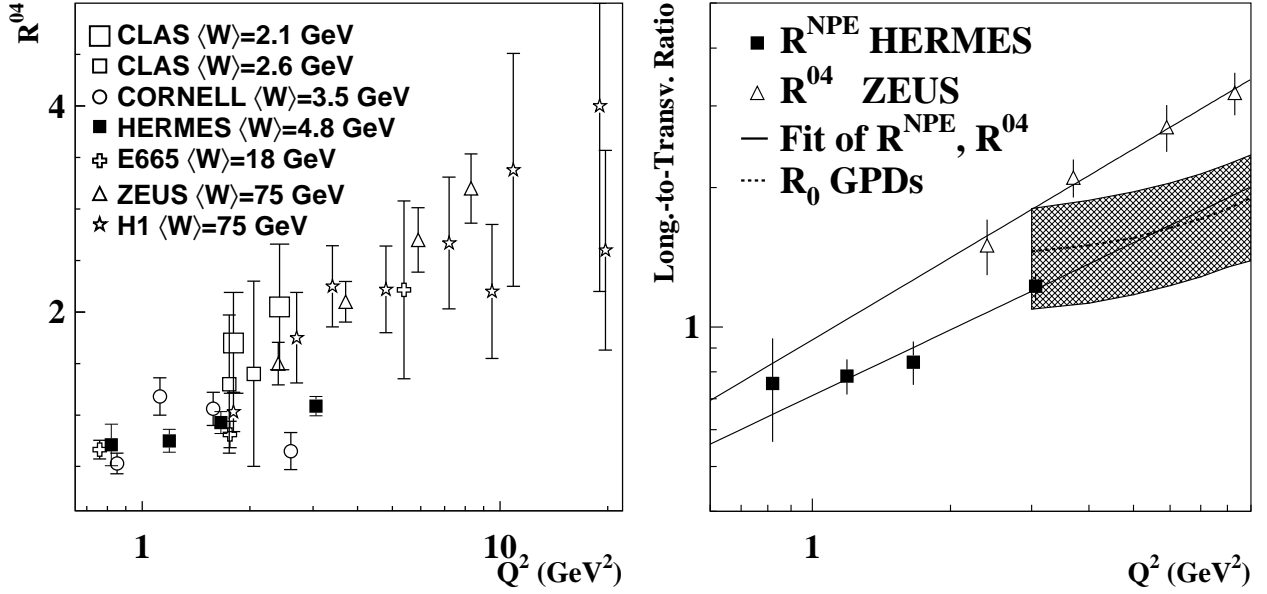}
\hspace{0.2cm}
\includegraphics[height=7.8cm]{./pln/sdme-combfit.epsi}
\caption{\small  
$Q^2$ dependence of the longitudinal-to-transverse cross section 
ratio for exclusive $\rho^0$ production on the proton. Left panel:
$R^{04}$ calculated from the SDME $r^{04}_{00}$ according to (\ref{eqr04}).
HERMES proton data (filled squares) are compared to measurements of
CLAS \cite{clas1,clas2}, Cornell \cite{CORNELL}, E665 \cite{E665}, 
H1~\cite{H1}, and ZEUS~\cite{zeussdme,zeussdme2}. The more recent CLAS 
data~\cite{clas2} (small squares) 
are from a narrow bin in $x_B$ with approximately the
same $\langle x_B\rangle$ as 
the HERMES data, which are integrated over the $x_B$ 
acceptance.
Right panel:
$R^{04}$ for ZEUS (triangles) and  $R^{NPE}$ for HERMES (squares), fitted separately
according to (\ref{eqfitr}). 
For all data points, total uncertainties are shown. 
Theoretical calculations~\cite{golos3} 
of $R_0 = |T_{00}|^2/|T_{11}|^2$
are shown as dashed line at $W$= 5 GeV; the
uncertainties arising from the uncertainties in the parton distribution functions are shown 
as a shaded band~\cite{golos3}. 
}
 \label{schcltratio}
 \end{center}
\end{figure*}

\subsection{HERMES Results on $\mathbf{R}$}

Evaluations of $R$ from HERMES data are
performed for the entire interval $0~{\rm GeV}^2<-t'<0.4$~GeV$^2$.
The $Q^2$ dependences of the quantities  $R^{04}$ (\ref{eqr04})
and  $R$ (\ref{truerschc},\ref{appreta}) 
are presented in Fig.~\ref{npeltratio}.
In the HERMES kinematic conditions, at $\epsilon \approx 0.8$, 
the value of  $\eta$ is about 0.1 ($-0.1$)  for the proton
(deuteron), and  the magnitude of the difference between
$R$ and $R^{04}$ is small, of the order of $0.1$.
 
In section~\ref{tot-crsec} it was shown
that by analyzing the amplitudes that comprise the SDMEs, 
a statistically significant, non-zero UPE contribution to the cross section
exists. At the intermediate energy of the
HERMES experiment this contribution is small.
If it is caused by 
exchange of $\pi$, $a_1$, or $b_1$, this contribution would be 
negligible at higher energies~\cite{iwing}.
In order to compare the HERMES results on $R$ with those of experiments 
at higher 
energy,
it is appropriate to correct $R^{04}$ for the UPE contribution and 
consider $R^{NPE}$.
The value of $\Delta R^{UPE}$ is about $-0.11$ ($-0.08$) for the proton
(deuteron) at HERMES kinematics.
The resulting values of $R^{NPE}$ are shown in Fig.~\ref{npeltratio}
and Tab.~\ref{rat-tab}.

\subsection{Comparison to World Data and Models}

Results for $R$ from different 
experiments can be compared only if either  $R$  is independent of $t'$,
or the $t'$ dependences of the cross sections $\frac{d\sigma_L}{dt}$ and
$\frac{d\sigma_T}{dt}$
and the 
$t'$ intervals of the measurements of $R$ are the same. 
The $t'$ dependence of $R$ is determined essentially by the $t'$ dependence of the  
SDME $r^{04}_{00}$ (see (\ref{a1})), which is found to be approximately flat in $t'$ 
both at 
HERMES (see Fig.~\ref{q2-class-a}) and at H1~\cite{H1} and ZEUS~\cite{zeussdme2} kinematics. 
For this case, the ratio of the total cross sections coincides with
the ratio of the cross sections that are differential in $t$ (see (\ref{eqr})).

The left panel of Fig.~\ref{schcltratio} shows HERMES results on the $Q^2$ 
dependence of $R^{04}$, 
as measured on the proton, in comparison to  world data. 
Given the experimental uncertainties, there is no discrepancy with 
the data  at lower energies
from CLAS~\cite{clas1,clas2} and CORNELL~\cite{CORNELL}.
The HERMES data at intermediate energies are not expected to 
agree exactly with those at high energies 
because of
the UPE contributions observed in the HERMES data, as discussed in 
sections~\ref{unpesec} and~\ref{tot-crsec}. 
We note that SCHC violating amplitudes
are also observed in the new CLAS data~\cite{clas2}.
Additional reasons may be the 
importance of valence-quark exchange for NPE amplitudes
and also a generally different $W$ dependence of the longitudinal and transverse cross sections, 
as recently discussed in Ref.~\cite{golos3} in the context of a GPD-based model. 

The right panel of Fig.~\ref{schcltratio}  presents the HERMES results on 
the longitudinal-to-transverse cross section ratio $R^{NPE}$, 
which is corrected for the UPE contributions shown in the previous section
to be of substantial size at intermediate energy.
The HERMES data are compared to the recent 
high energy data on  $R^{04}$ from ZEUS~\cite{zeussdme2}, 
for which the UPE contribution is expected to be strongly suppressed.

In order to investigate a possible $W$ dependence of the longitudinal-to-transverse 
cross section
ratio, the HERMES and ZEUS data are fitted separately to a $Q^2$ dependence 
suggested by VMD models~\cite{vmd2,brodsk,pichow}:
\begin{equation}\label{eqfitr}
R(Q^2)=c_0\left(\frac{Q^2}{M_V^2}\right)^{c_1},
\end{equation}
where $c_0$ and $c_1$ are free parameters and $M_V$ is the mass of 
the $\rho^0$ meson. 
The fit results are  $c_0 = 0.56 \pm 0.08$, $c_1 = 0.47 \pm 0.12$ for HERMES 
and  $c_0 = 0.69 \pm 0.22$, $c_1 = 0.59 \pm 0.15$ for ZEUS,
with  $\chi^2/d.o.f. = 0.45$ and 0.15 respectively. These  $\chi^2$ values indicate that
the fits are dominated by systematic uncertainties. 

A $W$ dependence of the $Q^2$ slope is consistent with 
recent  calculations using a GPD-based
model~\cite{golos3}. We note the agreement of these calculations
performed  at $W= 5$ GeV  for $Q^2$ values down to 3~GeV$^2$ 
(see dashed curve in  Fig.~\ref{schcltratio})
with the highest $Q^2=3$ GeV$^2$ point of HERMES.
Uncertainties in the model calculations originating 
from uncertainties  in the parton distributions employed are shown 
as a shaded band superimposed on the curve.

\section{Summary}

\begin{sloppypar}
HERMES has studied exclusive $\rho^0$ production on the proton and
deuteron at intermediate energies ($\langle W \rangle = 4.8$ GeV),
at the average  values of
$\langle Q^2 \rangle = 1.95$ GeV$^2$, $\langle -t' \rangle = 0.13$ GeV$^2$,
and $\langle x_B \rangle = 0.08$, using polarized beams and unpolarized targets. 
\end{sloppypar}

By performing a maximum likelihood fit, fifteen unpolarized 
SDMEs and, for the first time, eight polarized
SDMEs are obtained. 
The measured SDMEs are grouped according to their
theoretically expected hierarchy.  This facilitates the
investigation of the relative importance of various helicity amplitudes describing
different $\gamma^* \to \rho^0$ transitions.
Within the given experimental uncertainties, 
the expected hierarchy of relative sizes of helicity amplitudes is observed.

Non-zero values are observed  
for the  two helicity-flip
amplitudes $T_{01}$ and  $T_{10}$,
indicating a small but statistically significant deviation from the hypothesis of
$s$-channel helicity conservation.

The phase difference between the helicity-conser\-ving amplitudes $T_{11}$ and
$T_{00}$ is confirmed to be significantly non-zero and is also 
seen to have a possible $Q^2$ dependence.
For the first time, the sign of
the phase difference is determined using the polarized SDMEs. 

The kinematic dependences
of all 23 SDMEs are measured for both hydrogen and deuterium targets. 
Clear dependences on $Q^2$ and $t'$ are observed for certain SDMEs. 
No significant difference between proton and deuteron results is seen.

The evaluation of certain linear combinations of SDMEs 
provides an indication that at the intermediate energy of the HERMES experiment,
contributions of unnatural-parity-exchange amplitudes exist.
Such amplitudes are naturally generated by a quark-exchange mechanism.

\begin{sloppypar}
In order to determine the longitudinal-to-transverse 
cross section ratio with respect to
the polarization of the virtual photon, an approximation $R^{04}$ to the ratio~$R$
of the cross sections for longitudinal and transverse  $\rho^0$ 
polarizations 
is calculated from the SDME $r^{04}_{00}$
as a function of $Q^2$. 
The results obtained for
other SDMEs permit us to improve this approximation of $R$
by taking into account
transitions of natural parity that do not conserve s-channel helicity.
In order to facilitate  comparison with high-energy collider data, 
a correction is applied to $R^{04}$ to exclude contributions
from unnatural-parity exchange.
The comparison of the $Q^2$ dependences of $R$ at low and
high values of $W$ suggests a possible $W$ dependence of the ratio.
\end{sloppypar}

\begin{acknowledgement}
{\it Acknowledgments.} 
We gratefully acknowledge the DESY management for its support and the staff
at DESY and the collaborating institutions for their significant effort.
This work was supported by the FWO-Flanders and IWT, Belgium;
the Natural Sciences and Engineering Research Council of Canada;
the National Natural Science Foundation of China;
the Alexander von Humboldt Stiftung;
the German Bundesministerium f\"ur Bildung und Forschung (BMBF);
the Deutsche Forschun\-gsgemeinschaft (DFG);
the Italian Istituto Nazionale di Fisica Nucleare (INFN);
the MEXT, JSPS, and COE21 of Japan;
the Dutch Foundation for Fundamenteel Onderzoek der Materie (FOM);
the U. K. Engineering and Physical Sciences Research Council, the
Particle Physics and Astronomy Research Council and the
Scottish Universities Physics Alliance;
the U. S. Department of Energy (DOE) and the National Science Foundation (NSF);
the Russian Academy of Science, the Russian Federal Agency for 
Science and Innovations, and the Heisenberg-Landau program;
the Ministry of Trade and Economical Development and the Ministry
of Education and Science of Armenia;
and the European Community-Research Infrastructure Activity under the
FP6 ''Structuring the European Research Area'' program
(Hadron Physics, contract number RII3-CT-2004-506078).

It is a pleasure to thank M.~Diehl,  S.V.~Goloskokov,
D.Yu.~Ivanov, I.P.~Ivanov, P.~Kroll, N.N.~Nikolaev and
A.V.~Vinnikov for many useful discussions.

\end{acknowledgement}

\input{bibliography}
\section*{Appendix A. The 23 Spin-Density Matrix Elements Expressed in Terms of Helicity Amplitudes}
 
The basic expressions for the 23 spin density matrix elements measurable with a polarized 
lepton beam and an unpolarized target,
ordered according to the expected  hierarchy of amplitudes,
are given in (\ref{a1})--(\ref{a17}).
(COMMENT FOR THE EDITORS: these formulae appear on the next page).

\begin{figure*}
\begin{eqnarray} 
\mathbf{ A:} \;\;\; \gamma^*_L \to \rho^0_L  \;\;\; \mathrm{and} \;\;\;  \gamma^*_T \to \rho^0_T   \nonumber  \\
r_{00}^{04}=\widetilde{\sum} \{\epsilon |T_{00}|^2+|T_{01}|^2+|U_{01}|^2/\}/\mathcal{N} \label{a1} \,, \\
r_{1-1}^1 = \frac {1}{2} \widetilde{\sum}\{|T_{11}|^2+|T_{1-1}|^2-|U_{11}|^2-|U_{1-1}|^2 \}/\mathcal{N} \label{a7} \,,  \\
\mathrm{Im} \{r_{1-1}^2 \}=\frac {1}{2} \widetilde{\sum} \{ -|T_{11}|^2 +|T_{1-1}|^2
+|U_{11}|^2 -|U_{1-1}|^2\}/\mathcal{N} \label{a9} \,,  \\
 \vspace{0.750cm}
\mathbf{ B:}\;\;\;  \mathrm{interference} \;\;\; \mathrm{of} \;\;\; \gamma^*_L \to \rho^0_L \;\;\; \mathrm{and} \;\;\; \gamma^*_T \to \rho^0_T \nonumber  \\
\mathrm{Re} \{r_{10}^{5}\}=\frac{1}{\sqrt{8}} \widetilde{\sum}\mathrm{Re} \{2T_{10} T_{01}^*+(T_{11}-T_{1-1})T_{00}^*\}/\mathcal{N} \label{a12} \,,   \\
\mathrm{Im} \{r_{10}^{6}\}=\frac{1}{\sqrt{8}} \widetilde{\sum} \mathrm{Re} \{2U_{10} U_{01}^*-(T_{11}+T_{1-1})T_{00}^*\}/\mathcal{N} \label{a14} \,, \\
\mathrm{Im} \{r_{10}^{7}\}=\frac{1}{\sqrt{8}} \widetilde{\sum}
\mathrm{Im} \{2U_{10}U_{01}^*+(T_{11}+T_{1-1})T_{00}^*\}/\mathcal{N} \label{a18} \,, \\
\mathrm{Re} \{r_{10}^{8}\}=\frac{1}{\sqrt{8}} \widetilde{\sum}
\mathrm{Im} \{-2T_{10}T_{01}^*+(T_{11}-T_{1-1})T_{00}^*\}/\mathcal{N} \label{a22} \,, \\
\vspace{0.750cm}
{\mathbf C:} \;\;\;  \gamma^*_T \to \rho^0_L  \nonumber  \\
\mathrm{Re} \{r_{10}^{04} \} =
 \widetilde{\sum} \mathrm{Re} \{\epsilon T_{10}T_{00}^* +\frac {1}{2}T_{01}(T_{11}-T_{1-1})^*
+\frac {1}{2}U_{01}(U_{11}+U_{1-1})^*\}/\mathcal{N} \label{a2} \,, \\
\vspace{0.20cm}
\mathrm{Re} \{r_{10}^{1} \}= \frac{1}{2} \widetilde{\sum}
\mathrm{Re}\{-T_{01}(T_{11}-T_{1-1})^*+U_{01}(U_{11}+U_{1-1})^* \}/\mathcal{N} \label{a6} \,, \\
\mathrm{Im} \{r_{10}^2 \}=\frac{1}{2}\widetilde{\sum}
\mathrm{Re} \{T_{01}(T_{11}+T_{1-1})^*-U_{01}(U_{11}-U_{1-1})^* \}/\mathcal{N}  \label{a8} \,, \\
r_{00}^5=\sqrt{2} \widetilde{\sum} \mathrm{Re} \{ T_{01} T_{00}^*\}/\mathcal{N} \label{a11} \,, \\
r_{00}^1=\widetilde{\sum} \{- |T_{01}|^2+|U_{01}|^2\}/ \mathcal{N} \label{a5} \,, \\
\mathrm{Im} \{r_{10}^{3}\}= -\frac{1}{2} \widetilde{\sum}
\mathrm{Im} \{T_{01}(T_{11}+T_{1-1})^* +U_{01}(U_{11}-U_{1-1})^* \}/\mathcal{N}  \label{a16}   \,, \\
r_{00}^{8}=\sqrt{2} \widetilde{\sum} \mathrm{Im} \{ T_{01} T_{00}^* \}/\mathcal{N} \label{a21} \,, \\
\vspace{0.750cm}
{\mathbf D:} \;\;\;   \gamma^*_L \to \rho^0_T \nonumber \\
r_{11}^{5}=
\frac{1}{\sqrt{2}} \widetilde{\sum} \mathrm{Re} \{ T_{10} (T_{11}-T_{1-1})^*
+U_{10} (U_{11}-U_{1-1})^*\}/\mathcal{N} \label{a10} \,, \\
r_{1-1}^{5}=
\frac{1}{\sqrt{2}}\widetilde{\sum} \mathrm{Re} \{ -T_{10} (T_{11}-T_{1-1})^*
+U_{10} (U_{11}-U_{1-1})^*\}/\mathcal{N} \label{a13} \,, \\
\mathrm{Im} \{r^6_{1-1}\}= \frac{1}{\sqrt{2}} \widetilde{\sum}
\mathrm{Re} \{ T_{10} (T_{11}+T_{1-1})^*
-U_{10} (U_{11}+U_{1-1})^*\}/\mathcal{N} \label{a15} \,, \\
\mathrm{Im} \{r_{1-1}^{7}\}=\frac{1}{\sqrt{2}}  \widetilde{\sum}
\mathrm{Im} \{T_{10}(T_{11}+T_{1-1})^*-U_{10}(U_{11}+U_{1-1})^*\}/\mathcal{N} \label{a19}, \\
 r^8_{11}=-\frac{1}{\sqrt{2}} \widetilde{\sum}
\mathrm{Im} \{T_{10}(T_{11}-T_{1-1})^*+U_{10}(U_{11}-U_{1-1})^*\}/\mathcal{N} \label{a20} \,, \\
 r_{1-1}^{8}=\frac{1}{\sqrt{2}} \widetilde{\sum}
\mathrm{Im} \{T_{10}(T_{11}-T_{1-1})^*-U_{10}(U_{11}-U_{1-1})^*\}/\mathcal{N} \label{a23} \,, \\
\vspace{0.750cm}
{\mathbf E:} \;\;\;    \gamma^*_{-T} \to \rho^0_T  \nonumber \\
r_{1-1}^{04}=\widetilde{\sum} \mathrm{Re} \{ -\epsilon |T_{10}|^2 +\epsilon |U_{10}|^2
+T_{1-1}T_{11}^*-U_{1-1}U_{11}^* \}/\mathcal{N} \label{a3}\,, \\
r_{11}^{1}=\widetilde{\sum} \mathrm{Re} \{ T_{1-1}T_{11}^* +U_{1-1}U_{11}^*\}/\mathcal{N}
\label{a4} \,, \\
\mathrm{Im} \{r_{1-1}^{3}\}= -\widetilde{\sum}
 \mathrm{Im} \{T_{1-1}T_{11}^* -U_{1-1}U_{11}^*\} /\mathcal{N}\label{a17} \,, \\
\nonumber
\end{eqnarray} 
\end{figure*}

\section*{Appendix~B. Derivation of Formulae for {\boldmath$\cos \delta$} and 
{\boldmath$\sin \delta$}}

Neglecting small contributions of the products of spin-flip amplitudes $T_{10}T_{01}^*$
and $U_{10}U_{01}^*$ in (\ref{a12})-(\ref{a22})
we obtain the
approximate relations:
\begin{eqnarray} \label{appr5r6}
\sqrt{2}(\mathrm{Re}\{r^{5}_{10}\}-\mathrm{Im}\{r^{6}_{10}\}) \approx \mathrm{Re}(T_{11}T_{00}^*)/\mathcal{N}\nonumber  \\
=|T_{11}||T_{00}| \cos \delta/ \mathcal{N} \;,\\
\label{appr7r8}
\sqrt{2}(\mathrm{Re}\{r^{8}_{10}\}+\mathrm{Im}\{r^{7}_{10}\}) \approx
\mathrm{Im}(T_{11}T_{00}^*)/\mathcal{N} \nonumber  \\ 
=|T_{11}||T_{00}| \sin \delta/ \mathcal{N} \;.
\end{eqnarray}
Neglecting  $|T_{01}|^2$ and $|U_{01}|^2$ in the numerator
in relation  (\ref{a1})  we get
\begin{eqnarray} \label{appr04}
r^{04}_{00}/\epsilon \approx  |T_{00}|^2/ \mathcal{N} \;.
\end{eqnarray}
Recalling  the formulae  (\ref{sigmatrans}), (\ref{sigmalong}),
and (\ref{ntotal}) for $\mathcal{N}$,  
we obtain from (\ref{a1}) the approximate relation
\begin{eqnarray} \label{1r04}
1-r^{04}_{00}\approx\widetilde{\sum} \Bigl[|T_{11}|^2+|U_{11}|^2 \Bigr ]/ \mathcal{N}
\end{eqnarray}
if we neglect small contributions of spin-flip amplitudes in the numerator of (\ref{1r04}).
As seen from relation (\ref{1r04}) the difference $1-r^{04}_{00}$
contains $|U_{11}|^2$;  
to cancel this contribution,
the difference of two SDMEs defined by (\ref{a7}) and (\ref{a9}) is considered:
\begin{eqnarray} \label{r1r2}
r^{1}_{1-1}-\mathrm{Im}\{r^{2}_{1-1}\}=\widetilde{\sum} \Bigl [ |T_{11}|^2-|U_{11}|^2 \Bigr ]
/\mathcal{N}\;.\hspace*{0.5cm}
\end{eqnarray}
Adding (\ref{1r04}) and (\ref{r1r2}) together
the relation
\begin{eqnarray} \label{app1r04r1r2}
(1-r^{04}_{00}+r^{1}_{1-1}-\mathrm{Im}\{r^{2}_{1-1} \})/2 \approx
|T_{11}|^2/\mathcal{N}\;
\end{eqnarray}
is obtained.
Dividing (\ref{appr5r6}) and (\ref{appr7r8}) by the square root of the product of
(\ref{appr04})
and (\ref{app1r04r1r2}), formulae (\ref{eq:cosdelta}) and (\ref{eq:sindelta})
are obtained respectively.

\section*{Appendix C. Derivation of Formula for {\boldmath$\tau_{10}$}}

\label{intro}
\begin{sloppypar}
Here we derive the formula for $\tau_{10}$ only;
formulae for $\tau_{01}$ and $\tau_{1-1}$ can be derived in an analogous way.
If we retain only linear contributions of small $s$-channel helicity-flip amplitudes in the basic
formulae for SDMEs (\ref{a1})-(\ref{a17}),
neglecting unnatural-parity-exchange contributions and bilinear products of helicity-flip 
amplitudes we obtain:
\begin{equation}
r_{00}^{04}=\epsilon |T_{00}|^2/\mathcal{N}_0\;,
\label{eq:r04-00}
\end{equation}
\begin{equation}
\mathrm{Re}\{ r_{10}^{04} \}=
\mathrm{Re} \{\epsilon T_{10}T_{00}^* +\frac {1}{2}T_{01}T_{11}^*\}/\mathcal{N} _0\;,
\label{eq:r04-10}
\end{equation}
\begin{equation}
\mathrm{Re} \{ r_{10}^{1} \}=
- \frac{1}{2} \mathrm{Re} \{T_{01}T_{11}^* \}/\mathcal{N}_0 \;,
\label{r1-10}
\end{equation}
\begin{equation}
r_{1-1}^{1}=- \mathrm{Im} \{r_{1-1}^{2} \}=\frac {1}{2} |T_{11}|^2/\mathcal{N}_0 \;,
\label{r1-1-1}
\end{equation}
\begin{equation}
r_{11}^{5}= \mathrm{Im} \{ r_{1-1}^{6} \}=
\frac{1}{\sqrt{2}} \mathrm{Re} \{ T_{10} T_{11}^*\}/\mathcal{N}_0 \;,
\label{r5-11}
\end{equation}
\begin{equation}
 \mathrm{Im} \{r_{1-1}^{7} \}= -r_{11}^{8}=
\frac{1}{\sqrt{2}}
 \mathrm{Im} \{ T_{10}T_{11}^* \}/\mathcal{N}_0 \;,
\label{r7-1-1}
\end{equation}
where $\mathcal{N}_0 = \epsilon | T_{00}|^2 + |T_{11}|^2$ 
(see (\ref{sigmaschc})).
\end{sloppypar}

The parameter $\widetilde{\tau}_{10}$
was defined in \cite{zeussdme} by the relation
\begin{eqnarray}
\widetilde{\tau}_{10}
=|T_{10}|/\sqrt{|T_{00}|^2 +|T_{11}|^2}
\label{deftau10}
\end{eqnarray}
and estimated from (\ref{eq:r04-00},\ref{eq:r04-10},\ref{r1-10}) with the formula
\begin{eqnarray}
\nonumber
\widetilde{\tau}_{10} \approx  
\frac{  \mathrm{Re} \{r_{10}^{04}\} + \mathrm{Re} \{r_{10}^{1} \} }{\sqrt{r_{00}^{04}}} \\
=\frac{|T_{10}| \cos \delta _{10}}{\sqrt{|T_{00}|^2 +|T_{11}|^2/ \epsilon}}
\label{zeustau10}
\end{eqnarray}
where $\delta _{10} = arg(T_{10})-arg(T_{00})$.
Comparison of \\
(\ref{deftau10}) and $(\ref{zeustau10})$ shows that they are equal to each other
if $\epsilon \approx 1$ and $\delta_{10} \approx 0$.

\begin{sloppypar}
Instead of $(\ref{zeustau10})$ we derive a formula which is applicable for any values of 
$\delta _{10}$ and $\epsilon$. Combining $(\ref{r5-11})$ and 
$(\ref{r7-1-1})$, we obtain
\begin{eqnarray}
\nonumber
(r_{11}^{5}+ \mathrm{Im} \{r_{1-1}^{6} \})^2+( \mathrm{Im} \{r_{1-1}^{7}\}-r_{11}^{8} )^2 \\
\nonumber
= \frac{2}{\mathcal{N}_0^2} \Bigl [( \mathrm{Re} \{T_{10}T_{11}^* \})^2+( \mathrm{Im} \{ T_{10}T_{11}^* \})^2 \Bigr ] \\
=\frac{2}{\mathcal{N}_0^2}|T_{10}|^2|T_{11}|^2.\; \hspace*{3.3cm} 
\label{auxtau10}
\end{eqnarray}
Dividing $(\ref{auxtau10})$ 
by $2 r_{1-1}^{1}-2 \mathrm{Im} \{r_{1-1} ^{2}\}=2|T_{11}|^2/\mathcal{N}_0$ 
(see  $(\ref{r1-1-1})$) we get the final approximate formula
\begin{eqnarray}
\nonumber
\tau_{10} &\approx& |T_{10}|/\sqrt{\mathcal{N}_0} \\ 
 &\approx&  \frac{\sqrt{(r_{11}^{5}+ \mathrm{Im} \{r_{1-1}^{6} \} )^2
+( \mathrm{Im} \{ r_{1-1}^{7}\}-r_{11}^{8})^2} }
{\sqrt{2(r_{1-1}^{1}- \mathrm{Im} \{r_{1-1}^{2} \})}}. \nonumber  \\
\label{tau10app}
\end{eqnarray}
Since $\mathcal{N}_0=\mathcal{N}$ within the approximation under consideration, formula
$(\ref{tau10app})$ corresponds to definition $(\ref{tau000})$ of $\tau_{10}$.
In the case $\epsilon =1$, which was considered in Ref.~\cite{zeussdme}, the
estimate $(\ref{tau10app})$ for $\tau_{10}$ coincides with the 
definition given in (\ref{deftau10}).
\end{sloppypar}

\section*{Appendix D. Derivation of Relations \\
between  $\mathbf{R^{04}}$ and $\mathbf{R}$}

From (\ref{eqr04},\ref{a1},\ref{ntotal}) it follows that
\begin{eqnarray} 
\nonumber
\epsilon R^{04}= \frac{ \widetilde{\sum} \{\epsilon |T_{00}|^2+|T_{01}|^2+|U_{01}|^2 \}}
{\mathcal{N}_T + \epsilon \mathcal{N}_L-  \widetilde{\sum}  \{\epsilon |T_{00}|^2+|T_{01}|^2+|U_{01}|^2 \}} = \\
\frac{\epsilon \mathcal{N}_L+
\widetilde{\sum} \{-2\epsilon (|T_{10}|^2+|U_{10}|^2)+|T_{01}|^2+|U_{01}|^2 \}}
{\mathcal{N}_T -  \widetilde{\sum} \{-2\epsilon (|T_{10}|^2+|U_{10}|^2)+|T_{01}|^2+|U_{01}|^2 \}}, \nonumber \\
\label{eq1rschc}
\end{eqnarray}
where in the second step we have used the formula for $\mathcal{N}_L$ $(\ref{sigmalong})$ 
for the transformation 
of $(\ref{eq1rschc})$. 
Dividing both the numerator and 
denominator 
in $(\ref{eq1rschc})$ by $\mathcal{N}_T$ and remembering the definition $(\ref{eqr})$ 
of $R$ we get 
\begin{eqnarray}
\epsilon R^{04}=\frac{\epsilon R +\zeta}{1-\zeta}
\label{eq2rschc}
\end{eqnarray}
with
\begin{eqnarray}
\label{zeta}
\nonumber
\zeta = \widetilde{\sum}  \{-2\epsilon (|T_{10}|^2+|U_{10}|^2) 
+|T_{01}|^2+|U_{01}|^2 \} /\mathcal{N}_T \nonumber \\
= \frac{\eta(1+ \epsilon R)}{1+ \epsilon R^{04}}\;. 
\qquad \qquad \qquad \qquad \qquad \qquad \qquad \;\;\;\;\;
\end{eqnarray}
The last relation follows from comparison of $(\ref{zeta})$ and 
the definition (\ref{defeta}) of $\eta$. 
Equation (\ref{eq2rschc}) can be easily rewritten in the form
\begin{eqnarray}
R=R^{04}-\frac{\zeta}{\epsilon}(1+\epsilon R^{04})
\label{eq3rschc}
\end{eqnarray}
which is equivalent to (\ref{truerschc}) if we take into account 
the relation $(\ref{zeta})$ between $\eta$ and $\zeta$. \\

\input{app1nb-dp}

\newpage

\input{tab-corr-doublepage}

\end{document}

%% file: app1nb-dp.tex
\section*{Appendix E. Kinematic Intervals, Mean Values for Kinematic Variables
and SDMEs, for Proton and Deuteron } 

The resulting SDMEs with statistical and systematic
uncertainties are presented below in tabular form for hydrogen and deuterium targets. 
First, in Tab.~\ref{kinem-h} the mean kinematic values are presented for the entire kinematic 
region of the measurement and for each bin used in the $Q^2$, $t'$ and $x_B$ dependences. 
In Tabs.~\ref{tab15-h-q2}, \ref{tab15-h-tpr}, \ref{tab15-h-xbj} 
(\ref{tab15-d-q2}, \ref{tab15-d-tpr}, \ref{tab15-d-xbj}) the results of the measurement of
the $Q^2$, $t'$ and $x_B$-dependences, respectively, for the proton (deuteron) are listed.   
The values of the phase difference $\delta$ between $T_{11}$ and $T_{00}$ amplitudes, from 
proton and deuteron data,
are contained in Tab.~\ref{tab-phase}.
The kinematic dependences of the $u_1$ value used for the study of 
unnatural-parity-exchange amplitudes are in 
Tab.~\ref{tab-npe}. The SDMEs measured over the entire kinematic
region from proton data, but presented using the recent 
representation of Ref.~\cite{mdiehl} are listed in Tab.~\ref{mdiehl-tab}.
The correlation matrices of the 23 SDMEs measured for the proton and deuteron
over the entire kinematic region are presented in Tabs.~\ref{tab-corr-h} 
and~\ref{tab-corr-d}.

\begin{table*}
 \renewcommand{\arraystretch}{1.2} 
\begin{center}
\caption{ \label{kinem-h}
The definition of intervals and the mean values for kinematic variables for hydrogen (deuterium)
data. }
\newcommand{\m}{\hphantom{$-$}}
\newcommand{\cc}[1]{\multicolumn{1}{c}{#1}}
\renewcommand{\tabcolsep}{2pc}
\renewcommand{\arraystretch}{1.2}
\begin{footnotesize}
\begin{tabular}{|c|c|}
\hline
bin  &$ \langle Q^{2} \rangle$, GeV$^2$   \\
\hline
 $1~{\rm GeV}^2 < Q^{2} < 7$~GeV$^2$  & 1.95 (1.94) \\
\hline
 \hline
$0.5~{\rm GeV}^2 < Q^{2} < 1.0$~GeV$^2$ & 0.82 (0.82) \\
$1.0~{\rm GeV}^2 < Q^{2} < 1.4$~GeV$^2$   & 1.19 (1.18) \\
$1.4~{\rm GeV}^2 < Q^{2} < 2 $~GeV$^2$  & 1.66 (1.66) \\
$2~{\rm GeV}^2 < Q^{2} < 7$~GeV$^2$   & 3.06 (3.04) \\
\hline

 & $ \langle t' \rangle $, GeV$^2$    \\
\hline
$0.0~{\rm GeV}^2 < -t' < 0.04$~GeV$^2$ & 0.019 (0.018)   \\
$0.04~{\rm GeV}^2 < -t' < 0.10$~GeV$^2$  &  0.068 (0.068) \\
$0.10~{\rm GeV}^2 < -t' < 0.20$~GeV$^2$  &  0.146 (0.145) \\
$0.20~{\rm GeV}^2 < -t' < 0.40$~GeV$^2$  &  0.281 (0.283) \\
\hline
 & $ \langle x_B \rangle $ \\   
\hline
$0.0 < x_B < 0.05$  & 0.042 (0.042) \\
$0.05 < x_B < 0.08$ & 0.064 (0.064) \\
$0.08 < x_B < 0.35$ & 0.120 (0.119) \\
\hline
\hline
 $1~{\rm GeV}^2 < Q^{2} < 7$~GeV$^2$  & $ \langle \epsilon \rangle = 0.80 \pm 0.01$ \\
\hline  
\end{tabular} \\[2pt]
\end{footnotesize}
\end{center}
\end{table*}

\begin{table*}
 \renewcommand{\arraystretch}{1.2} 
\begin{center}
\caption{The 23 unpolarized and polarized SDMEs for $\rho^0$ production
from the proton in 
$Q^2$  bins defined by the 
limits 0.5, 1.0, 1.4, 2.0, and~7.0 GeV$^2$. The first uncertainties are 
statistical, the second  systematic.}
\label{tab15-h-q2}
\newcommand{\m}{\hphantom{$-$}}
\newcommand{\cc}[1]{\multicolumn{1}{c}{#1}}
\renewcommand{\arraystretch}{1.2} 
\begin{footnotesize}
\begin{tabular}{|l|c|c|c|c|}
\hline
Element & $\langle Q^2 \rangle =0.82$~GeV$^2$ & $\langle Q^2 \rangle =1.19$~GeV$^2$ & $\langle Q^2 \rangle =1.66$~GeV$^2$ &$\langle Q^2 \rangle =3.06$~GeV$^2$ \\
\hline
$r^{04}_{00}$    &  0.349 $\pm$  0.026 $\pm$  0.061 &  0.368 $\pm$  0.018 $\pm$  0.011 &  0.397 $\pm$  0.017 $\pm$  0.018 &  0.454 $\pm$  0.014 $\pm$  0.011\\
$r^1_{1-1}$      &  0.283 $\pm$  0.023 $\pm$  0.049 &  0.262 $\pm$  0.018 $\pm$  0.024 &  0.274 $\pm$  0.019 $\pm$  0.024 &  0.204 $\pm$  0.017 $\pm$  0.012\\
Im $r^2_{1-1}$   & $-$0.294 $\pm$  0.019 $\pm$  0.038 & $-$0.255 $\pm$  0.016 $\pm$  0.022 & $-$0.239 $\pm$  0.017 $\pm$  0.011 & $-$0.197 $\pm$  0.017 $\pm$  0.012\\
Re $r^5_{10}$    &  0.151 $\pm$  0.028 $\pm$  0.026 &  0.171 $\pm$  0.007 $\pm$  0.000 &  0.161 $\pm$  0.006 $\pm$  0.004 &  0.141 $\pm$  0.006 $\pm$  0.008\\
Im $r^6_{10}$    & $-$0.149 $\pm$  0.015 $\pm$  0.010 & $-$0.167 $\pm$  0.007 $\pm$  0.003 & $-$0.167 $\pm$  0.006 $\pm$  0.005 & $-$0.156 $\pm$  0.006 $\pm$  0.010\\
Im $r^7_{10}$    &  0.079 $\pm$  0.068 $\pm$  0.011 &  0.092 $\pm$  0.038 $\pm$  0.010 &  0.039 $\pm$  0.036 $\pm$  0.004 &  0.187 $\pm$  0.034 $\pm$  0.018\\
Re $r^8_{10}$    &  0.040 $\pm$  0.043 $\pm$  0.011 &  0.020 $\pm$  0.031 $\pm$  0.008 &  0.074 $\pm$  0.034 $\pm$  0.002 &  0.098 $\pm$  0.032 $\pm$  0.005\\
Re $r^{04}_{10}$ &  0.028 $\pm$  0.028 $\pm$  0.020 &  0.029 $\pm$  0.007 $\pm$  0.003 &  0.035 $\pm$  0.007 $\pm$  0.011 &  0.026 $\pm$  0.007 $\pm$  0.003\\
Re $r^1_{10}$    & $-$0.037 $\pm$  0.044 $\pm$  0.032 & $-$0.043 $\pm$  0.012 $\pm$  0.006 & $-$0.036 $\pm$  0.012 $\pm$  0.012 & $-$0.009 $\pm$  0.013 $\pm$  0.010\\
Im $r^2_{10}$    &  0.023 $\pm$  0.019 $\pm$  0.007 &  0.022 $\pm$  0.012 $\pm$  0.018 &  0.005 $\pm$  0.012 $\pm$  0.024 &  0.022 $\pm$  0.013 $\pm$  0.008\\
$r^5_{00}$       &  0.121 $\pm$  0.038 $\pm$  0.039 &  0.094 $\pm$  0.017 $\pm$  0.017 &  0.057 $\pm$  0.015 $\pm$  0.019 &  0.151 $\pm$  0.015 $\pm$  0.007\\
$r^1_{00}$       & $-$0.054 $\pm$  0.039 $\pm$  0.013 &  0.011 $\pm$  0.032 $\pm$  0.018 &  0.007 $\pm$  0.031 $\pm$  0.009 &  0.037 $\pm$  0.034 $\pm$  0.002\\
Im $r^3_{10}$    &  0.002 $\pm$  0.041 $\pm$  0.008 & $-$0.041 $\pm$  0.026 $\pm$  0.005 & $-$0.074 $\pm$  0.025 $\pm$  0.005 &  0.048 $\pm$  0.024 $\pm$  0.006\\
$r^8_{00}$       &  0.022 $\pm$  0.079 $\pm$  0.026 &  0.040 $\pm$  0.084 $\pm$  0.014 &  0.054 $\pm$  0.086 $\pm$  0.011 &  0.010 $\pm$  0.085 $\pm$  0.016\\
$r^5_{11}$       & $-$0.015 $\pm$  0.010 $\pm$  0.007 & $-$0.011 $\pm$  0.006 $\pm$  0.006 & $-$0.008 $\pm$  0.006 $\pm$  0.011 & $-$0.021 $\pm$  0.006 $\pm$  0.016\\
$r^5_{1-1}$      &  0.009 $\pm$  0.011 $\pm$  0.019 &  0.008 $\pm$  0.007 $\pm$  0.006 & $-$0.013 $\pm$  0.007 $\pm$  0.003 &  0.020 $\pm$  0.007 $\pm$  0.008\\
Im $r^6_{1-1}$   & $-$0.011 $\pm$  0.010 $\pm$  0.013 &  0.002 $\pm$  0.007 $\pm$  0.007 &  0.002 $\pm$  0.007 $\pm$  0.004 & $-$0.010 $\pm$  0.007 $\pm$  0.007\\
Im $r^7_{1-1}$   & $-$0.003 $\pm$  0.078 $\pm$  0.021 &  0.023 $\pm$  0.056 $\pm$  0.013 & $-$0.005 $\pm$  0.055 $\pm$  0.010 & $-$0.109 $\pm$  0.047 $\pm$  0.004\\
$r^8_{11}$       &  0.019 $\pm$  0.053 $\pm$  0.007 &  0.056 $\pm$  0.045 $\pm$  0.004 &  0.051 $\pm$  0.044 $\pm$  0.006 & $-$0.002 $\pm$  0.035 $\pm$  0.005\\
$r^8_{1-1}$      &  0.013 $\pm$  0.062 $\pm$  0.008 &  0.072 $\pm$  0.053 $\pm$  0.011 & $-$0.018 $\pm$  0.054 $\pm$  0.004 &  0.004 $\pm$  0.045 $\pm$  0.014\\
$r^{04}_{1-1}$   & $-$0.024 $\pm$  0.013 $\pm$  0.021 & $-$0.014 $\pm$  0.010 $\pm$  0.010 & $-$0.019 $\pm$  0.010 $\pm$  0.003 &  0.001 $\pm$  0.009 $\pm$  0.007\\
$r^1_{11}$       & $-$0.039 $\pm$  0.017 $\pm$  0.018 & $-$0.034 $\pm$  0.013 $\pm$  0.013 & $-$0.023 $\pm$  0.013 $\pm$  0.008 & $-$0.018 $\pm$  0.012 $\pm$  0.010\\
Im $r^3_{1-1}$   &  0.021 $\pm$  0.051 $\pm$  0.010 &  0.000 $\pm$  0.033 $\pm$  0.004 & $-$0.031 $\pm$  0.032 $\pm$  0.007 & $-$0.026 $\pm$  0.028 $\pm$  0.005\\
\hline
\end{tabular} \\[2pt]
\end{footnotesize}
\end{center}
\end{table*}

\begin{table*}
\begin{center}

 \renewcommand{\arraystretch}{1.2} 
\caption{The 23 unpolarized and polarized SDMEs for $\rho^0$ production
from the proton in  
$-t'$ bins defined by the limits 0.0, 0.04, 0.1, 0.2, and~0.4 GeV$^2$. 
The first uncertainties are statistical, the 
second systematic.}
\label{tab15-h-tpr}
\newcommand{\m}{\hphantom{$-$}}
\newcommand{\cc}[1]{\multicolumn{1}{c}{#1}}
\renewcommand{\arraystretch}{1.2} 
\begin{footnotesize}
\begin{tabular}{|l|c|c|c|c|}
\hline
Element & $ \langle -t' \rangle =0.019$ ~GeV$^2$ & $\langle -t' \rangle =0.068$~GeV$^2$ & $\langle -t' \rangle =0.146$~GeV$^2$ & $\langle -t' \rangle =0.281$ ~GeV $^2$  \\
\hline
$r^{04}_{00}$    &  0.393 $\pm$  0.018 $\pm$  0.019 &  0.394 $\pm$  0.018 $\pm$  0.025 &  0.415 $\pm$  0.019 $\pm$  0.021 &  0.481 $\pm$  0.019 $\pm$  0.028\\
$r^1_{1-1}$      &  0.235 $\pm$  0.021 $\pm$  0.014 &  0.265 $\pm$  0.021 $\pm$  0.012 &  0.260 $\pm$  0.020 $\pm$  0.031 &  0.203 $\pm$  0.020 $\pm$  0.030\\
Im $r^2_{1-1}$   & $-$0.204 $\pm$  0.020 $\pm$  0.008 & $-$0.243 $\pm$  0.019 $\pm$  0.011 & $-$0.232 $\pm$  0.019 $\pm$  0.041 & $-$0.231 $\pm$  0.019 $\pm$  0.021\\
Re $r^5_{10}$    &  0.156 $\pm$  0.007 $\pm$  0.005 &  0.156 $\pm$  0.007 $\pm$  0.010 &  0.153 $\pm$  0.007 $\pm$  0.007 &  0.153 $\pm$  0.008 $\pm$  0.009\\
Im $r^6_{10}$    & $-$0.162 $\pm$  0.007 $\pm$  0.005 & $-$0.170 $\pm$  0.007 $\pm$  0.010 & $-$0.155 $\pm$  0.007 $\pm$  0.006 & $-$0.153 $\pm$  0.008 $\pm$  0.008\\
Im $r^7_{10}$    &  0.103 $\pm$  0.040 $\pm$  0.007 &  0.112 $\pm$  0.040 $\pm$  0.007 &  0.081 $\pm$  0.042 $\pm$  0.011 &  0.163 $\pm$  0.047 $\pm$  0.033\\
Re $r^8_{10}$    &  0.042 $\pm$  0.037 $\pm$  0.011 &  0.059 $\pm$  0.037 $\pm$  0.006 &  0.100 $\pm$  0.036 $\pm$  0.013 &  0.114 $\pm$  0.039 $\pm$  0.015\\
Re $r^{04}_{10}$ &  0.018 $\pm$  0.008 $\pm$  0.004 &  0.027 $\pm$  0.008 $\pm$  0.011 &  0.035 $\pm$  0.008 $\pm$  0.007 &  0.038 $\pm$  0.008 $\pm$  0.003\\
Re $r^1_{10}$    & $-$0.009 $\pm$  0.014 $\pm$  0.005 & $-$0.045 $\pm$  0.014 $\pm$  0.024 & $-$0.013 $\pm$  0.015 $\pm$  0.009 & $-$0.046 $\pm$  0.016 $\pm$  0.015\\
Im $r^2_{10}$    & $-$0.001 $\pm$  0.014 $\pm$  0.011 &  0.030 $\pm$  0.014 $\pm$  0.019 &  0.015 $\pm$  0.013 $\pm$  0.019 &  0.030 $\pm$  0.015 $\pm$  0.013\\
$r^5_{00}$       &  0.039 $\pm$  0.016 $\pm$  0.001 &  0.068 $\pm$  0.016 $\pm$  0.030 &  0.136 $\pm$  0.018 $\pm$  0.010 &  0.219 $\pm$  0.020 $\pm$  0.022\\
$r^1_{00}$       &  0.019 $\pm$  0.035 $\pm$  0.020 &  0.015 $\pm$  0.035 $\pm$  0.027 & $-$0.026 $\pm$  0.036 $\pm$  0.021 & $-$0.005 $\pm$  0.041 $\pm$  0.013\\
Im $r^3_{10}$    & $-$0.035 $\pm$  0.028 $\pm$  0.001 & $-$0.044 $\pm$  0.027 $\pm$  0.004 &  0.018 $\pm$  0.029 $\pm$  0.007 &  0.009 $\pm$  0.032 $\pm$  0.025\\
$r^8_{00}$       & $-$0.013 $\pm$  0.103 $\pm$  0.011 &  0.128 $\pm$  0.097 $\pm$  0.012 & $-$0.028 $\pm$  0.095 $\pm$  0.028 &  0.066 $\pm$  0.097 $\pm$  0.012\\
$r^5_{11}$       & $-$0.009 $\pm$  0.007 $\pm$  0.004 & $-$0.011 $\pm$  0.006 $\pm$  0.009 & $-$0.009 $\pm$  0.007 $\pm$  0.011 & $-$0.031 $\pm$  0.006 $\pm$  0.024\\
$r^5_{1-1}$      & $-$0.009 $\pm$  0.008 $\pm$  0.003 &  0.005 $\pm$  0.008 $\pm$  0.004 &  0.003 $\pm$  0.008 $\pm$  0.008 &  0.012 $\pm$  0.008 $\pm$  0.008\\
Im $r^6_{1-1}$   &  0.010 $\pm$  0.008 $\pm$  0.002 & $-$0.010 $\pm$  0.008 $\pm$  0.008 &  0.006 $\pm$  0.008 $\pm$  0.003 & $-$0.009 $\pm$  0.008 $\pm$  0.002\\
Im $r^7_{1-1}$   & $-$0.034 $\pm$  0.064 $\pm$  0.011 & $-$0.040 $\pm$  0.060 $\pm$  0.004 & $-$0.076 $\pm$  0.060 $\pm$  0.015 & $-$0.005 $\pm$  0.058 $\pm$  0.008\\
$r^8_{11}$       &  0.018 $\pm$  0.050 $\pm$  0.003 & $-$0.010 $\pm$  0.048 $\pm$  0.009 &  0.061 $\pm$  0.045 $\pm$  0.006 &  0.068 $\pm$  0.042 $\pm$  0.006\\
$r^8_{1-1}$      &  0.021 $\pm$  0.062 $\pm$  0.005 &  0.024 $\pm$  0.058 $\pm$  0.009 & $-$0.019 $\pm$  0.055 $\pm$  0.007 &  0.051 $\pm$  0.053 $\pm$  0.006\\
$r^{04}_{1-1}$   &  0.008 $\pm$  0.011 $\pm$  0.003 &  0.008 $\pm$  0.011 $\pm$  0.010 & $-$0.033 $\pm$  0.011 $\pm$  0.008 & $-$0.029 $\pm$  0.010 $\pm$  0.003\\
$r^1_{11}$       & $-$0.022 $\pm$  0.015 $\pm$  0.016 &  0.002 $\pm$  0.015 $\pm$  0.009 & $-$0.036 $\pm$  0.014 $\pm$  0.003 & $-$0.034 $\pm$  0.014 $\pm$  0.012\\
Im $r^3_{1-1}$   & $-$0.038 $\pm$  0.036 $\pm$  0.008 & $-$0.015 $\pm$  0.035 $\pm$  0.001 & $-$0.014 $\pm$  0.036 $\pm$  0.006 & $-$0.036 $\pm$  0.036 $\pm$  0.010\\
\hline
\end{tabular} \\[2pt]
\end{footnotesize}
\end{center}
\end{table*}

\begin{table*}
 \renewcommand{\arraystretch}{1.2} 
\begin{center}
\caption{The 23 unpolarized and polarized SDMEs for $\rho^0$ production from
the proton in 
$x_B$ bins defined by the limits 0.0, 0.05, 0.08, and~0.35. 
The first uncertainties are statistical, the second systematic.}
\label{tab15-h-xbj}
\newcommand{\m}{\hphantom{$-$}}
\newcommand{\cc}[1]{\multicolumn{1}{c}{#1}}
\renewcommand{\arraystretch}{1.2} 
\begin{footnotesize}
\begin{tabular}{|l|c|c|c|}
\hline
Element & $ \langle x_B \rangle$ =0.042 & $\langle x_B \rangle$=0.064 & $\langle x_B \rangle$=0.120  \\
\hline
$r^{04}_{00}$    &  0.349 $\pm$  0.044 $\pm$  0.031 &  0.405 $\pm$  0.017 $\pm$  0.013 &  0.448 $\pm$  0.014 $\pm$  0.012 \\
$r^1_{1-1}$      &  0.297 $\pm$  0.036 $\pm$  0.006 &  0.253 $\pm$  0.014 $\pm$  0.023 &  0.216 $\pm$  0.015 $\pm$  0.019 \\
Im $r^2_{1-1}$   & $-$0.308 $\pm$  0.032 $\pm$  0.024 & $-$0.229 $\pm$  0.031 $\pm$  0.008 & $-$0.198 $\pm$  0.014 $\pm$  0.015 \\
Re $r^5_{10}$    &  0.151 $\pm$  0.027 $\pm$  0.014 &  0.176 $\pm$  0.008 $\pm$  0.007 &  0.146 $\pm$  0.006 $\pm$  0.012 \\
Im $r^6_{10}$    & $-$0.172 $\pm$  0.015 $\pm$  0.006 & $-$0.169 $\pm$  0.005 $\pm$  0.003 & $-$0.157 $\pm$  0.006 $\pm$  0.012 \\
Im $r^7_{10}$    &  0.132 $\pm$  0.067 $\pm$  0.015 &  0.045 $\pm$  0.061 $\pm$  0.016 &  0.160 $\pm$  0.032 $\pm$  0.018 \\
Re $r^8_{10}$    &  0.021 $\pm$  0.041 $\pm$  0.002 &  0.067 $\pm$  0.042 $\pm$  0.000 &  0.092 $\pm$  0.032 $\pm$  0.011 \\
Re $r^{04}_{10}$ &  0.023 $\pm$  0.021 $\pm$  0.005 &  0.038 $\pm$  0.017 $\pm$  0.010 &  0.033 $\pm$  0.006 $\pm$  0.005 \\
Re $r^1_{10}$    & $-$0.013 $\pm$  0.035 $\pm$  0.014 & $-$0.050 $\pm$  0.011 $\pm$  0.010 & $-$0.021 $\pm$  0.011 $\pm$  0.019 \\
Im $r^2_{10}$    &  0.035 $\pm$  0.021 $\pm$  0.017 &  0.009 $\pm$  0.022 $\pm$  0.015 &  0.027 $\pm$  0.011 $\pm$  0.017 \\
$r^5_{00}$       &  0.079 $\pm$  0.051 $\pm$  0.034 &  0.056 $\pm$  0.013 $\pm$  0.017 &  0.130 $\pm$  0.013 $\pm$  0.014 \\
$r^1_{00}$       &  0.044 $\pm$  0.064 $\pm$  0.016 &  0.000 $\pm$  0.029 $\pm$  0.005 &  0.067 $\pm$  0.030 $\pm$  0.007 \\
Im $r^3_{10}$    & $-$0.025 $\pm$  0.042 $\pm$  0.001 & $-$0.067 $\pm$  0.033 $\pm$  0.001 &  0.014 $\pm$  0.022 $\pm$  0.005 \\
$r^8_{00}$       &  0.100 $\pm$  0.092 $\pm$  0.011 &  0.030 $\pm$  0.074 $\pm$  0.003 & $-$0.005 $\pm$  0.090 $\pm$  0.031 \\
$r^5_{11}$       &  0.005 $\pm$  0.022 $\pm$  0.011 & $-$0.014 $\pm$  0.011 $\pm$  0.005 & $-$0.019 $\pm$  0.004 $\pm$  0.016 \\
$r^5_{1-1}$      & $-$0.014 $\pm$  0.020 $\pm$  0.002 &  0.017 $\pm$  0.007 $\pm$  0.004 &  0.000 $\pm$  0.006 $\pm$  0.005 \\
Im $r^6_{1-1}$   &  0.000 $\pm$  0.017 $\pm$  0.003 &  0.003 $\pm$  0.014 $\pm$  0.007 & $-$0.001 $\pm$  0.006 $\pm$  0.007 \\
Im $r^7_{1-1}$   &  0.019 $\pm$  0.099 $\pm$  0.012 &  0.071 $\pm$  0.135 $\pm$  0.019 & $-$0.142 $\pm$  0.044 $\pm$  0.010 \\
$r^8_{11}$       &  0.042 $\pm$  0.060 $\pm$  0.007 &  0.081 $\pm$  0.079 $\pm$  0.008 & $-$0.002 $\pm$  0.033 $\pm$  0.009 \\
$r^8_{1-1}$      &  0.089 $\pm$  0.065 $\pm$  0.003 &  0.031 $\pm$  0.077 $\pm$  0.003 &  0.005 $\pm$  0.043 $\pm$  0.009 \\
$r^{04}_{1-1}$   & $-$0.025 $\pm$  0.020 $\pm$  0.007 & $-$0.001 $\pm$  0.012 $\pm$  0.010 & $-$0.016 $\pm$  0.008 $\pm$  0.003 \\
$r^1_{11}$       & $-$0.067 $\pm$  0.030 $\pm$  0.008 & $-$0.018 $\pm$  0.014 $\pm$  0.011 & $-$0.032 $\pm$  0.010 $\pm$  0.009 \\
Im $r^3_{1-1}$   & $-$0.020 $\pm$  0.067 $\pm$  0.014 & $-$0.017 $\pm$  0.061 $\pm$  0.009 & $-$0.001 $\pm$  0.026 $\pm$  0.004 \\
\hline
\end{tabular} \\[2pt]
\end{footnotesize}
\end{center}
\end{table*}

\newpage
%
\begin{table*}
 \renewcommand{\arraystretch}{1.2} 
\begin{center}
\caption{The 23 unpolarized and polarized SDMEs for $\rho^0$ production
from the deuteron in 
$Q^2$  bins defined by the 
limits 0.5, 1.0, 1.4, 2.0, and~7.0 GeV$^2$. The first uncertainties are 
statistical, the second  systematic.}
\label{tab15-d-q2}
\newcommand{\m}{\hphantom{$-$}}
\newcommand{\cc}[1]{\multicolumn{1}{c}{#1}}
\renewcommand{\arraystretch}{1.2} 
\begin{footnotesize}
\begin{tabular}{|l|c|c|c|c|}
\hline
SDME  &$ \langle Q^{2} \rangle =0.82$~GeV$^{2}$ & $\langle Q^{2} \rangle =1.18$~GeV$^{2}$ & $\langle Q^{2} \rangle =1.66$~GeV$^{2}$ & $\langle Q^{2} \rangle =3.04$~GeV$^{2}$  \\
\hline
$r^{04}_{00}$    &  0.365 $\pm$  0.015 $\pm$  0.058 &  0.386 $\pm$  0.017 $\pm$  0.010 &  0.445 $\pm$  0.012 $\pm$  0.022 &  0.407 $\pm$  0.011 $\pm$  0.013\\
$r^1_{1-1}$      &  0.294 $\pm$  0.013 $\pm$  0.060 &  0.278 $\pm$  0.016 $\pm$  0.022 &  0.235 $\pm$  0.013 $\pm$  0.012 &  0.216 $\pm$  0.014 $\pm$  0.008\\
Im $r^2_{1-1}$   & $-$0.287 $\pm$  0.014 $\pm$  0.047 & $-$0.267 $\pm$  0.013 $\pm$  0.034 & $-$0.194 $\pm$  0.014 $\pm$  0.020 & $-$0.219 $\pm$  0.014 $\pm$  0.017\\
Re $r^5_{10}$    &  0.159 $\pm$  0.008 $\pm$  0.025 &  0.173 $\pm$  0.005 $\pm$  0.002 &  0.160 $\pm$  0.005 $\pm$  0.006 &  0.154 $\pm$  0.005 $\pm$  0.005\\
Im $r^6_{10}$    & $-$0.159 $\pm$  0.008 $\pm$  0.015 & $-$0.165 $\pm$  0.005 $\pm$  0.001 & $-$0.155 $\pm$  0.005 $\pm$  0.003 & $-$0.139 $\pm$  0.005 $\pm$  0.007\\
Im $r^7_{10}$    &  0.070 $\pm$  0.042 $\pm$  0.004 &  0.080 $\pm$  0.028 $\pm$  0.011 &  0.102 $\pm$  0.030 $\pm$  0.004 &  0.125 $\pm$  0.026 $\pm$  0.007\\
Re $r^8_{10}$    &  0.080 $\pm$  0.028 $\pm$  0.012 &  0.080 $\pm$  0.023 $\pm$  0.003 &  0.132 $\pm$  0.026 $\pm$  0.006 &  0.117 $\pm$  0.024 $\pm$  0.012\\
Re $r^{04}_{10}$ &  0.021 $\pm$  0.007 $\pm$  0.019 &  0.036 $\pm$  0.006 $\pm$  0.004 &  0.026 $\pm$  0.006 $\pm$  0.016 &  0.023 $\pm$  0.005 $\pm$  0.006\\
Re $r^1_{10}$    & $-$0.025 $\pm$  0.011 $\pm$  0.031 & $-$0.017 $\pm$  0.009 $\pm$  0.002 & $-$0.014 $\pm$  0.010 $\pm$  0.015 & $-$0.031 $\pm$  0.010 $\pm$  0.013\\
Im $r^2_{10}$    & $-$0.006 $\pm$  0.011 $\pm$  0.017 &  0.016 $\pm$  0.009 $\pm$  0.018 &  0.015 $\pm$  0.010 $\pm$  0.038 &  0.004 $\pm$  0.010 $\pm$  0.018\\
$r^5_{00}$       &  0.097 $\pm$  0.017 $\pm$  0.035 &  0.088 $\pm$  0.013 $\pm$  0.011 &  0.113 $\pm$  0.012 $\pm$  0.018 &  0.119 $\pm$  0.012 $\pm$  0.006\\
$r^1_{00}$       &  0.019 $\pm$  0.028 $\pm$  0.014 & $-$0.018 $\pm$  0.024 $\pm$  0.019 & $-$0.031 $\pm$  0.025 $\pm$  0.020 & $-$0.036 $\pm$  0.026 $\pm$  0.009\\
Im $r^3_{10}$    &  0.005 $\pm$  0.029 $\pm$  0.004 &  0.003 $\pm$  0.019 $\pm$  0.008 &  0.024 $\pm$  0.021 $\pm$  0.003 &  0.062 $\pm$  0.018 $\pm$  0.003\\
$r^8_{00}$       &  0.138 $\pm$  0.061 $\pm$  0.010 &  0.221 $\pm$  0.066 $\pm$  0.021 &  0.058 $\pm$  0.069 $\pm$  0.016 & $-$0.098 $\pm$  0.063 $\pm$  0.007\\
$r^5_{11}$       & $-$0.009 $\pm$  0.006 $\pm$  0.005 & $-$0.013 $\pm$  0.004 $\pm$  0.009 & $-$0.017 $\pm$  0.004 $\pm$  0.013 & $-$0.027 $\pm$  0.004 $\pm$  0.015\\
$r^5_{1-1}$      &  0.008 $\pm$  0.007 $\pm$  0.016 &  0.009 $\pm$  0.005 $\pm$  0.002 &  0.006 $\pm$  0.006 $\pm$  0.004 &  0.021 $\pm$  0.006 $\pm$  0.011\\
Im $r^6_{1-1}$   & $-$0.007 $\pm$  0.007 $\pm$  0.018 & $-$0.003 $\pm$  0.005 $\pm$  0.004 & $-$0.003 $\pm$  0.006 $\pm$  0.006 & $-$0.013 $\pm$  0.006 $\pm$  0.005\\
Im $r^7_{1-1}$   & $-$0.066 $\pm$  0.052 $\pm$  0.008 & $-$0.040 $\pm$  0.040 $\pm$  0.016 & $-$0.026 $\pm$  0.044 $\pm$  0.001 & $-$0.100 $\pm$  0.037 $\pm$  0.013\\
$r^8_{11}$       &  0.007 $\pm$  0.039 $\pm$  0.003 & $-$0.011 $\pm$  0.035 $\pm$  0.010 &  0.047 $\pm$  0.033 $\pm$  0.008 &  0.037 $\pm$  0.028 $\pm$  0.001\\
$r^8_{1-1}$      & $-$0.015 $\pm$  0.047 $\pm$  0.014 & $-$0.055 $\pm$  0.040 $\pm$  0.015 & $-$0.083 $\pm$  0.041 $\pm$  0.003 & $-$0.072 $\pm$  0.036 $\pm$  0.017\\
$r^{04}_{1-1}$   &  0.000 $\pm$  0.009 $\pm$  0.021 &  0.003 $\pm$  0.008 $\pm$  0.012 & $-$0.006 $\pm$  0.008 $\pm$  0.009 & $-$0.008 $\pm$  0.007 $\pm$  0.003\\
$r^1_{11}$       & $-$0.029 $\pm$  0.012 $\pm$  0.019 & $-$0.002 $\pm$  0.009 $\pm$  0.013 & $-$0.003 $\pm$  0.010 $\pm$  0.011 & $-$0.012 $\pm$  0.010 $\pm$  0.007\\
Im $r^3_{1-1}$   &  0.006 $\pm$  0.031 $\pm$  0.010 & $-$0.017 $\pm$  0.024 $\pm$  0.008 & $-$0.023 $\pm$  0.026 $\pm$  0.003 &  0.029 $\pm$  0.022 $\pm$  0.004\\
\hline
\end{tabular} \\[2pt]
\end{footnotesize}
\end{center}
\end{table*}
\normalsize

\begin{table*}
 \renewcommand{\arraystretch}{1.2} 
\begin{center}
\caption{The 23 unpolarized and polarized SDMEs for $\rho^0$ production
from the deuteron in  
$-t'$ bins defined by the limits 0.0, 0.04, 0.1, 0.2, and~0.4 GeV$^2$. 
The first uncertainties are statistical, the 
second systematic.}
\label{tab15-d-tpr}
\newcommand{\m}{\hphantom{$-$}}
\newcommand{\cc}[1]{\multicolumn{1}{c}{#1}}
\renewcommand{\arraystretch}{1.2} 
\begin{footnotesize}
\begin{tabular}{|l|c|c|c|c|}
\hline
Element &$ \langle -t'\rangle =0.018$~GeV$^{2}$ & $\langle -t'\rangle =0.068$~GeV$^{2}$ & $\langle -t'\rangle =0.145$~GeV$^{2}$ & $\langle -t'\rangle =0.283$~GeV$^2$  \\
\hline
$r^{04}_{00}$    &  0.440 $\pm$  0.014 $\pm$  0.017 &  0.396 $\pm$  0.015 $\pm$  0.024 &  0.389 $\pm$  0.015 $\pm$  0.017 &  0.434 $\pm$  0.017 $\pm$  0.032\\
$r^1_{1-1}$      &  0.225 $\pm$  0.015 $\pm$  0.014 &  0.261 $\pm$  0.017 $\pm$  0.011 &  0.272 $\pm$  0.016 $\pm$  0.029 &  0.247 $\pm$  0.018 $\pm$  0.029\\
Im $r^2_{1-1}$   & $-$0.208 $\pm$  0.014 $\pm$  0.015 & $-$0.258 $\pm$  0.015 $\pm$  0.019 & $-$0.239 $\pm$  0.015 $\pm$  0.028 & $-$0.229 $\pm$  0.016 $\pm$  0.026\\
Re $r^5_{10}$    &  0.155 $\pm$  0.005 $\pm$  0.006 &  0.153 $\pm$  0.006 $\pm$  0.010 &  0.172 $\pm$  0.006 $\pm$  0.009 &  0.154 $\pm$  0.007 $\pm$  0.009\\
Im $r^6_{10}$    & $-$0.154 $\pm$  0.005 $\pm$  0.003 & $-$0.150 $\pm$  0.005 $\pm$  0.004 & $-$0.150 $\pm$  0.006 $\pm$  0.006 & $-$0.154 $\pm$  0.007 $\pm$  0.010\\
Im $r^7_{10}$    &  0.084 $\pm$  0.029 $\pm$  0.006 &  0.122 $\pm$  0.028 $\pm$  0.008 &  0.108 $\pm$  0.034 $\pm$  0.007 &  0.086 $\pm$  0.039 $\pm$  0.015\\
Re $r^8_{10}$    &  0.161 $\pm$  0.026 $\pm$  0.010 &  0.095 $\pm$  0.028 $\pm$  0.007 &  0.081 $\pm$  0.027 $\pm$  0.006 &  0.099 $\pm$  0.030 $\pm$  0.012\\
Re $r^{04}_{10}$ &  0.016 $\pm$  0.006 $\pm$  0.005 &  0.025 $\pm$  0.006 $\pm$  0.011 &  0.034 $\pm$  0.006 $\pm$  0.005 &  0.046 $\pm$  0.007 $\pm$  0.005\\
Re $r^1_{10}$    & $-$0.010 $\pm$  0.010 $\pm$  0.005 & $-$0.002 $\pm$  0.011 $\pm$  0.021 & $-$0.020 $\pm$  0.011 $\pm$  0.010 & $-$0.047 $\pm$  0.013 $\pm$  0.018\\
Im $r^2_{10}$    &  0.012 $\pm$  0.010 $\pm$  0.010 &  0.018 $\pm$  0.010 $\pm$  0.018 &  0.005 $\pm$  0.011 $\pm$  0.018 &  0.036 $\pm$  0.013 $\pm$  0.017\\
$r^5_{00}$       &  0.053 $\pm$  0.012 $\pm$  0.003 &  0.085 $\pm$  0.013 $\pm$  0.033 &  0.108 $\pm$  0.014 $\pm$  0.003 &  0.215 $\pm$  0.018 $\pm$  0.032\\
$r^1_{00}$       & $-$0.077 $\pm$  0.026 $\pm$  0.025 & $-$0.001 $\pm$  0.028 $\pm$  0.034 & $-$0.053 $\pm$  0.029 $\pm$  0.028 & $-$0.040 $\pm$  0.035 $\pm$  0.030\\
Im $r^3_{10}$    &  0.020 $\pm$  0.020 $\pm$  0.003 &  0.043 $\pm$  0.020 $\pm$  0.003 &  0.022 $\pm$  0.023 $\pm$  0.004 &  0.033 $\pm$  0.027 $\pm$  0.006\\
$r^8_{00}$       &  0.157 $\pm$  0.073 $\pm$  0.013 &  0.111 $\pm$  0.076 $\pm$  0.011 & $-$0.147 $\pm$  0.070 $\pm$  0.013 &  0.078 $\pm$  0.079 $\pm$  0.012\\
$r^5_{11}$       & $-$0.018 $\pm$  0.005 $\pm$  0.007 & $-$0.014 $\pm$  0.005 $\pm$  0.009 & $-$0.012 $\pm$  0.005 $\pm$  0.005 & $-$0.025 $\pm$  0.006 $\pm$  0.023\\
$r^5_{1-1}$      &  0.001 $\pm$  0.006 $\pm$  0.003 &  0.007 $\pm$  0.006 $\pm$  0.004 &  0.006 $\pm$  0.006 $\pm$  0.007 &  0.017 $\pm$  0.007 $\pm$  0.008\\
Im $r^6_{1-1}$   & $-$0.007 $\pm$  0.006 $\pm$  0.002 & $-$0.005 $\pm$  0.006 $\pm$  0.005 &  0.001 $\pm$  0.006 $\pm$  0.009 & $-$0.003 $\pm$  0.007 $\pm$  0.002\\
Im $r^7_{1-1}$   & $-$0.037 $\pm$  0.043 $\pm$  0.006 & $-$0.041 $\pm$  0.045 $\pm$  0.004 & $-$0.060 $\pm$  0.046 $\pm$  0.002 & $-$0.080 $\pm$  0.048 $\pm$  0.008\\
$r^8_{11}$       & $-$0.030 $\pm$  0.036 $\pm$  0.002 & $-$0.003 $\pm$  0.037 $\pm$  0.004 &  0.078 $\pm$  0.035 $\pm$  0.001 &  0.056 $\pm$  0.037 $\pm$  0.008\\
$r^8_{1-1}$      & $-$0.095 $\pm$  0.044 $\pm$  0.010 & $-$0.055 $\pm$  0.044 $\pm$  0.008 & $-$0.030 $\pm$  0.044 $\pm$  0.008 & $-$0.091 $\pm$  0.045 $\pm$  0.006\\
$r^{04}_{1-1}$   &  0.016 $\pm$  0.008 $\pm$  0.004 & $-$0.005 $\pm$  0.009 $\pm$  0.006 & $-$0.021 $\pm$  0.009 $\pm$  0.001 & $-$0.013 $\pm$  0.009 $\pm$  0.008\\
$r^1_{11}$       &  0.026 $\pm$  0.011 $\pm$  0.013 & $-$0.004 $\pm$  0.012 $\pm$  0.015 & $-$0.036 $\pm$  0.012 $\pm$  0.004 & $-$0.016 $\pm$  0.012 $\pm$  0.011\\
Im $r^3_{1-1}$   &  0.003 $\pm$  0.025 $\pm$  0.002 & $-$0.036 $\pm$  0.026 $\pm$  0.005 & $-$0.033 $\pm$  0.028 $\pm$  0.002 &  0.052 $\pm$  0.031 $\pm$  0.005\\
\hline
\end{tabular} \\[2pt]
\end{footnotesize}
\end{center}
\end{table*}

\normalsize
\begin{table*}
 \renewcommand{\arraystretch}{1.2} 
\begin{center}
\caption{The 23 unpolarized and polarized SDMEs for $\rho^0$ production from
the deuteron in 
$x_B$ bins defined by the limits 0.0, 0.05, 0.08, and~0.35. 
The first uncertainties are statistical, the second systematic.}
\label{tab15-d-xbj}
\newcommand{\m}{\hphantom{$-$}}
\newcommand{\cc}[1]{\multicolumn{1}{c}{#1}}
\renewcommand{\arraystretch}{1.2} 
\begin{footnotesize}
\begin{tabular}{|l|c|c|c|}
\hline
Element & $ \langle x_B \rangle$=0.042 & $\langle x_B \rangle$=0.064 & $\langle x_B \rangle$=0.119  \\
\hline
$r^{04}_{00}$    &  0.410 $\pm$  0.017 $\pm$  0.032 &  0.413 $\pm$  0.012 $\pm$  0.014 &  0.411 $\pm$  0.011 $\pm$  0.010 \\
$r^1_{1-1}$      &  0.233 $\pm$  0.019 $\pm$  0.018 &  0.263 $\pm$  0.013 $\pm$  0.022 &  0.223 $\pm$  0.011 $\pm$  0.012 \\
Im $r^2_{1-1}$   & $-$0.226 $\pm$  0.020 $\pm$  0.039 & $-$0.234 $\pm$  0.011 $\pm$  0.019 & $-$0.218 $\pm$  0.012 $\pm$  0.018 \\
Re $r^5_{10}$    &  0.162 $\pm$  0.005 $\pm$  0.021 &  0.168 $\pm$  0.005 $\pm$  0.007 &  0.148 $\pm$  0.004 $\pm$  0.011 \\
Im $r^6_{10}$    & $-$0.162 $\pm$  0.003 $\pm$  0.017 & $-$0.160 $\pm$  0.004 $\pm$  0.004 & $-$0.132 $\pm$  0.005 $\pm$  0.009 \\
Im $r^7_{10}$    &  0.102 $\pm$  0.038 $\pm$  0.028 &  0.106 $\pm$  0.026 $\pm$  0.006 &  0.099 $\pm$  0.025 $\pm$  0.005 \\
Re $r^8_{10}$    &  0.080 $\pm$  0.017 $\pm$  0.018 &  0.080 $\pm$  0.020 $\pm$  0.005 &  0.145 $\pm$  0.024 $\pm$  0.018 \\
Re $r^{04}_{10}$ &  0.044 $\pm$  0.005 $\pm$  0.009 &  0.027 $\pm$  0.005 $\pm$  0.012 &  0.026 $\pm$  0.005 $\pm$  0.005 \\
Re $r^1_{10}$    & $-$0.017 $\pm$  0.013 $\pm$  0.027 & $-$0.016 $\pm$  0.010 $\pm$  0.009 & $-$0.025 $\pm$  0.009 $\pm$  0.018 \\
Im $r^2_{10}$    &  0.013 $\pm$  0.008 $\pm$  0.031 &  0.024 $\pm$  0.009 $\pm$  0.019 &  0.011 $\pm$  0.010 $\pm$  0.017 \\
$r^5_{00}$       &  0.138 $\pm$  0.014 $\pm$  0.043 &  0.075 $\pm$  0.011 $\pm$  0.014 &  0.120 $\pm$  0.010 $\pm$  0.014 \\
$r^1_{00}$       & $-$0.010 $\pm$  0.023 $\pm$  0.061 & $-$0.056 $\pm$  0.022 $\pm$  0.011 & $-$0.019 $\pm$  0.024 $\pm$  0.009 \\
Im $r^3_{10}$    &  0.048 $\pm$  0.026 $\pm$  0.020 &  0.010 $\pm$  0.018 $\pm$  0.005 &  0.040 $\pm$  0.017 $\pm$  0.002 \\
$r^8_{00}$       &  0.199 $\pm$  0.043 $\pm$  0.024 &  0.074 $\pm$  0.058 $\pm$  0.005 & $-$0.059 $\pm$  0.062 $\pm$  0.016 \\
$r^5_{11}$       &  0.010 $\pm$  0.007 $\pm$  0.021 & $-$0.006 $\pm$  0.004 $\pm$  0.005 & $-$0.030 $\pm$  0.004 $\pm$  0.015 \\
$r^5_{1-1}$      &  0.018 $\pm$  0.009 $\pm$  0.011 &  0.000 $\pm$  0.005 $\pm$  0.002 &  0.020 $\pm$  0.004 $\pm$  0.007 \\
Im $r^6_{1-1}$   & $-$0.010 $\pm$  0.011 $\pm$  0.002 &  0.001 $\pm$  0.005 $\pm$  0.008 & $-$0.012 $\pm$  0.005 $\pm$  0.005 \\
Im $r^7_{1-1}$   & $-$0.081 $\pm$  0.033 $\pm$  0.032 & $-$0.028 $\pm$  0.037 $\pm$  0.003 & $-$0.081 $\pm$  0.036 $\pm$  0.010 \\
$r^8_{11}$       & $-$0.028 $\pm$  0.026 $\pm$  0.007 &  0.020 $\pm$  0.029 $\pm$  0.001 &  0.049 $\pm$  0.026 $\pm$  0.004 \\
$r^8_{1-1}$      & $-$0.116 $\pm$  0.042 $\pm$  0.028 & $-$0.046 $\pm$  0.035 $\pm$  0.004 & $-$0.073 $\pm$  0.034 $\pm$  0.017 \\
$r^{04}_{1-1}$   &  0.004 $\pm$  0.008 $\pm$  0.020 & $-$0.003 $\pm$  0.007 $\pm$  0.012 & $-$0.007 $\pm$  0.006 $\pm$  0.002 \\
$r^1_{11}$       & $-$0.023 $\pm$  0.016 $\pm$  0.020 & $-$0.007 $\pm$  0.009 $\pm$  0.011 & $-$0.010 $\pm$  0.009 $\pm$  0.010 \\
Im $r^3_{1-1}$   & $-$0.001 $\pm$  0.013 $\pm$  0.019 & $-$0.032 $\pm$  0.022 $\pm$  0.004 &  0.027 $\pm$  0.019 $\pm$  0.004 \\
\hline
\end{tabular} \\[2pt]
\end{footnotesize}
\end{center}
\end{table*}
\normalsize

\begin{table*}
 \renewcommand{\arraystretch}{1.2} 
\begin{center}
\caption{The values of the phase difference $\delta$ between $T_{11}$ and $T_{00}$ amplitudes   
calculated according to (\ref{eq:cosdelta}) for the proton  and 
deuteron in $Q^2$ bins  
defined by the limits 0.5, 1.0, 1.4, 2.0, and~7.0 GeV$^2$.
The first uncertainties are statistical, the second systematic.}
\label{tab-phase}
\newcommand{\m}{\hphantom{$-$}}
\newcommand{\cc}[1]{\multicolumn{1}{c}{#1}}
\renewcommand{\arraystretch}{1.2} 
\begin{footnotesize}
\begin{tabular}{|l|c|c|c|c|}
\hline
 target &  $\langle Q^2 \rangle =0.82$~GeV$^2$ & $\langle Q^2 \rangle =1.19$~GeV$^2$ & $\langle Q^2 \rangle =1.66$~GeV$^2$ &$\langle Q^2 \rangle =3.06$~GeV$^2$ \\
\hline 
proton &  36.17 $\pm$ 12.33 $\pm$  7.09  &  18.76 $\pm$ 6.99 $\pm$ 5.62   &  26.52 $\pm$  3.92 $\pm$ 0.91  & 36.53 $\pm$ 2.79 $\pm$ 3.65   \\
deuteron    & 33.18  $\pm$  4.55  $\pm$  9.88  &  22.55 $\pm$ 4.07  $\pm$ 2.83  & 30.37  $\pm$  3.13  $\pm$  0.54 & 36.58 $\pm$ 2.28 $\pm$ 3.99  \\
  \hline
\end{tabular} \\[2pt]
\end{footnotesize}
\end{center}
\end{table*}

\begin{table*}
 \renewcommand{\arraystretch}{1.2} 
\begin{center}
\caption{Values in different kinematic bins of the  variable 
$u_1=1-r_{00}^{04} + 2 r_{1-1}^{04} - 2 r_{11}^{1} - 2 r_{1-1}^{1}$, used for the test
of NPE dominance, for proton and 
deuteron data.  The first uncertainties are statistical, the second systematic.}
\label{tab-npe}
\newcommand{\m}{\hphantom{$-$}}
\newcommand{\cc}[1]{\multicolumn{1}{c}{#1}}
\renewcommand{\arraystretch}{1.2} 
\begin{footnotesize}
\begin{tabular}{|c|c|c|c|}
\hline
bin  & $u_1$ proton & $u_1$ deuteron    \\
\hline
$0.5~{\rm GeV}^2 < Q^{2} < 1$ GeV$^2$  &   0.114   $\pm$   0.053   $\pm$   0.045 & 0.104   $\pm$    0.035   $\pm$   0.061 \\
$1.0~{\rm GeV}^2 < Q^{2} < 1.4$ GeV$^2$  &   0.148   $\pm$   0.035   $\pm$   0.044 & 0.069   $\pm$    0.026   $\pm$   0.048 \\
$1.4~{\rm GeV}^2 < Q^{2} < 2.0 $ GeV$^2$ &   0.063   $\pm$   0.037   $\pm$   0.077 & 0.078   $\pm$    0.028   $\pm$   0.028 \\
$2.0~{\rm GeV}^2 < Q^{2} < 7$ GeV$^2$  &   0.178   $\pm$   0.038   $\pm$   0.040 & 0.169   $\pm$    0.032   $\pm$   0.024 \\
\hline                                                                                         
$0.0~{\rm GeV}^2 < -t' < 0.04 $ GeV$^2$   &   0.197   $\pm$   0.043   $\pm$   0.035 & 0.091   $\pm$    0.029   $\pm$   0.024 \\
$0.04~{\rm GeV}^2 < -t' < 0.10$ GeV$^2$   &   0.090   $\pm$   0.040   $\pm$   0.041 & 0.082   $\pm$    0.032   $\pm$   0.039 \\
$0.10~{\rm GeV}^2 < -t' < 0.20$ GeV$^2$   &   0.073   $\pm$   0.041   $\pm$   0.078 & 0.097   $\pm$    0.033   $\pm$   0.068 \\
$0.20~{\rm GeV}^2 < -t' < 0.40$ GeV$^2$   &   0.125   $\pm$   0.040   $\pm$   0.107 & 0.077   $\pm$    0.036   $\pm$   0.095 \\
\hline                                                                                         
$0.0 < x_B < 0.05$   &             0.142   $\pm$   0.099   $\pm$   0.005 & 0.180   $\pm$    0.063   $\pm$   0.075 \\
$0.05 < x_B < 0.08$  &             0.123   $\pm$   0.037   $\pm$   0.029 & 0.070   $\pm$    0.027   $\pm$   0.031 \\
$0.08 < x_B < 0.35$  &             0.152   $\pm$   0.031   $\pm$   0.055 & 0.149   $\pm$    0.027   $\pm$   0.039 \\
\hline
\end{tabular} \\[2pt]
\end{footnotesize}
\label{npe-u1-kinem}
\end{center}
\end{table*}
%

\begin{table*}[hbtc!]
 \renewcommand{\arraystretch}{1.2} 
\begin{center}
\caption{\label{npe-tests} 
Results on $u_1$, $u_2$ and $u_3$, 
calculated according to (\ref{eq:req-npe}-\ref{eq:u3def}), 
shown together with average value or range in $Q^2$ and $W$. 
Top section: HERMES results from proton and deuteron data, 
shown with statistical and systematic uncertainties
separately. Bottom section: results from other experiments 
calculated from published SDMEs, with 
statistical and systematic uncertainties 
combined in quadrature without accounting for correlations
between the SDMEs. 
}
\begin{tabular}{|c|c|c|c|c|c|}
\hline
 Experiment & $Q^2$, GeV$^2$ & W, GeV & $u_1$ & $u_2$ & $u_3$  \\ 
\hline
HERMES $p$ & 1.95 & 4.8 & 0.125 $\pm$ 0.021 $\pm$ 0.050 & $-0.011 \pm$ 0.004 $\pm$ 0.012 & 0.055 $\pm$ 0.045 $\pm$ 0.006 \\
HERMES $d$ & 1.94 & 4.8 & 0.091 $\pm$ 0.016 $\pm$ 0.046 & $-0.008 \pm$ 0.003 $\pm$ 0.010 & $-0.040 \pm$ 0.035 $\pm$ 0.007 \\
\hline
ZEUS DIS~\cite{zeussdme2} & 2.4  & 90  & 0.018 $\pm$ 0.066   & 0.018  $\pm$  0.011  & \\
ZEUS BPC~\cite{zeussdme}  & 0.41  & 45  &  0.058 $\pm$ 0.078     & $-0.002 \pm$  0.016 & \\
H1~\cite{H1}         &  2.5--3.5 & 30--100 & 0.065 $\pm$ 0.15  & $-0.017  \pm$  0.034 & \\
 SLAC~\cite{SLAC1}    &  0.9          & 3.14           & 0.85 $\pm$ 0.32   & $-0.050  \pm$  0.072 & \\
 SLAC~\cite{SLAC3}   &  0.9          & 3.14        &  1.174 $\pm$ 0.379     & 0.039  $\pm$  0.082  & \\
  DESY~\cite{DESY2}  & 1.05            & 2  --  2.8   & 0.73 $\pm$ 0.33 &  $-0.040  \pm$  0.064  & \\
  \hline
\end{tabular}  \\[2pt]
\end{center}
\end{table*}


\begin{table*}[hbtc!]
  \renewcommand{\arraystretch}{1.2} 
\begin{center}
\caption{ \label{tau-tests} 
Ratios of certain helicity-flip amplitudes to the 
square root of the sum of all amplitudes squared:  
$\tau_{01}$  for the transition  $\gamma^*_T \to \rho^0_L$,
 $\tau_{10}$  for the transition  $\gamma^*_L \to \rho^0_T$, and
 $\tau_{1-1}$  for the transition  $\gamma^*_{-T} \to \rho^0_T$.
Top section: HERMES results from proton and deuteron data 
calculated according to (\ref{tau01f}-\ref{tau1m1f}),
shown with statistical and systematic uncertainties separately.
Bottom section: results from other experiments calculated according 
to (\ref{tau01}-\ref{tau1m1}), with 
statistical and 
systematic uncertainties combined in quadrature without accounting for correlations
between the SDMEs. 
}
\begin{tabular}{|c|c|c|c|}
\hline
HERMES & $\tau_{01}$ & $\tau_{10}$  &   $\tau_{1-1}$ \\
\hline 
proton & 0.114 $\pm$ 0.007 $\pm$ 0.010 & 0.075 $\pm$  0.030 $\pm$  0.003 & 0.051 $\pm$ 0.029 $\pm$ 0.010 \\
deuteron & 0.122 $\pm$ 0.007 $\pm$ 0.006 & 0.090 $\pm$ 0.022 $\pm$  0.011 & 0.007 $\pm$ 0.025 $\pm$ 0.015 \\
  \hline
Experiment &  $\widetilde{\tau}_{01}$ & $\widetilde{\tau}_{10}$  &  $\widetilde{\tau}_{1-1}$ \\ 
\hline 
ZEUS BPC\cite{zeussdme} & 0.069  $\pm$  0.027 & 0.003   $\pm$  0.029 & 0.048   $\pm$  0.028 \\
ZEUS DIS\cite{zeussdme2} & 0.078  $\pm$  0.016 & $-0.010 \pm$ 0.028 & 0.013   $\pm$ 0.032  \\
H1~\cite{H1}         &   0.088 $\pm$  0.036   & 0.019 $\pm$  0.065 & 0.035 $\pm$  0.109  \\
 SLAC~\cite{SLAC1}  & 0.095   $\pm$  0.165  & 0.030   $\pm$  0.133  & 0.112   $\pm$  0.231  \\
 SLAC~\cite{SLAC3}  & 0.084   $\pm$  0.177  & 0.412   $\pm$  0.430  & 0.042  $\pm$  0.389 \\
  DESY~\cite{DESY2} & 0.041   $\pm$  0.247  & 0.335   $\pm$  0.436  &    \\
  \hline
\end{tabular} \\[2pt]
\end{center}
\end{table*}

\begin{table*}
 \renewcommand{\arraystretch}{1.2} 
\begin{center}
\caption{
The longitudinal-to-transverse cross section ratios $R^{04}$, $R$, and $R^{NPE}$ 
for the proton and deuteron in $Q^2$ bins defined by the limits 
0.5, 1.0, 1.4, 2.0, and~7.0 GeV$^2$. The total uncertainties are shown.}
 \label{rat-tab}
\newcommand{\m}{\hphantom{$-$}}
\newcommand{\cc}[1]{\multicolumn{1}{c}{#1}}
\renewcommand{\arraystretch}{1.2} 
\begin{footnotesize}

\begin{tabular}{|c|c|c|c|c|c|}
\hline

Ratio & Target &  $\langle Q^2 \rangle =0.82$~GeV$^2$ & $\langle Q^2 \rangle =1.19$~GeV$^2$ & $\langle Q^2 \rangle =1.66$~GeV$^2$ &$\langle Q^2 \rangle =3.06$~GeV$^2$ \\
\hline 
$R^{04}$ &  proton & 0.694 $\pm$ 0.187 & 0.701 $\pm$ 0.063 & 0.798 $\pm$ 0.080 & 1.053 $\pm$ 0.074   \\
   & deuteron    & 0.748 $\pm$ 0.179 & 0.755 $\pm$ 0.063 &  0.973 $\pm$ 0.098 & 0.870 $\pm$ 0.061  \\
  \hline
$R$ &  proton   & 0.649 $\pm$ 0.188 & 0.671 $\pm$ 0.065 & 0.794 $\pm$ 0.083 & 1.068 $\pm$  0.087 \\
   & deuteron & 0.677 $\pm$ 0.180 & 0.583 $\pm$ 0.067 & 0.958 $\pm$  0.102 & 0.922 $\pm$ 0.074 \\
  \hline

$R^{NPE}$ & proton &  0.755 $\pm$  0.190 & 0.783 $\pm$ 0.068 & 0.840 $\pm$ 0.090 &  1.225  $\pm$ 0.084 \\
   & deuteron & 0.809 $\pm$   0.182 & 0.798 $\pm$ 0.067 & 1.041 $\pm$ 0.102 & 0.994 $\pm$  0.065 \\

  \hline
\end{tabular} \\[2pt]
\end{footnotesize}
\end{center}
\end{table*}


\begin{table*}
 \renewcommand{\arraystretch}{1.2} 
\begin{center}
\caption{  \label{mdiehl} The SDME results from the proton data,  
integrated over the entire HERMES kinematic range, presented in the
notation of Ref.~\cite{mdiehl}.
The first uncertainties are statistical and the second systematic.
}
\begin{footnotesize}  
\begin{tabular}{|r@{\;=\;}l|r|} \hline

$u^{00}_{++}+\epsilon u^{00}_{00}$ & $r^{04}_{00}$                    &  0.412 $\pm$  0.010 $\pm$  0.010  \\
$ \mathrm{ Re \; }(u^{0+}_{0+}- u^{-0}_{0+})$ & $\sqrt{2}( \mathrm{ Im \;} r^6_{10}- r^5_{10})$ & $-$0.464 $\pm$  0.005 $\pm$  0.028 \\
$u^{++}_{++}+u^{--}_{++}+2 \epsilon u^{++}_{00}$ & $1-r^{04}_{00}$    &  0.588 $\pm$  0.010 $\pm$  0.010 \\
$u^{-+}_{-+} $ & $r^1_{1-1}- \mathrm{ Im \;} r^2_{1-1}$                      &  0.473  $\pm$  0.012 $\pm$  0.029 \\
$\mathrm{ Re \; }u^{00}_{0+} $ & $-r^5_{00}/\sqrt{2}$                       & $-$0.077 $\pm$ 0.006 $\pm$  0.006 \\
$u^{0+}_{++}-u^{-0}_{++}+2 \epsilon u^{0+}_{00}$ & $2 \mathrm{ Re \;}r^{04}_{10}$  &  0.062 $\pm$  0.008 $\pm$  0.016 \\
$\mathrm{ Re \;} u^{0+}_{-+} $ & $\mathrm{ Re\; } r^1_{10}- \mathrm{ Im \;} r^2_{10}$                           & $-$0.054 $\pm$  0.009 $\pm$  0.027\\
$\mathrm{ Re \;} (u^{0-}_{0+}-u^{+0}_{0+})$ & $ \sqrt{2}( \mathrm{ Im \;} r^6_{10}+ \mathrm{ Re \;} r^5_{10})$   & $-$0.008 $\pm$  0.005 $\pm$  0.015  \\
$\mathrm{ Re \;} (u^{-+}_{++}+\epsilon u^{-+}_{00})$ & $r^{04}_{1-1}$               & $-$0.011 $\pm$  0.005 $\pm$  0.005  \\
$\mathrm{ Re \;} u^{++}_{-+}$ & $r^{1}_{11}$                                        & $-$0.025 $\pm$  0.007 $\pm$  0.008  \\
$\mathrm{ Re \;} (u^{++}_{0+}+u^{--}_{0+}) $ & $ -\sqrt{2} r^{5}_{11}$                & $-$0.023 $\pm$  0.004 $\pm$  0.018  \\
$\mathrm{ Re \;} u^{-+}_{0+} $ & $ ( \mathrm{ Im \;} r^6_{1-1}- r^5_{1-1})/\sqrt{2}$    & $-$0.005 $\pm$  0.005 $\pm$  0.008  \\
$u^{00}_{-+}$ & $ r^1_{00}$                                            & 0.011 $\pm$  0.019 $\pm$  0.008  \\
$\mathrm{ Re \;} u^{+0}_{-+}$ & $ \mathrm{ Re \;} r^1_{10}+\mathrm{ Im \;} r^2_{10}$                         & 0.009 $\pm$  0.009 $\pm$  0.004  \\
$\mathrm{ Re \;} u^{+-}_{0+}$ & $ -( \mathrm{ Im \;} r^6_{1-1}+ r^5_{1-1})/\sqrt{2}$               & $-$0.002 $\pm$  0.005 $\pm$  0.001  \\
$\mathrm{ Re \;} u^{+-}_{-+}$ & $ r^1_{1-1}+\mathrm{ Im \;} r^{2}_{1-1}$                          & 0.018 $\pm$ 0.012 $\pm$  0.011  \\
\hline
$\mathrm{ Im \;} (u^{0+}_{0+}-u^{-0}_{0+})$ & $\mathrm{ Im \; } r^7_{10}+\mathrm{ Re \;} r^8_{10}$            & 0.264 $\pm$  0.030 $\pm$  0.023  \\
$\mathrm{ Im \;} u^{00}_{0+}$ & $r^8_{00}/\sqrt{2}$                               & 0.025 $\pm$  0.035 $\pm$  0.007  \\
$\mathrm{ Im \;} (u^{0+}_{++}-u^{-0}_{=+}) $ & $ -2\mathrm{ Im \;} r^3_{10}$                      & 0.034 $\pm$  0.030 $\pm$  0.008  \\
$\mathrm{ Im \;} (u^{0-}_{0+}-u^{+0}_{0+})$ & $\sqrt{2}(\mathrm{ Im \;} r^7_{10}-\mathrm{ Re \;} r^8_{10})$   & 0.054 $\pm$  0.035 $\pm$  0.008  \\
$\mathrm{ Im \;} u^{-+}_{++}$ & $-\mathrm{ Im \;} r^3_{1-1}$                                   & 0.024 $\pm$  0.018  $\pm$  0.001 \\
$\mathrm{ Im \;} (u^{++}_{0+}+u^{--}_{0+})$ & $\sqrt{2}r^8_{11} $                   & 0.51 $\pm$  0.034 $\pm$  0.001 \\
$u^{-+}_{0+}$ & $(r^8_{1-1}+\mathrm{ Im \;} r^7_{1-1})/\sqrt{2}$                  & $-$0.012 $\pm$  0.039 $\pm$  0.007 \\
$u^{+-}_{0+}$ & $ (r^8_{1-1}-\mathrm{ Im \;} r^7_{1-1})/\sqrt{2}$                    & 0.038 $\pm$  0.039 $\pm$  0.001\\
\hline
\end{tabular}
\end{footnotesize}
\label{mdiehl-tab}
\vspace*{120.cm}
\end{center}
\end{table*}

%% file: tab-corr-doublepage.tex
\begin{table*}
\caption{\small The correlation matrix for the 23 SDMEs
obtained from the hydrogen target data. The column headings do not indicate the real and imaginary parts of any
SDMEs in order to keep the table compact.
}
\renewcommand{\arraystretch}{1.3}
\setlength{\tabcolsep}{3pt}
\hspace*{2.0cm}
\begin{sideways}
\begin{footnotesize}
\begin{tabular}{|l|c|c|c|c|c|c|c|c|c|c|c|c|c|c|c|c|c|c|c|c|c|c|c| }
\hline
SDME & $r^{04}_{00}$  & $r^{04}_{10}$ & $r^{04}_{1-1}$  & $r^1_{11}$ & $r^1_{00}$ & $r^1_{10}$ & $r^1_{1-1}$ & $r^2_{10}$ & $r^2_{1-1}$ & $r^5_{11}$ & $r^5_{00}$ & $r^5_{10}$ & $r^5_{1-1}$ & $r^6_{10}$ & $r^6_{1-1}$ & $r^3_{10}$ & $r^3_{1-1}$ & $r^7_{10}$ & $r^7_{1-1}$ & $r^8_{11}$ & $r^8_{00}$ & $r^8_{10}$ & $r^8_{1-1}$ \\
\hline
$r^{04}_{00}$   & 1.00& \\
Re $r^{04}_{10}$& 0.16& 1.00& \\
$r^{04}_{1-1}$  &-0.05&-0.01& 1.00& \\
$r^1_{11}$      &-0.04&-0.01& 0.62& 1.00& \\
$r^1_{00}$      &-0.02& 0.12&-0.18&-0.48& 1.00& \\
Re $r^1_{10}$   & 0.00&-0.41&-0.08&-0.03& 0.02& 1.00& \\
$r^1_{1-1}$     &-0.65&-0.22&-0.05& 0.00& 0.00& 0.02& 1.00& \\
Im $r^2_{10}$   & 0.02& 0.33&-0.03&-0.03& 0.07&-0.17&-0.10& 1.00& \\
Im $r^2_{1-1}$  & 0.54& 0.23& 0.00&-0.03& 0.01&-0.11&-0.32& 0.07& 1.00& \\
$r^5_{11}$      &-0.17& 0.15& 0.02&-0.04& 0.09&-0.03& 0.10& 0.07&-0.07& 1.00& \\
$r^5_{00}$      & 0.37& 0.01& 0.06& 0.09&-0.26& 0.02&-0.27& 0.00& 0.20&-0.50& 1.00& \\
Re $r^5_{10}$   &-0.10& 0.08&-0.13&-0.12& 0.06&-0.18&-0.03& 0.08& 0.04&-0.01&-0.12& 1.00& \\
$r^5_{1-1}$     &-0.05&-0.15& 0.10& 0.10&-0.09& 0.01& 0.04& 0.02&-0.08&-0.43& 0.08&-0.04& 1.00& \\
Im $r^6_{10}$   & 0.15&-0.02&-0.08&-0.05&-0.06& 0.20&-0.01&-0.21& 0.00&-0.05& 0.19&-0.34& 0.04& 1.00& \\
Im $r^6_{1-1}$  & 0.04& 0.16&-0.07&-0.13& 0.02& 0.06&-0.09& 0.06& 0.05& 0.46&-0.09&-0.05&-0.27&-0.04& 1.00& \\
Im $r^3_{10}$   & 0.07& 0.00& 0.03& 0.00&-0.05& 0.04&-0.05&-0.03& 0.01&-0.02& 0.06&-0.05& 0.02& 0.07& 0.00& 1.00& \\
Im $r^3_{1-1}$  & 0.00& 0.03& 0.02& 0.00& 0.00& 0.00&-0.01&-0.01& 0.00&-0.01& 0.00&-0.06& 0.00& 0.01& 0.04&-0.02& 1.00& \\
Im $r^7_{10}$   & 0.08& 0.06& 0.07& 0.00&-0.09&-0.02&-0.12& 0.03& 0.04&-0.02& 0.00&-0.08&-0.03& 0.08& 0.00& 0.51&-0.11& 1.00& \\
Im $r^7_{1-1}$  &-0.05&-0.05& 0.01& 0.01&-0.01& 0.05& 0.00& 0.00& 0.00&-0.01& 0.01& 0.02& 0.02& 0.00& 0.02&-0.13& 0.24& 0.04& 1.00& \\
$r^8_{11}$      &-0.11&-0.04&-0.01& 0.05&-0.03& 0.00&-0.02&-0.02& 0.00& 0.01&-0.01& 0.04&-0.01&-0.05& 0.00&-0.19&-0.10& 0.03& 0.39& 1.00& \\
$r^8_{00}$      & 0.11& 0.07&-0.01&-0.03& 0.05& 0.01&-0.03&-0.03& 0.04& 0.00&-0.01&-0.05& 0.00& 0.05& 0.00&-0.05&-0.02& 0.03&-0.12&-0.51& 1.00&\\
Re $r^8_{10}$   & 0.12&-0.02&-0.07&-0.03& 0.10&-0.03&-0.11&-0.05& 0.06& 0.02& 0.00&-0.10&-0.03& 0.09& 0.08& 0.10& 0.15&-0.07&-0.04&-0.06& 0.08& 1.00& \\
Im $r^8_{1-1}$     &-0.06&-0.06& 0.00& 0.03&-0.01&-0.02&-0.04&-0.05&-0.01& 0.00& 0.01& 0.00&-0.01&-0.01& 0.03&-0.19&-0.06& 0.04& 0.17& 0.46&-0.16&-0.10& 1.00 \\
\hline
SDME & $r^{04}_{00}$  & $r^{04}_{10}$ & $r^{04}_{1-1}$  & $r^1_{11}$ & $r^1_{00}$ & $r^1_{10}$ & $r^1_{1-1}$ & $r^2_{10}$ & $r^2_{1-1}$ & $r^5_{11}$ & $r^5_{00}$ & $r^5_{10}$ & $r^5_{1-1}$ & $r^6_{10}$ & $r^6_{1-1}$ & $r^3_{10}$ & $r^3_{1-1}$ & $r^7_{10}$ & $r^7_{1-1}$ & $r^8_{11}$ & $r^8_{00}$ & $r^8_{10}$ & $r^8_{1-1}$  \\
\hline
\end{tabular}
\end{footnotesize}
\end{sideways}
\label{tab-corr-h}
\end{table*}


\begin{table*}
\caption{\small The correlation matrix for the 23 SDMEs
obtained from the deuterium target data. The column headings do not indicate the real and imaginary parts of any
SDMEs in order to keep the table compact.
}
\renewcommand{\arraystretch}{1.3}
\setlength{\tabcolsep}{3pt}
\hspace*{2.0cm}
\begin{sideways}
\begin{footnotesize}
\begin{tabular}{|l|c|c|c|c|c|c|c|c|c|c|c|c|c|c|c|c|c|c|c|c|c|c|c| }
\hline
SDME & $r^{04}_{00}$  & $r^{04}_{10}$ & $r^{04}_{1-1}$  & $r^1_{11}$ & $r^1_{00}$ & $r^1_{10}$ & $r^1_{1-1}$ & $r^2_{10}$ & $r^2_{1-1}$ & $r^5_{11}$ & $r^5_{00}$ & $r^5_{10}$ & $r^5_{1-1}$ & $r^6_{10}$ & $r^6_{1-1}$ & $r^3_{10}$ & $r^3_{1-1}$ & $r^7_{10}$ & $r^7_{1-1}$ & $r^8_{11}$ & $r^8_{00}$ & $r^8_{10}$ & $r^8_{1-1}$  \\
\hline
$r^{04}_{00}$   & 1.00& \\
Re $r^{04}_{10}$& 0.12& 1.00& \\
$r^{04}_{1-1}$  &-0.07&-0.01& 1.00& \\
$r^1_{11}$      &-0.05& 0.00& 0.61& 1.00& \\
$r^1_{00}$      &-0.03& 0.10&-0.16&-0.48& 1.00& \\
Re $r^1_{10}$   & 0.04&-0.39&-0.10&-0.02& 0.05& 1.00& \\
$r^1_{1-1}$     &-0.60&-0.18& 0.03& 0.04& 0.02&-0.04& 1.00& \\
Im $r^2_{10}$   &-0.01& 0.32&-0.04&-0.02& 0.04&-0.14&-0.06& 1.00& \\
Im $r^2_{1-1}$  & 0.52& 0.19&-0.04&-0.03&-0.03&-0.06&-0.27& 0.04& 1.00& \\
$r^5_{11}$      &-0.17& 0.15& 0.04&-0.05& 0.12&-0.02& 0.12& 0.07&-0.05& 1.00& \\
$r^5_{00}$      & 0.39& 0.04& 0.06& 0.10&-0.29&-0.01&-0.26&-0.01& 0.22&-0.51& 1.00& \\
Re $r^5_{10}$   &-0.12& 0.06&-0.15&-0.11& 0.00&-0.20&-0.02& 0.04& 0.01& 0.02&-0.14& 1.00& \\
$r^5_{1-1}$     &-0.06&-0.17& 0.08& 0.08&-0.09& 0.01& 0.00&-0.01&-0.08&-0.42& 0.08&-0.04& 1.00& \\
Im $r^6_{10}$   & 0.17& 0.00&-0.04&-0.03& 0.00& 0.19&-0.03&-0.19& 0.02&-0.05& 0.18&-0.36& 0.05& 1.00& \\
Im $r^6_{1-1}$  & 0.02& 0.15&-0.06&-0.09& 0.01& 0.04&-0.08& 0.08& 0.11& 0.45&-0.09&-0.02&-0.24&-0.07& 1.00& \\
Im $r^3_{10}$   & 0.05& 0.01& 0.02&-0.01&-0.01& 0.02&-0.05&-0.02& 0.02& 0.00& 0.03&-0.04& 0.02& 0.06&-0.04& 1.00& \\
Im $r^3_{1-1}$  &-0.03&-0.01&-0.03&-0.04& 0.00&-0.01&-0.01& 0.03& 0.02&-0.01&-0.01&-0.06&-0.01&-0.04& 0.07&-0.04& 1.00& \\
Im $r^7_{10}$   & 0.04& 0.02& 0.05& 0.00&-0.06&-0.01&-0.06&-0.01& 0.04& 0.03&-0.03&-0.03&-0.02& 0.04&-0.01& 0.48&-0.11& 1.00& \\
Im $r^7_{1-1}$  &-0.04&-0.04&-0.01&-0.02&-0.01& 0.05&-0.02& 0.03& 0.02& 0.02&-0.01&-0.03&-0.03& 0.00& 0.03&-0.14& 0.24& 0.01& 1.00& \\
$r^8_{11}$      &-0.08&-0.02& 0.00& 0.05&-0.05& 0.02& 0.00&-0.03& 0.02& 0.00& 0.00& 0.02& 0.00&-0.02&-0.01&-0.16&-0.09& 0.03& 0.40& 1.00& \\
$r^8_{00}$      & 0.11& 0.08&-0.02&-0.05& 0.12& 0.03&-0.06& 0.00& 0.02& 0.00&-0.03&-0.08&-0.01& 0.08&-0.01&-0.04&-0.04& 0.04&-0.12&-0.50& 1.00& \\
Re $r^8_{10}$   & 0.10& 0.07&-0.09&-0.03& 0.13& 0.03&-0.13& 0.01& 0.06&-0.01&-0.01&-0.12& 0.00& 0.10&-0.01& 0.09& 0.11&-0.10&-0.03&-0.02& 0.04& 1.00& \\
Im $r^8_{1-1}$     &-0.04&-0.05& 0.03& 0.04&-0.05&-0.03& 0.03&-0.05& 0.00& 0.02& 0.01& 0.02&-0.01&-0.02& 0.00&-0.16&-0.07& 0.06& 0.20& 0.47&-0.13&-0.09& 1.00 \\
\hline
SDME & $r^{04}_{00}$  & $r^{04}_{10}$ & $r^{04}_{1-1}$  & $r^1_{11}$ & $r^1_{00}$ & $r^1_{10}$ & $r^1_{1-1}$ & $r^2_{10}$ & $r^2_{1-1}$ & $r^5_{11}$ & $r^5_{00}$ & $r^5_{10}$ & $r^5_{1-1}$ & $r^6_{10}$ & $r^6_{1-1}$ & $r^3_{10}$ & $r^3_{1-1}$ & $r^7_{10}$ & $r^7_{1-1}$ & $r^8_{11}$ & $r^8_{00}$ & $r^8_{10}$ & $r^8_{1-1}$  \\
\hline
\end{tabular}
\end{footnotesize}
\end{sideways}
\label{tab-corr-d}
\end{table*}